\documentclass[11pt, a4paper]{article}
\usepackage{fullpage}
\usepackage{amsfonts}
\usepackage{amssymb}
\usepackage{amsmath}
\usepackage{amsthm}
\usepackage{graphicx}
\usepackage{bm}
\usepackage[makeroom]{cancel}
\usepackage{enumitem}
\usepackage{url}
\usepackage[margin=.9in]{geometry}
\usepackage{amsopn}
\usepackage{mathtools}
\usepackage{hyperref}
\usepackage{doi}
\usepackage{cite}
\usepackage[symbol]{footmisc}
\usepackage{subcaption}

\newcommand{\nekpp}{{\em Nektar++} }
\newcommand{\spectralhp}{spectral/$hp$ element }

\usepackage{listings}
\usepackage{xcolor}
\definecolor{codegreen}{rgb}{0,0.6,0}
\definecolor{codegray}{rgb}{0.5,0.5,0.5}
\definecolor{codepurple}{rgb}{0.58,0,0.82}
\definecolor{backcolour}{rgb}{0.95,0.95,0.92}
\lstdefinestyle{mystyle}{
    backgroundcolor=\color{backcolour},   
    commentstyle=\color{codegreen},
    keywordstyle=\color{magenta},
    numberstyle=\tiny\color{codegray},
    stringstyle=\color{codepurple},
    basicstyle=\ttfamily\normalsize,
    breakatwhitespace=false,         
    breaklines=true,                 
    captionpos=b,                    
    keepspaces=true,                 
    numbers=left,                    
    numbersep=3pt,                  
    showspaces=false,                
    showstringspaces=false,
    showtabs=false,                  
    tabsize=4,
}
\begin{document}
\begin{center}
    \LARGE \bf Scalable Modal Analysis of the\\ Imperial Front Wing Flow Field
\end{center}
\begin{center}
    Parv Khurana$^{a,b}$\footnote[1]{co-first author}\footnote[2]{Corresponding author: \href{mailto:p.khurana22@imperial.ac.uk}{p.khurana22@imperial.ac.uk}},
    Benet Eiximeno$^{c*}$,
    Alexandra Liosi$^{a,b}$,
   Spencer Sherwin$^{a}$,
   Athanasios Chatzopoulos$^{b}$,
   Julien Hoessler$^{b}$,
   Arnau Miró$^{c,d}$,
    Oriol Lehmkuhl$^{c}$,
    Ivette Rodriguez$^{d}$,
    J. Nathan Kutz$^{e}$
\end{center}
\begin{center}
    \scriptsize{
    ${}^a$ Department of Aeronautics, Imperial College London, SW7 2AZ London, UK \\
    ${}^b$ CFD Technology Group, McLaren Racing Limited, GU21 4YH Woking, UK \\
    ${}^c$ Barcelona Supercomputing Center, Plaça d'Eusebi Güell 1-3, 08034 Barcelona, Spain\\
   ${}^d$ Universitat Politècnica de Catalunya, Carrer de Jordi Girona 31, 08034 Barcelona, Spain \\
    ${}^e$ Autodesk Research, 6 Agar Street, WC2N 4HN, London, UK 
}
\end{center}

\begin{abstract}
%
The aerodynamic performance of motorsport vehicles relies on understanding the unsteady flow dynamics around complex geometries at high Reynolds numbers. Extracting the dominant 3D coherent structures from high-fidelity simulations is often intractable due to the high dimensionality of the state-space, and standard modal decompositions, such as \emph{proper orthogonal decomposition} (POD), do not give interpretable meaning to the temporal dynamics of the modes. This study demonstrates a scalable POD and \emph{Dynamic Mode Decomposition} (DMD) method, implemented in the open-source pyLOM framework, that can identify coherent structures and their temporal dynamics in high-fidelity simulations of Formula 1 aerodynamics, specifically the Imperial Front Wing (IFW), a publicly available benchmark based on the McLaren MP4-17D, at Re = $2.2 \times 10^5$. DMD is applied to large eddy simulation (LES) results of the flow field that is resolved using spectral/hp element method via Nektar++, producing $49.72 \times 10^6$ degrees of freedom at polynomial order p=4. The simulation captures four dominant vortical structures (endplate, canard, main, and footplate), with time-averaged force coefficients showing 2.24\% and 2.87\% differences in $C_L$ and $C_D$ compared to the reference data. DMD extracts the frequencies, amplitudes and damping ratios of the coherent structures, and modes at chord-based Strouhal numbers of 5.1, 7.5 and 9.3 are associated, through point-probe spectra and mode slices, with the main-element laminar separation bubble, motion of the canard, main and endplate vortex cores, and lateral meander of the endplate vortex, respectively. The DMD allows for a compression scheme, here reducing approximately 6~TB of snapshots to 200~GB of modal data.
\end{abstract}


\section{Introduction}
Scalable modal decomposition methods are critical for  identifying the coherent flow-field structures in motorsport aerodynamics. However, \textit{direct numerical simulations} (DNS) of the incompressible flows at high Reynolds numbers around complex geometries are computationally challenging if not intractable, leading to approximations of the flow physics by low-fidelity methods such as Reynolds-Averaged Navier-Stokes (RANS) using \textit{finite volume} (FV) discretisations.  RANS models, unfortunately,  often fail to capture laminar to turbulent transition and 
boundary layer separation, along with vortex creation, interaction and stability (bursting)\cite{Katz2021} required for design updates. In contrast, {\em large eddy simulations} (LES) are a high-fidelity modelling approach that accurately approximate unsteady flow features and resolve a broader range of flow scales.  As a consequence, LES are useful for understanding the intricacies of complex unsteady flow dynamics and vortical systems around typical motorsport geometries.  The LES presented here results in a state-space dimension of $O(10^7)$, which makes standard modal decomposition methods intractable.  As such, analysis is typically limited to judiciously selected planes of the flow dynamics and/or time-averaged dynamics as the significantly smaller dimensions are tractable.  Our proposed scalable modal decomposition method allows for an analysis of the full 3D spatio-temporal dynamics, allowing for improved insight into the flow physics and the dominant vortical features induced by the geometry.    

\begin{figure}
    \centering
    \includegraphics[width=\linewidth]{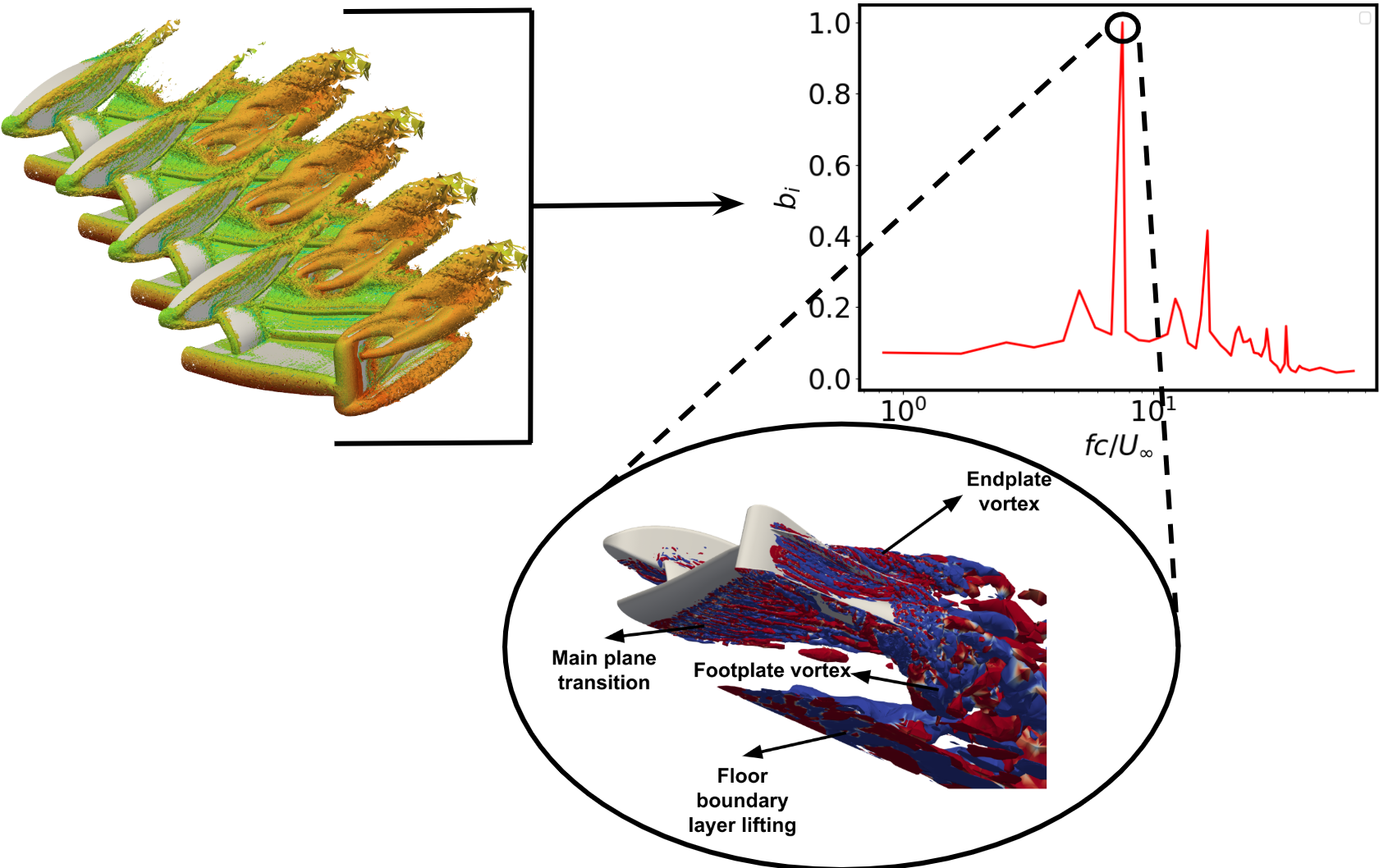}
    \caption{Illustration of the workflow followed in DMD analysis: collection of instantaneous snapshots, DMD computation to detect the dominant frequencies in the flow field and visualisation of the spatial correlations linked to the frequencies.}
    \label{fig:thefigure1}
\end{figure}

\begin{figure}[t]
    \centering
    \subfloat[]{\includegraphics[width=0.5\textwidth]{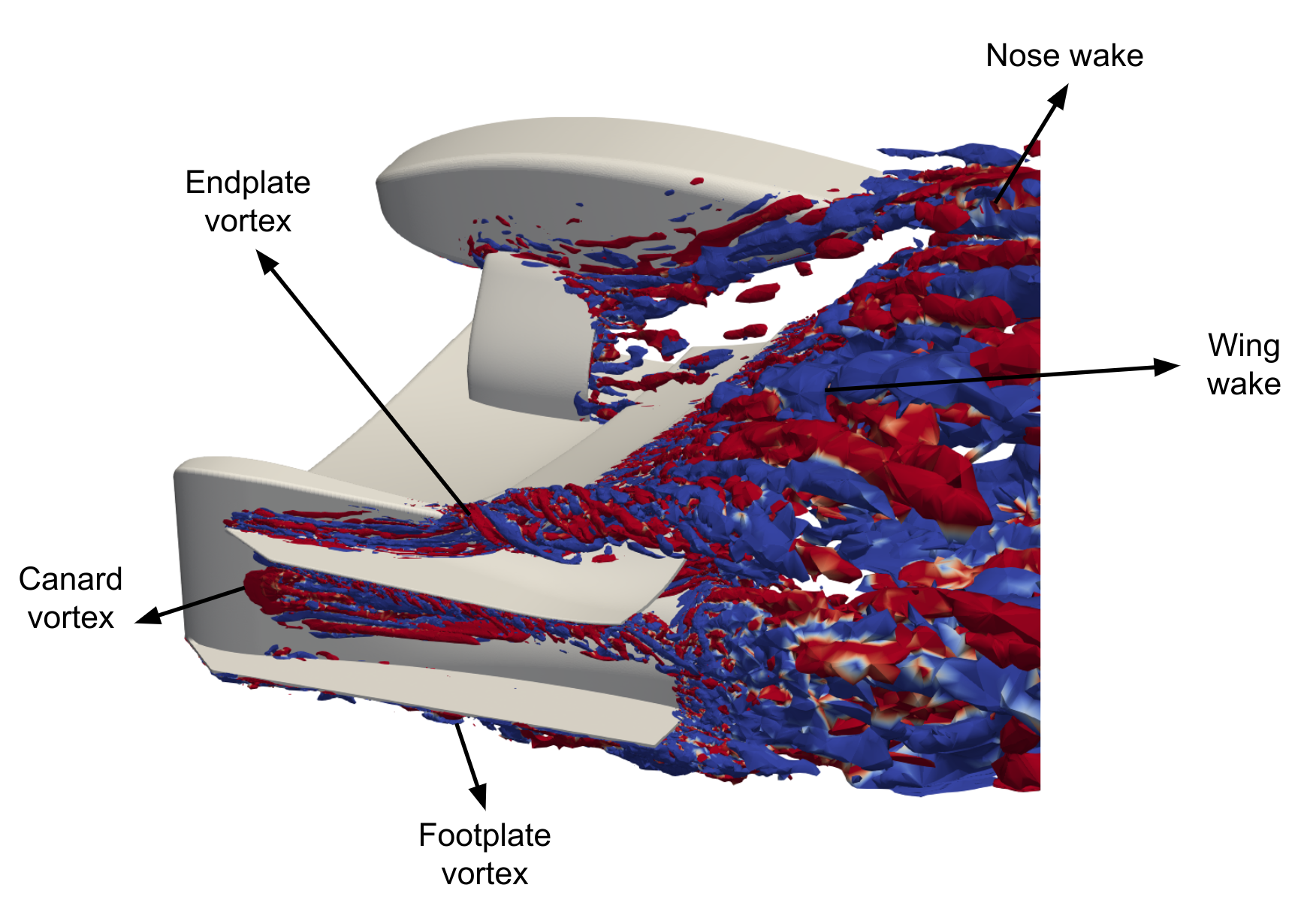}}
    \subfloat[]{\includegraphics[width=0.5\textwidth]{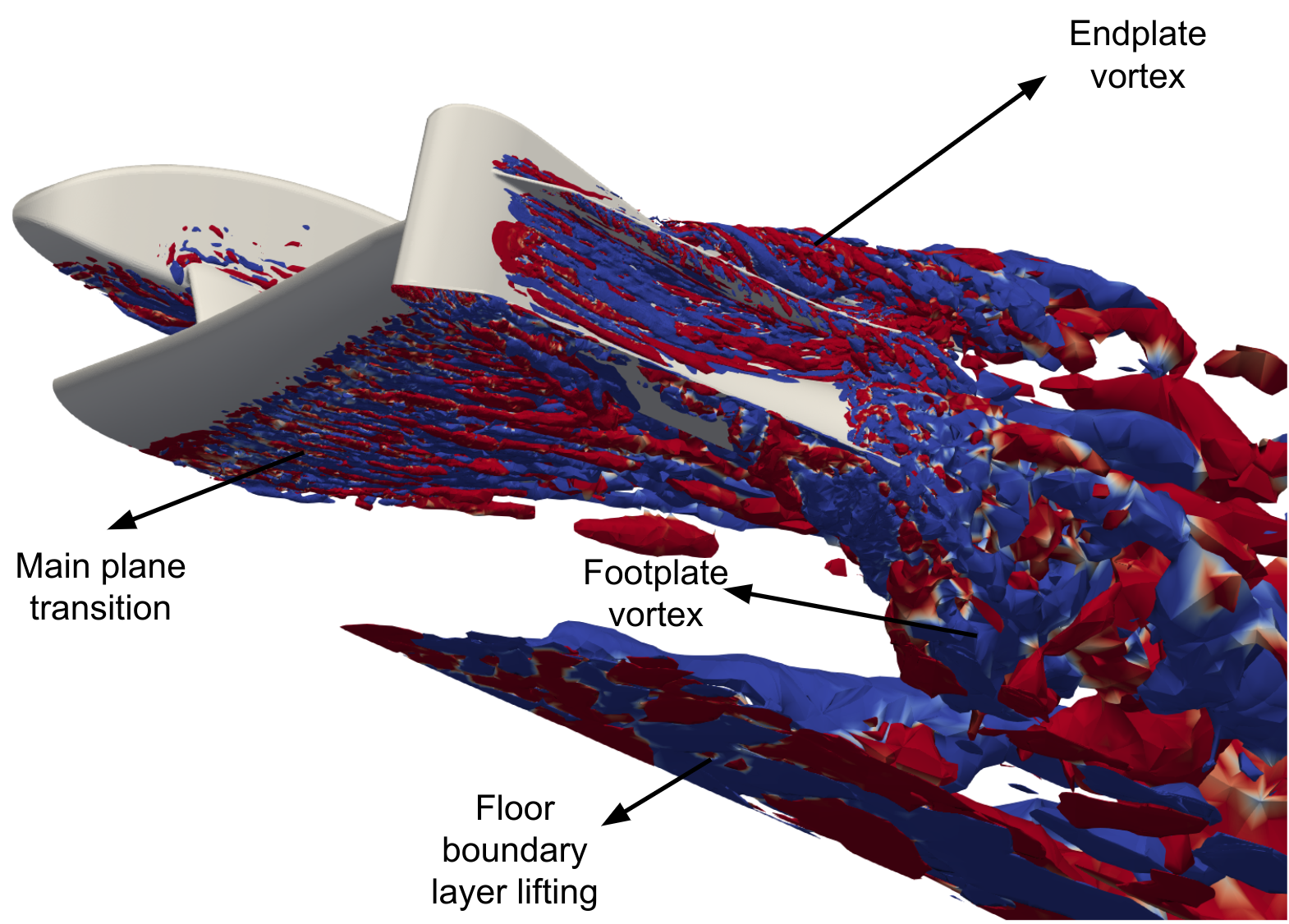}}
    \caption{Spatial structure of a DMD mode of the streamwise velocity, $u$, around the Imperial Front Wing, shown by red and blue isosurfaces. (a) View from the outboard side, showing the nose and wing wakes and the endplate, canard and footplate vortices. (b) View of the underside, showing transition on the main plane, the endplate and footplate vortices, and lifting of the floor boundary layer.}
    \label{fig:dmdfu}
\end{figure}

Integrating high-fidelity numerical methods within an industrial \textit{computational fluid dynamics} (CFD) pipeline can be a performance differentiator in the motorsport context. Not only for its ability to produce more accurate predictions of the mean-flow than any RANS simulation, but also because of the possibility of understanding the transient dynamics of the flow. The latter objective is crucial for adding value to large eddy simulations as its higher accuracy comes together with a significant increase in the computational cost in both solution time and amount of storage needed to save the results, as they typically require several terabytes of disk space.  But even with the capability to simulate at such scales, the analysis for the flow fields remains challenging as the dominant features of the flow must be extracted for human understanding.  What is required is a scalable and systematic methodology for coherent structure identification that relates every correlated region in the domain with its dominant frequency. To resolve this issue, the present work introduces the spatio-temporal decomposition method of {\em dynamic mode decomposition} (DMD)~\cite{schmid_2010} that extracts dominant features of the velocity fluctuations by a principled algorithmic workflow which is represented in \autoref{fig:thefigure1}. This scalable decomposition technique allows one to couple the modal analysis with full state LES, enabling an improved understanding of the critical dominant flow features around a given geometry. It is especially relevant to the analysis of the spatial correlations as the ones represented in \autoref{fig:dmdfu}.

The DMD algorithm \cite{schmid_2010, tu_2014, dmd_book, schmid2022dynamic} seeks to decompose time-varying data sets into a low-rank coherent set of spatiotemporal structures.  Specifically, the temporal evolution is assumed to be approximated by a linear dynamical system $\dot{\mathbf{x}}(t) = \mathbf{A} \mathbf{x}(t)$ whose evolution dynamics are exponentials of the form
\begin{equation}
    \mathbf{x}(t) = \sum_{j=1}^N b_j \boldsymbol{\phi}_j \exp(\omega_j t)
    \label{eq:DMDsolution}
\end{equation}
with the imaginary component of the exponent $\omega_j$ modelling periodic temporal oscillations.  There are now many well-developed variants of the DMD algorithm that can be exploited~\cite{ichinaga2024pydmd}.  However, directly regressing to the solution form \autoref{eq:DMDsolution} through optimised DMD~\cite{optdmd,bopdmd} has many benefits, and it robustly allows for tasks such as model reduction and future-state prediction.  Moreover, a computationally tractable and parsimonious representation~\cite{kutz2022parsimony} of the dynamics in a low-rank latent space can typically be leveraged with a singular value decomposition~\cite{brunton_2016,kutz2026data}.    This insight paired with the equation-free nature of the algorithm has led to the widespread application of DMD in fluid dynamics \cite{schmid_2008, schmid_2009, schmid_2010, Noack2016jfm} and across many other scientific fields \cite{dmd_book, kutz2026data}. The method has additionally been connected to Koopman operator theory and has since become the standard approach for approximating the Koopman operator from data \cite{rowley_2009, tu_2014, modern_koopman}.
In the case of the data from high-fidelity flow simulations around motorsport geometries, the DMD results are typically for representing flows on meshes with tens of millions of degrees of freedom.


\begin{figure}[t]
\centering
\includegraphics[width=.55\textwidth]{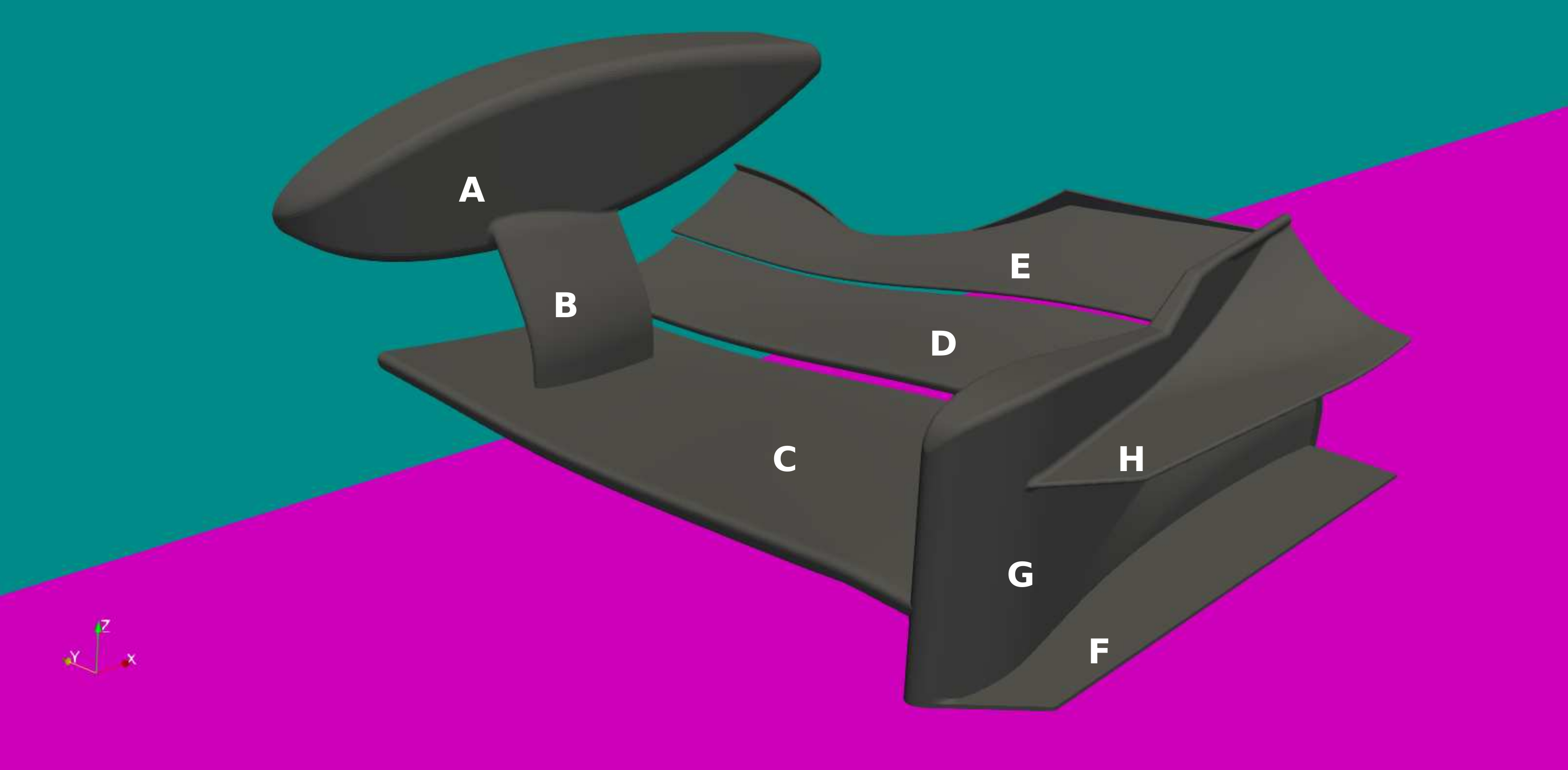}
\includegraphics[width=.36\textwidth]{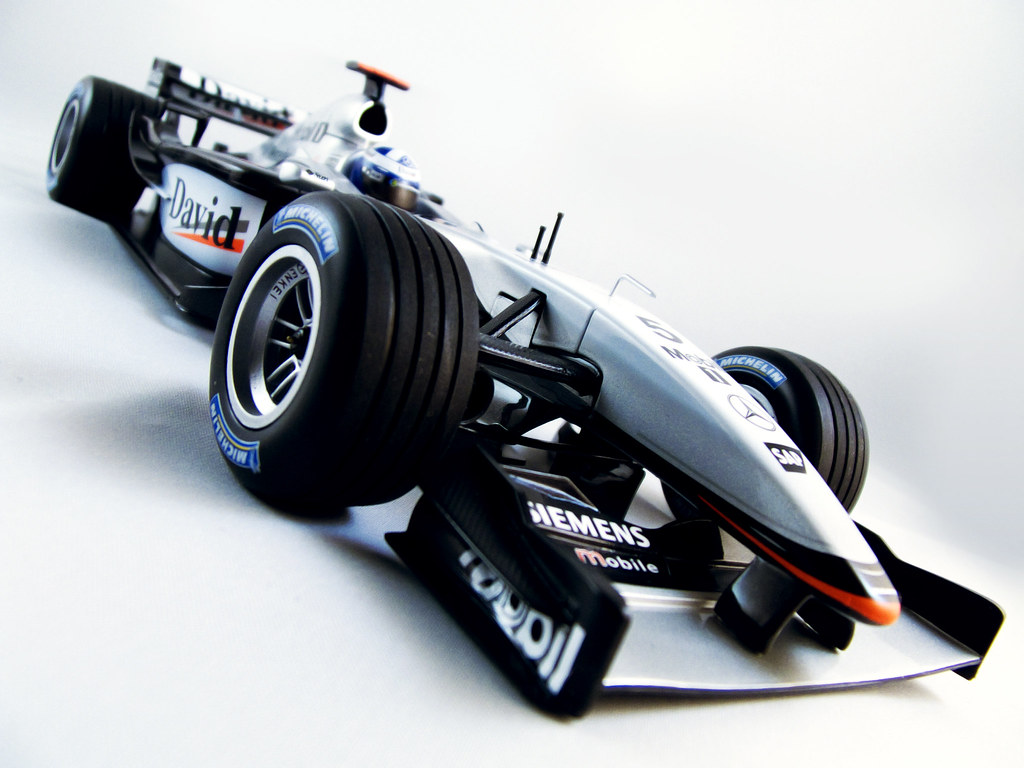}
\caption{(left panel) Imperial Front Wing  A: nose cone, B: hanger, C: mainplane, D: first flap, E: second flap, F: footplate, G: endplate, H: canard, Pink: rolling road, green: symmetry Plane. (right panel) The McLaren MP4-17D from the 2003 Formula 1 season.}
\label{fig:ifw}
\end{figure}

The scalable DMD analysis here is computed using pyLOM \cite{eiximeno2025}, which is an open-source Python tool that enables SVD-based modal decompositions in parallel and exploiting the high performance computing (HPC) resources available either on CPU or GPU \cite{Miro2025billionaire}. The DMD implementation in pyLOM has been extensively used for other large computational fluid dynamics simulations. For instance, it was used in the analysis of the pressure fluctuations in the subsonic wake of a circular cylinder at $Re_D=U_{\infty}D/\nu=10^4$ \cite{eiximeno2023hybrid}, where $U_{\infty}$ is the free-stream velocity, $D$ the diameter of the cylinder and $\nu$ stands for the kinematic viscosity, and in the large-scale motion identification in the flow in the Stanford diffuser at $Re_b=U_bh/\nu=10^4$ \cite{miro2023self}, where $U_b$ is the bulk velocity and $h$ is the height of the diffuser.

In order to prove that such scalable modal decompositions can be useful for the motorsport industry, the whole process is tested on the \href{https://data.hpc.imperial.ac.uk/resolve/?doi=6049}{\textit{Imperial front wing} (IFW)}, the Formula 1 front wing shown in \autoref{fig:ifw}. The IFW was introduced as an industrial benchmark for methodology comparison by  \cite{Pegrum2006} through a detailed experimental study at high Reynolds number of $Re = 2.2 \times 10^5$, where the characteristic length scale is the chord length of the main-plane of the IFW (see ``C", \autoref{fig:ifw}). Its relevance lies in its realism, as it derives directly from the unraced front-wing configuration of the McLaren MP4-17D race car from 2003 and is publicly available, providing a valuable resource for aerodynamic studies. IFW is a multi-element wing operating in ground effect, generating a complex system of interacting vortices downstream. This industrial benchmark is valuable because it encompasses multiscale phenomena. The larger scales are associated with vortex generation and interaction, while the smaller scales are related to flow transition or separation, depending on the examined conditions.

Various high-fidelity numerical simulation studies of the IFW and its variants have followed at the high $Re = 2.2 \times 10^{5}$. Initial demonstrations of viability were done by \cite{lombard2017high}, followed by improved implicit LES modelling studies leading to good agreement with time-averaged experimental fields \cite{Buscariolo2022,liosi2024DLES}. Benchmarks were also created on a simpler extruded variant of the IFW (eIFW), reducing computational complexity and allowing for unsteady aerodynamics analysis and comparison \cite{Slaughter2022, ref_andrew_aiaa2024, Ntoukas2024}. Improvements in numerical methodology followed in future studies, establishing a reliable toolchain for LES of complex F1 geometries \cite{ref_andrew_ssrn2025, khurana2025industrialization}.
In the present work, the flow is resolved with \nekpp \cite{Moxey2020Nektar}, an open source framework for solving unsteady partial differential equations using the \spectralhp method \cite{karniadakis2005spectral}. This method has shown promise for wall-resolved implicit LES (iLES) on complex motorsports geometries as it offers good numerical convergence properties and provides higher accuracy with fewer degrees of freedom, while retaining the geometric flexibility that is characteristic of traditional linear finite element methods (FEM). 
This method has been previously applied to automotive \cite{Hambli2022, Mengaldo2021, Filipe2021Ahmed} and motorsport \cite{lombard2017high, Buscariolo2022, Slaughter2022, liosi2024DLES, khurana2025industrialization} geometries.

\section{Imperial front wing dataset} 
\label{sec:methods}
This section describes the Imperial front wing dataset decomposed using the DMD algorithm from pyLOM. It includes a definition of the numerical method used for the simulations as well as a validation of the results with experimental data and a description of the key flow features identified.

\subsection{Numerical methodology}
We consider an incompressible flow regime in these simulations, as density variations and compressibility effects are negligible at the flow speeds considered. The flow is therefore governed by the incompressible Navier–Stokes equations, which describe viscous Newtonian fluid motion in a domain $\Omega \subset \mathbb{R}^{d}$ with boundary $\partial\Omega$ and are given by
\begin{subequations}
\begin{equation}
   \frac{\partial \mathbf{u}}{\partial t} + \mathbf{u} \cdot \nabla \mathbf{u} = -\nabla p + \nu \nabla^2 \mathbf{u} +  \mathbf{f} \label{eqn.NSmom}
 \end{equation}
 \begin{equation}
    \nabla \cdot \mathbf{u} = 0
    \end{equation}
    \label{eqn:IncNS}
\end{subequations}
where $\mathbf{u}$ is the velocity, $p$ is the specific pressure (including density), $f$ is the forcing term and $\nu$ the kinematic viscosity. An initial condition $\mathbf{u}(\mathbf{x},0)=\mathbf{u}_{0}(\mathbf{x})$ is prescribed, together with appropriate boundary conditions on $\partial\Omega$.
\subsubsection{Spectral/$hp$ element methods}
The incompressible Navier-Stokes (IncNS) equations, \autoref{eqn:IncNS}, are discretised in space using the \spectralhp method, which is a flavour of Galerkin high-order finite element methods. It combines the idea of resolving complex geometries with geometric elements of characteristic mesh size $h$ and higher-order basis functions with polynomial order $P$.
In this method, the solution $u$ at a given point in space $(x_i, y_i, z_i)$ is evaluated as
\begin{equation}
    u(x_i,y_i,z_i) = \sum_{pqr}^{P-1} \hat{u} \phi_{pqr}(x_i, y_i, z_i) 
    ,
\end{equation}
where $\phi_{pqr}$ are polynomials, calculated as tensor products of one-dimensional basis functions.
The numerical operations, such as integration or differentiation, are evaluated using Gaussian quadrature, and the solution $u$ is approximated with Jacobi polynomials $\phi$ of order $P$. The chosen numerical framework for the \spectralhp method is the open-source software \nekpp \cite{Moxey2020Nektar}, which implements this method for a variety of flow problems.

\subsubsection{Simulation setup}
The simulations on the Imperial front wing are run for a Reynolds number based on the chord of the main element of $Re = U_\infty L_c/\nu = 2.2 \times 10^5$, where $L_c = 0.25$~m is the chord length of the mainplane and $U_\infty = 1$~m\,s$^{-1}$ is the freestream velocity of the simulation, with unit density. The Reynolds number is imposed through the kinematic viscosity, $\nu = U_\infty L_c/Re$. Frequencies are reported as the Strouhal number $St = fL_c/U_\infty$, which is the same for the wind-tunnel conditions of the force and PIV measurements of \cite{Buscariolo2022} ($U_\infty = 12.5$~m\,s$^{-1}$) at the same Reynolds number. The IFW geometry is simulated for a polynomial order of $P=4$ for velocity and $P=3$ for pressure, following the simulation methodology from Liosi et al. \cite{liosi2024DLES}. The IncNS solver in \nekpp uses a velocity correction scheme (VCS) or a high-order splitting scheme \cite{KARNIADAKIS1991414}, which decouples the velocity and the pressure system. The VCS is a semi-implicit scheme in which the diffusion terms are treated implicitly, and the advection terms are handled explicitly. The chosen time step ($\Delta t_{\mathrm{sim}} = 5 \times 10^{-6}$~s) to run the simulation is bounded by the simulation's stability, governed by the Courant–Friedrichs–Lewy (CFL) criterion imposed by the explicit treatment of advection terms. Time integration is performed using a second-order accurate implicit-explicit (IMEX) scheme to balance accuracy and computational efficiency. Stabilisation is achieved using the spectral vanishing viscosity DG kernel \cite{rosh06} and global de-aliasing. The discretised system of equations is solved iteratively using the conjugate gradient (CG) algorithm, with static condensation applied to the system matrix \cite{karniadakis2005spectral} to reduce the number of algebraic degrees of freedom.

\begin{figure}[t]
    \centering
    \includegraphics[width=\textwidth]{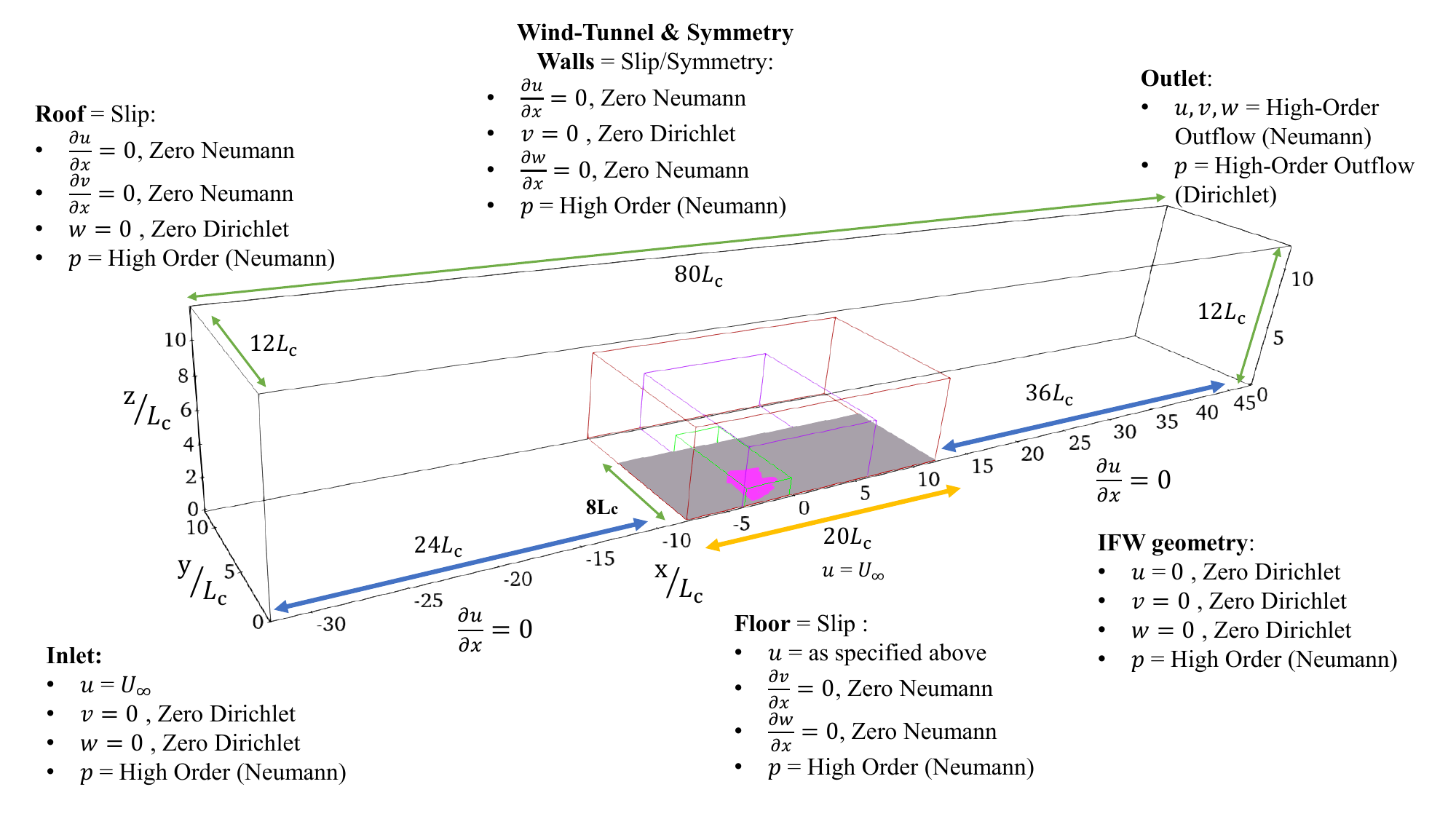}
    \caption{
    Schematic of the computational domain and boundary conditions. The vehicle geometry and the rolling road are indicated in pink and grey, respectively. Nested refinement zones are denoted by coloured outlines. The coordinate system is defined such that $x$ represents the streamwise direction, $y$ the spanwise direction, and $z$ the wall-normal direction.}
    \label{fig:DomainBCs}
\end{figure}

The dimensions of the computational domain can be seen in \autoref{fig:DomainBCs}, designed to emulate a wind tunnel with a rolling road. The default chosen ride height ($h$), which is the distance between the ground and the lowermost point of the wing is of $h = 0.36 L_c$. In this case, the height is measured from the footplate of the geometry. The pitch angle of the IFW with respect to the ground is $1.094^{\circ}$. The boundary conditions for IFW geometry are no-slip, i.e. the pressure is zero Neumann, and all velocity components are Dirichlet. A uniform inflow is imposed at the inlet in a Dirichlet manner. The floor consists of a moving wall with an inflow-like velocity to replicate the rolling road in a wind tunnel, surrounded by two patches of stationary ground, with the slip condition imposed on the velocity. The slip condition is also applied to the domain ceiling. The side walls are subject to the symmetry condition, allowing zero cross-flow.  For pressure, a high-order Neumann boundary condition \cite{KARNIADAKIS1991414, GuermondandShen2003} is applied along all computational domain walls except the outlet. For the outlet, a high-order Dirichlet convective boundary condition is applied \cite{DoKa14}. The boundary conditions for the computational domain can be seen in \autoref{fig:DomainBCs}.

The mesh is generated in two stages. First, the input geometry file \cite{IFWSource} is imported to a traditional finite volume (FV) mesh generator, and a linear (straight-sided), unstructured, and conformal mesh is created. This linear mesh contains a single macro-prism layer around the entire wing and tetrahedra within the domain. The total thickness of the macro-prism layer is $0.01 L_c$. Appropriate refinement boxes are defined in the linear mesh to ensure accurate spatial resolution of the critical flow structures. 

The smallest box is tightly concentrated around the flow-separation region (from $x/L_c=-2.60$ to $x/L_c=0.80$) and extends to a height of $z/L_c=1.20$, ensuring accurate resolution of the base pressure and recirculation zone. The intermediate and large boxes nest the smaller box and smoothly transition the mesh resolution. The intermediate box spans from $x/L_c=-2.60$ to $x/L_c=7.20$, and extends to $z/L_c=4.0$ to capture the vortex evolution. The largest box spans from $x/L_c=-6.80$ to $x/L_c=13.20$ and reaches a maximum height of $z/L_c=6.00$, thereby adequately capturing the full evolution of the large-scale wake structures as they dissipate. The smallest box spans the same width as the rolling road $y/L_c=8.00$, and the other boxes span $y/L_c=12.00$. The resulting linear mesh contains $1.81 \times 10^{6}$ elements.

The second stage involves several steps, all performed within \textit{NekMesh} \cite{NekMeshGREEN2024}, the mesh-generation utility of \nekpp. Within \textit{NekMesh}, the straight-sided linear surface mesh is curved to the CAD geometry using fourth-order polynomials, providing sufficient geometrical accuracy for the wing surfaces. The macro-prism layer is split in the wall-normal direction using the iso-parametric splitting technique. For the present case, 8 prism layers are employed with a progression ratio of 1.2, while preserving element quality and geometric fidelity. These procedures follow the established industrialisation pipeline \cite{Kirilov2026}, and have been used for motorsport geometries \cite{liosi2024DLES, khurana2025industrialization}. The resulting curved higher-order mesh has $49.72 \times 10^{6}$ global degrees of freedom (DoF) for each velocity component at $P=4$, and $21.04 \times 10^{6}$ global DoF for pressure at $P=3$.

\subsection{Validation}
The simulation is time-averaged based on two criteria: a) the moving standard deviation
of the force traces of the geometry to be below 1\% and b) the temporal derivative of the sampled window ($0.25$~s) for the force traces be lower than $10^{-6}$. The comparison of the force coefficients in \autoref{tab:IFWresults} shows a good agreement between the current results and the experimental reference from \cite{Buscariolo2022}.
\begin{figure}[t]
    \centering
    \includegraphics[width=0.85\textwidth]{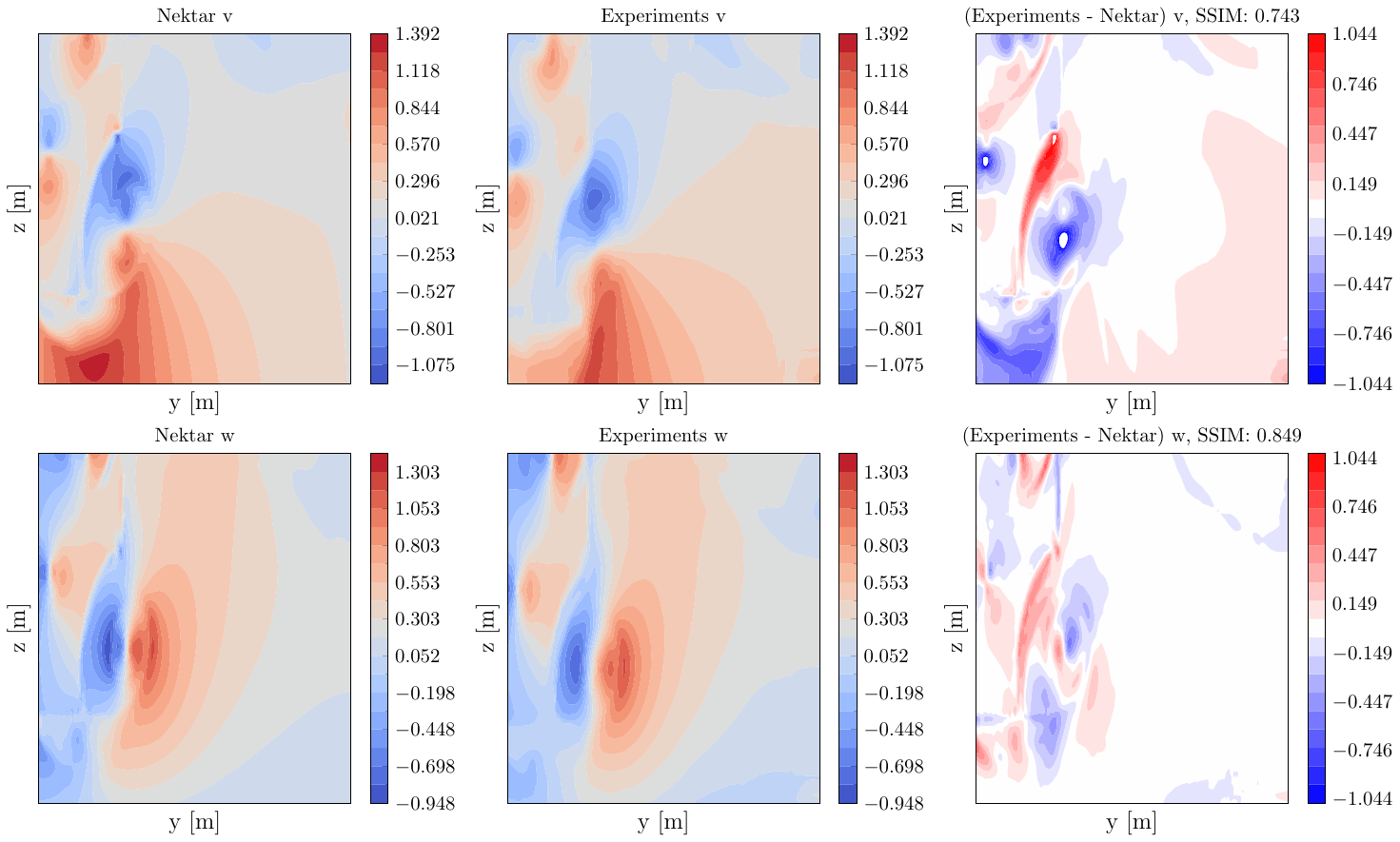} 
    \caption{Comparison of the averaged velocity components between the \nekpp results (left) and the experimental PIV data (right) for averaged data at the location of PIV plane 1 from \cite{Buscariolo2022} at $x/L_c =-1.176$. The SSIM is the structural similarity index \cite{Wang2004} used to quantify the correlation between the results.}
    \label{fig:PIVplanes}
\end{figure}

To validate the development of flow features for the IFW, the current time-averaged IFW results are compared with the available experimental PIV plane data from \cite{Buscariolo2022}. The comparison of the spanwise and normal velocity components for PIV plane 1 is seen in \autoref{fig:PIVplanes}, where the main vortical structures for the current study qualitatively agree with the experiments. For a more detailed discussion of the validation, the reader is referred to the work by \cite{liosi2024DLES}.
\begin{table}[t]
\caption{Comparison of the time-averaged force coefficients with the experimental reference from \cite{Buscariolo2022}}
\label{tab:IFWresults}
 \centering
 \begin{tabular}{cccccc}
 \hline
Scheme & time step ($\Delta t_{\mathrm{sim}}$) & $C_L$ & $C_D$ & $C_L$ Diff, $\%$ & $C_D$ Diff, $\%$\\\hline
Experiment &  & -5.540 & 0.575 & N/A & N/A \\
Current results & $5 \times 10^{-6}$~s  &  $-5.676 \pm 0.001$  & $0.593 \pm 0.002$ & 2.24 & 2.87 \\
 \hline
\end{tabular}
\end{table}

\begin{figure}[t]
\centering
\includegraphics[width=0.7\textwidth]{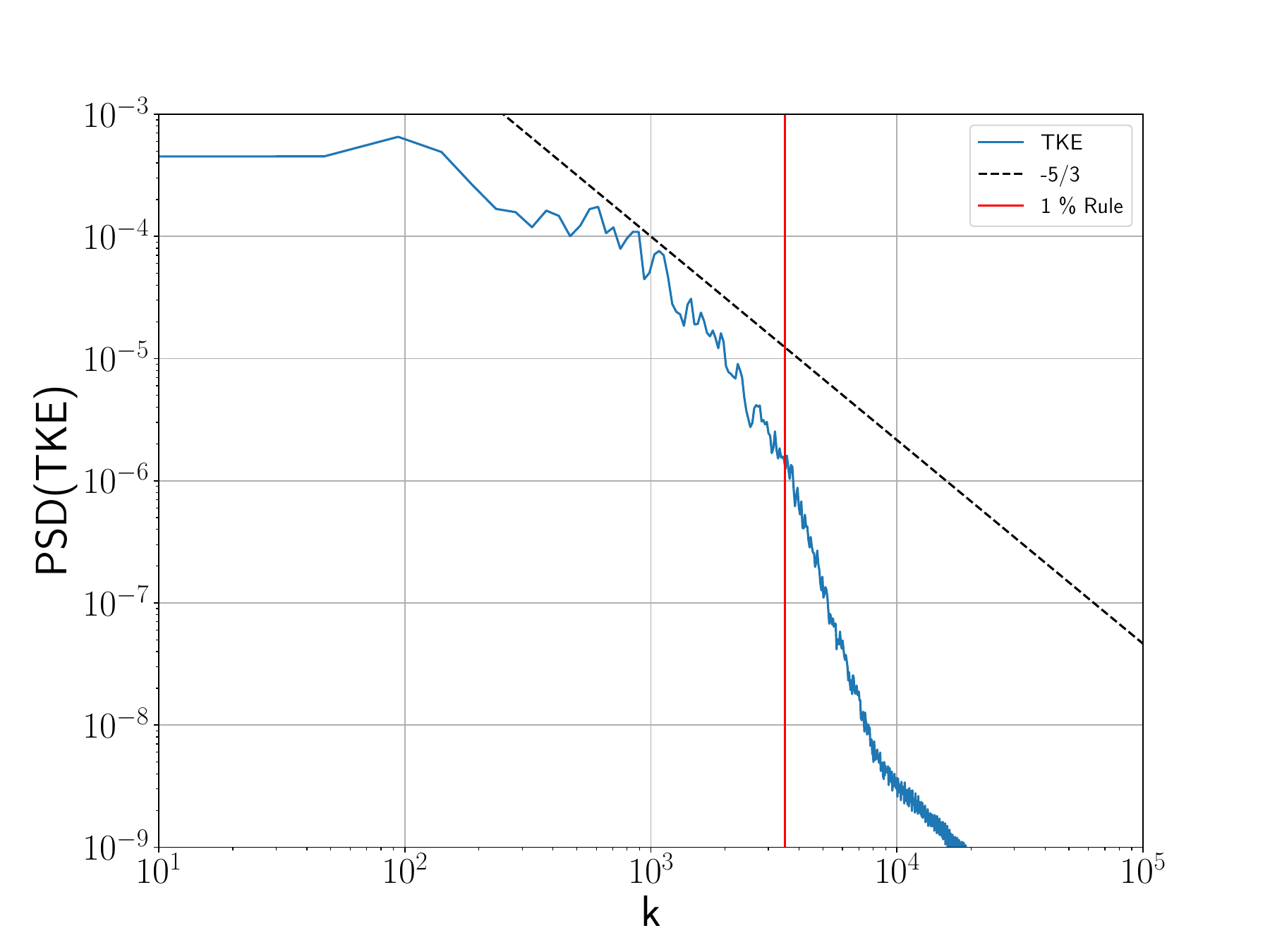}
\caption{Turbulent kinetic energy (TKE) spectra w.r.t wavenumbers, and limit of the resolved spectrum defined from the $1 \%$ rule\cite{moura2015linear, moura2017eddy}. The probe point is located on the suction side of the mainplane at $(x/L_c, y/L_c, z/L_c) = (-2.724,0.000,0.272)$ (Probe 5 in \autoref{fig:probes}), and following the process described in \cite{beck2014high} to create the power spectral density (PSD) plot of the TKE.}
\label{fig:PSD_TKE}
\end{figure}

For implicit LES using high-order discretisations, the inertial range of the turbulent energy spectrum is explicitly resolved, and the $1\%$ rule~\cite{moura2015linear,moura2017eddy} provides a criterion for identifying the highest wavenumber that remains accurately represented. This rule estimates the onset of the numerically induced dissipation range, beyond which artificial diffusion dominates the measured energy spectrum~\cite{moura2017eddy}. \autoref{fig:PSD_TKE} shows the turbulent kinetic energy (TKE) spectra for a probe on the suction side of the mainplane. The red vertical line marks the $1 \%$ Rule limit where the TKE is resolved and aligns with the slope change from the inertial to the dissipation range of the TKE. 

\subsection{Flow description}

\begin{figure}[t]
\centering
\includegraphics[width=0.7\textwidth]{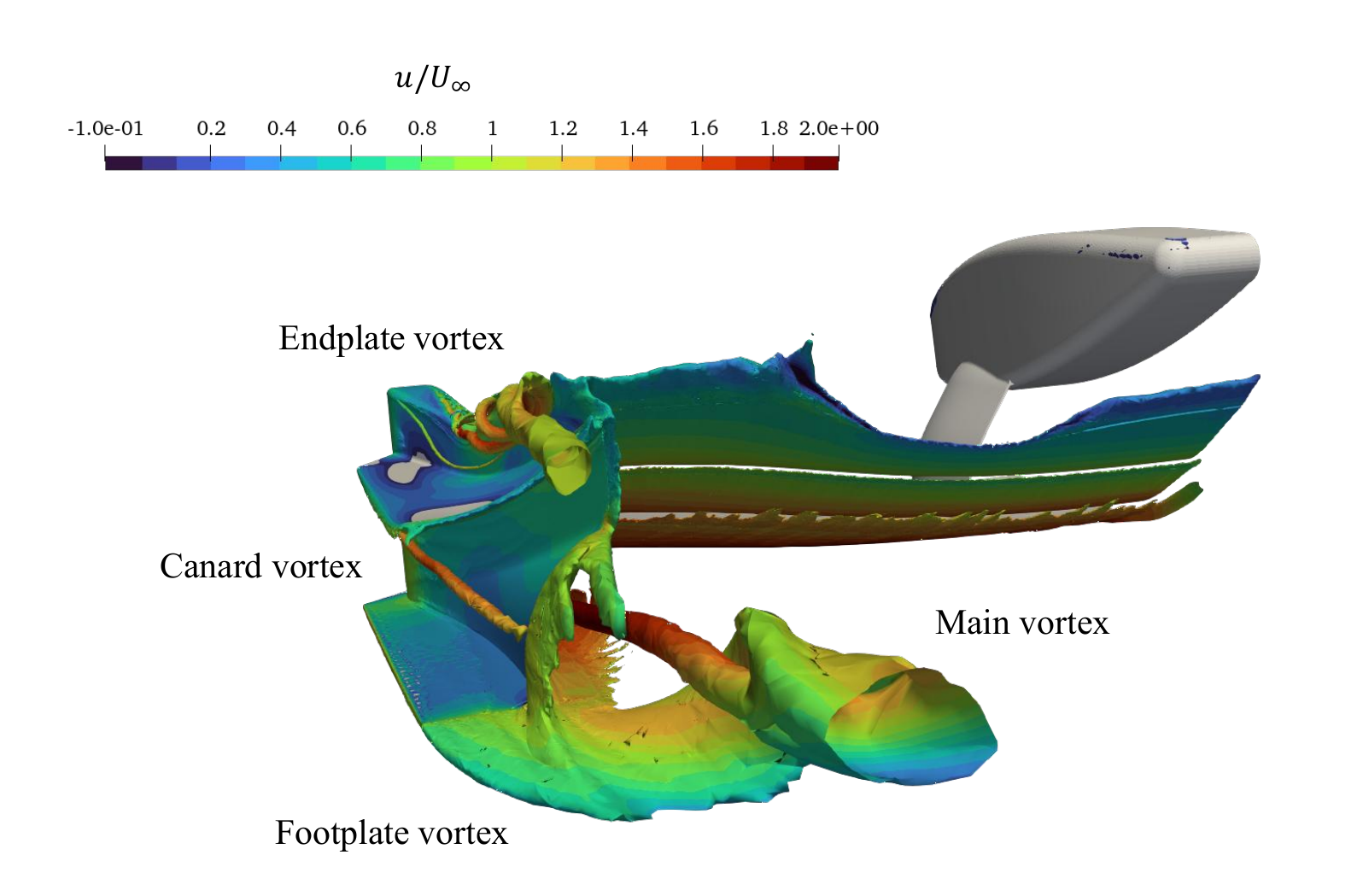}
\caption{3D View from the rear for the IFW - iso-contour of zero total pressure coefficient ($C_{p0} = 0$) for the average flow field, coloured by the streamwise velocity component ($u/U_\infty$). Here $C_{p0} = (p_0 - p_\infty)/(\tfrac{1}{2}\rho U_\infty^2)$, where $p_0 = p + \tfrac{1}{2}\rho|\bm{u}|^2$ is the total pressure, $p$ the static pressure, $\bm{u}$ the velocity, $\rho$ the density, and $p_\infty$ and $U_\infty$ the free-stream static pressure and velocity. The four vortices are named: endplate (top) vortex, canard vortex, main vortex and footplate vortex.}
\label{fig:vortices}
\end{figure} 

\subsubsection{Mean flow description}
Four dominant vortical structures emerge from different components of the IFW. These are illustrated in \autoref{fig:vortices} and are named according to the device that generates them, i.e., the endplate (also referred to as top) vortex, canard vortex, main vortex, and footplate vortex. Both the main and top vortices are generated by the endplate's geometric features. In particular, the pressure difference between the upper and lower surfaces of the IFW elements creates flow circulation around the wing edges, which generates the vortical structures for each element (top, canard, footplate).

The developed pressure and shear-stress fields on the wing surface directly contribute to the formation of vortices from different components. \autoref{App:IFW_mean_surface} details the mean surface flow fields to provide further context on the formation of the vortical system around the IFW. The main vortex is produced by the shear layer that is formed at the LE of the endplate. Then, this shear-layer rolls up following an anti-clockwise trajectory based on the inboard shape of the endplate. The top vortex is also produced at the upper tip of the endplate, where two helical structures are created due to the airfoil-like profile of the endplate's vertical area. These helical structures rotate around each other and merge into a single vortex downstream. On the canard, the pressure difference between its upper and lower surfaces creates a flow circulation around its edges, generating the canard vortex. Also, its curved profile enhances the pressure difference, which helps power the vortex. The pressure difference between the pressure and the suction side of the mainplane and the flaps is very important for the strength of the main vortex as it impacts the flow acceleration at the shortest area below the mainplane, hence affecting the strength of the shear-layer that rotates around the inboard side of the endplate. On the 
footplate, two shear layers are formed from its triangular LE. Each shear-layer rolls up towards the inboard and the outboard side of the footplate, respectively. They interact with the main vortex, pulling each other together until they merge into a single structure.

\subsubsection{Instantaneous flow description}
\label{sec:inst}
\begin{figure}[htbp]
    \centering
    \includegraphics[width=\textwidth]{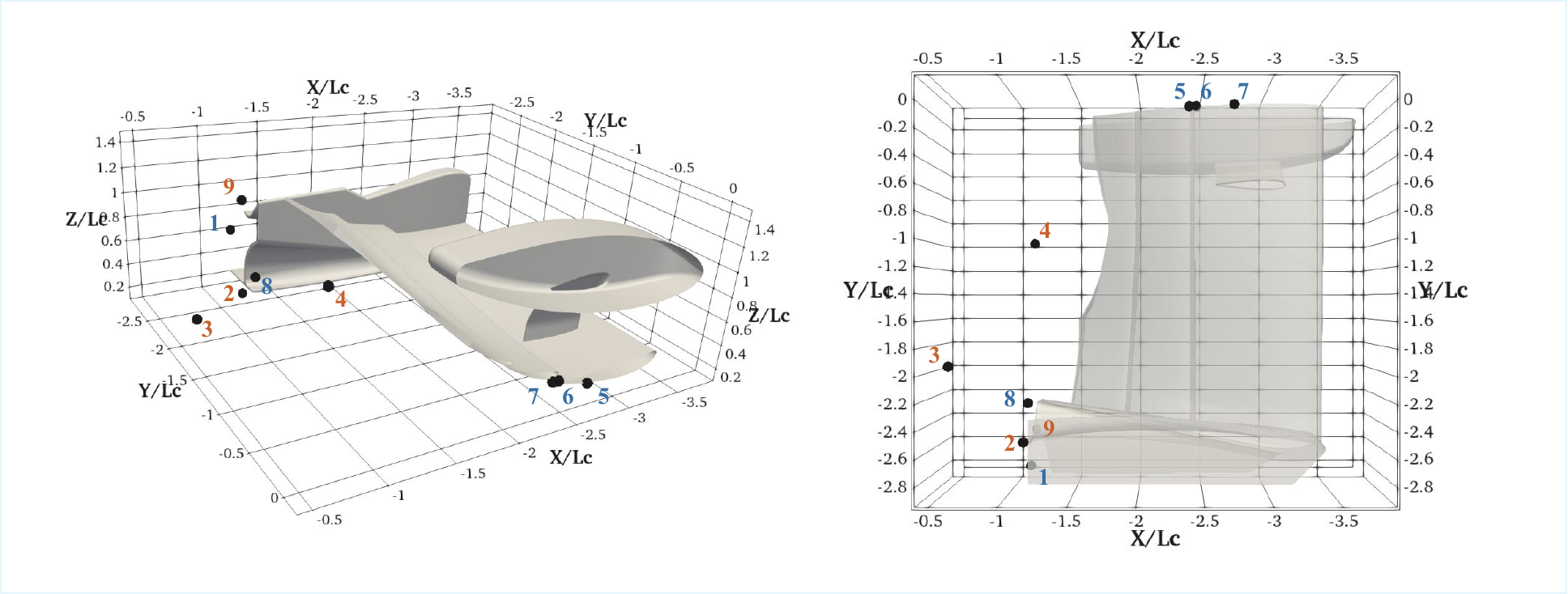}
    \caption{Probe locations. Red: 250~Hz probes. Blue: 20~kHz probes.}
    \label{fig:probes}
\end{figure}

\begin{figure}[ht]
    \centering
    \subfloat[]{\includegraphics[width=0.49\textwidth]{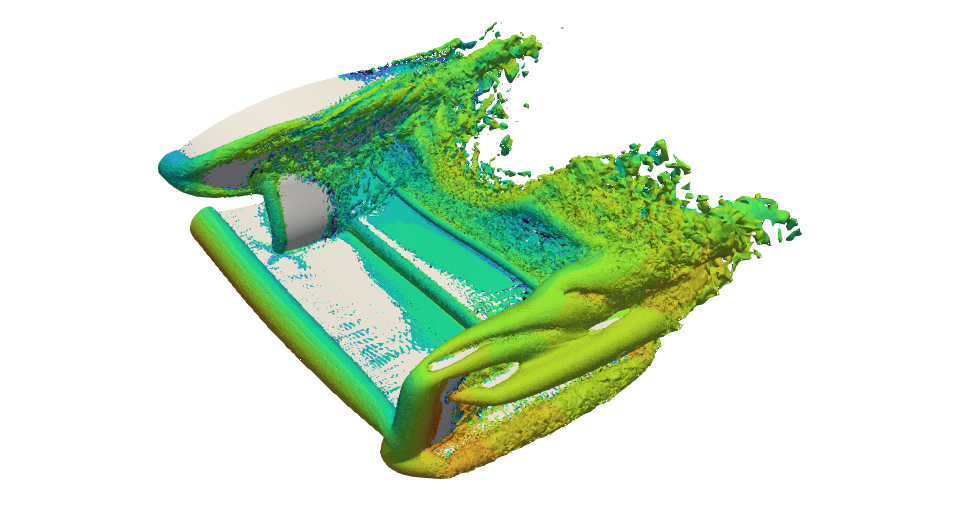}}
    \hfill 
    \subfloat[]{\includegraphics[width=0.49\textwidth]{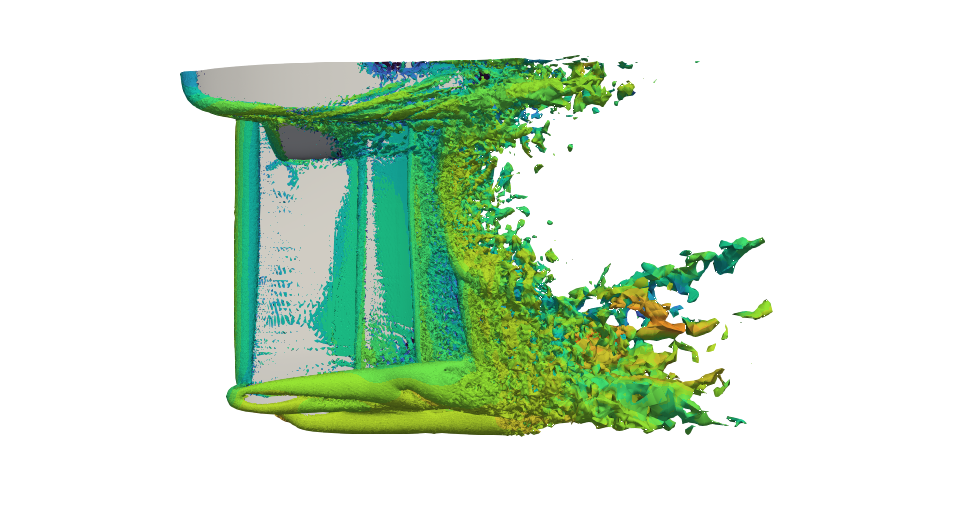}}

    \subfloat[]{\includegraphics[width=0.49\textwidth]{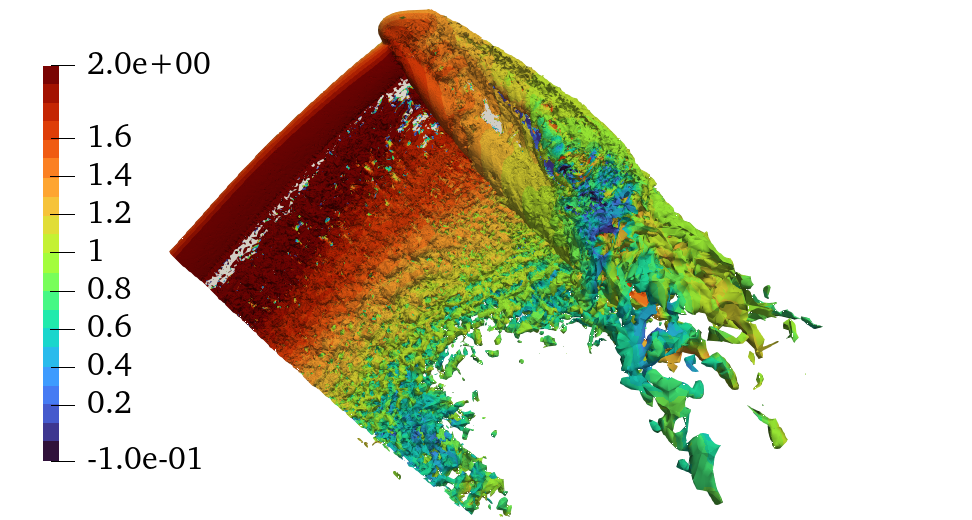}}
    
    \caption{$\lambda_2$ criterion isocontours ($\lambda_2 = 100$) of the instantaneous IFW flow field, coloured by the streamwise velocity component ($u/U_\infty$). (a) iso view, (b) top view, and (c) bottom view.}
    \label{fig:L2_100}
\end{figure}

The instantaneous flow field analysis utilises data acquired from the numerical probes shown in \autoref{fig:probes}, with precise coordinates listed in \autoref{tab:coords}. These probes were placed after the mean flow analysis, at approximate locations chosen to capture the dominant flow dynamics identified there, as well as the coherent structures visualised via $\lambda_2$ criterion isocontours (\autoref{fig:L2_100}). All probes are assigned an ID for representation purposes. The high-frequency probes ($20$~kHz) were sampled directly during the simulation to resolve rapid fluctuations, whereas the low-frequency probes ($250$~Hz) were extracted during post-processing from the stored flow snapshots. The dominant frequencies of the velocity spectra for these probes are reported in \autoref{tab:freqs}, while the spectra plots can be found in \autoref{App:IFW_probes}. A broadband low-frequency background near $St = fL_c/U_{\infty}= 2.0 - 2.7$ is present at every probe. Because this range lies immediately above the $St = 2.0$ resolvability limit and coincides with the rising low-frequency end of the turbulent spectrum, it is interpreted as a shared low-frequency floor rather than a discrete modal tone.

\begin{table}[htbp]
    \centering
    \begin{tabular}{cp{5.2cm}cccc}
        \hline
        \textbf{ID} & \textbf{Location} & \textbf{$x/L_c$} & \textbf{$y/L_c$} & \textbf{$z/L_c$} & \textbf{$f_s$ (Hz)} \\
        \hline
        1 & Canard vortex - PIV plane 1     & -1.176 & -2.724 & 0.664 & 20000 \\
        2 & Footplate vortex - PIV plane 1  & -1.176 & -2.484 & 0.196 & 250   \\
        3 & Main vortex - PIV plane 3       & -0.600 & -1.940 & 0.324 & 250   \\
        4 & Y250 plane - PIV plane 1        & -1.176 & -1.000 & 0.884 & 250   \\
        5 & Main element suction            & -2.724 & 0.000  & 0.272 & 20000 \\
        6 & Main element pressure           & -2.444 & 0.000  & 0.360 & 20000 \\
        7 & Main element trailing           & -2.392 & 0.000  & 0.364 & 20000 \\
        8 & Main vortex - PIV plane 1       & -1.176 & -2.220 & 0.448 & 20000 \\
        9 & Endplate vortex - PIV plane 1   & -1.176 & -2.484 & 1.000 & 250   \\
        \hline
    \end{tabular}
    \caption{Probe location coordinates and sampling frequencies.}
    \label{tab:coords}
\end{table}

\begin{table}[ht]
    \centering
    \small
    \caption{\small Prominent local non-dimensional frequencies ($St = fL_c/U_{\infty}$) by velocity component. Probe data acquired at 20~kHz were anti-alias decimated to $f_s=250$~Hz before spectral analysis. Each spectrum uses $N=300$ samples over $T=1.2$~s, giving $\Delta St \approx 0.21$; reported frequencies are therefore rounded to one decimal.  Peaks are reported for $2 < St \leq 25$, with the lower bound set by the approximately ten oscillation periods and the upper bound set by the $-1$~dB point of the composite decimation filter. (Nyquist limit $St=31.25$). All listed peaks exceed a uniform log-PSD prominence threshold of 1.5, applied identically to every probe and velocity component; the number of reported peaks therefore varies by probe. The $St$ are reported in decreasing order of prominence.}
    \begin{tabular}{cp{3.5cm}p{3cm}p{3cm}p{3cm}}
        \hline
        \textbf{ID} & \textbf{Location} & \textbf{Streamwise ($u$)} & \textbf{Cross-stream ($v$)} & \textbf{Vertical ($w$)} \\
        \hline
        \textbf{1} & Canard vortex (PIV Plane 1)
        & 4.3, 7.4, 2.2
        & 7.4, 4.0
        & 7.4, 2.7, 7.8, 3.9 \\
        \hline
        \textbf{2} & Footplate vortex (PIV Plane 1)
        & 9.6
        & 2.8, 6.2, 4.0, 16.1, 9.4
        & 6.3, 3.5, 2.2, 3.2, 11.7 \\
        \hline
        \textbf{3} & Main vortex (PIV Plane 3)
        & 2.4, 4.0, 5.5, 3.4
        & 2.5
        & 4.3 \\
        \hline
        \textbf{4} & Y250 plane (PIV Plane 1)
        & 2.1
        & 2.9, 5.9, 5.3
        & 4.0, 2.1 \\
        \hline
        \textbf{5} & Main element suction
        & 5.1, 2.5, 14.8, 10.3
        & --
        & 5.1, 2.5, 14.8, 10.2, 7.7, 5.8 \\
        \hline
        \textbf{6} & Main element pressure
        & 2.6, 5.2, 10.2
        & --
        & 3.2, 5.2 \\
        \hline
        \textbf{7} & Main element trailing
        & 4.7, 5.4, 12.4, 2.3
        & --
        & 2.5, 10.3, 3.9, 3.2, 4.5 \\
        \hline
        \textbf{8} & Main vortex (PIV Plane 1)
        & 3.3, 4.6, 6.1
        & 2.6, 7.2
        & 3.3, 7.2, 4.2, 8.6 \\
        \hline
        \textbf{9} & Endplate vortex (PIV Plane 1)
        & 12.2, 5.5, 4.0, 4.9, 3.5, 5.9, 7.3, 2.5
        & 3.9, 4.9, 7.3, 2.7
        & 3.6, 9.3, 6.2, 5.2 \\
        \hline
    \end{tabular}
    \label{tab:freqs}
\end{table}

Probes 1, 2, 8 and 9 are located at the same streamwise position as the trailing edge of the endplate. Probe 1 is placed outboard of the wing, between the canard and the foot plate and is meant to capture the canard vortex. This vortex is a coherent, positionally stable structure with a strong anti-clockwise rotating core. Probe 2 is positioned under the footplate, and it tracks the footplate vortex. \autoref{fig:L2_100} reveals a spatially diffuse, wide-envelope structure undergoing lateral wandering directed inboard, with visible core breakdown and no sustained coherent rotation. Probe 8 is positioned inboard near the core of the main vortex, a strong anti-clockwise rotating structure. Probe 3 follows the same vortex, but at a downstream position. Probe 9 tracks the endplate vortex. Together, these probes are aimed to capture local frequencies close to the vortex core of the four-vortex system.

A non-dimensional frequency band of $St \approx 4$ is present in all the five vortex core probes, and it is the only frequency common in at least one velocity component at every core probe. This common frequency across probes is necessary but not sufficient evidence of a system-wide mechanism; this will be analysed in Section \ref{sec:modal_analysis}. Another prominent frequency, $St\approx7.4$, is carried by the canard, endplate and the main-vortex station (probe 8), but is absent from both the footplate and the downstream main-vortex station (probe 3). Canard vortex sees this frequency band coincidence in all three velocity components. The footplate probe (Probe 2) has a wide range of frequency peaks in the cross-stream and vertical velocity components, along with high distinct tones of $St= 11.7$ and $16.1$. The endplate vortex has the richest streamwise signature, with a dense ladder of $u$-peaks, and the most prominent tone at $St = 12.2$. The endplate and footplate vortex probes also produce distinct signatures around $St \approx 9$.

Probes 5--7 are located on the wing's symmetry plane at the main element of the IFW. Their cross-stream velocity component has no spectrum, as seen in \autoref{tab:freqs}, due to the imposed zero Dirichlet boundary condition. Probe 5 is located on the suction side, while Probe 6 is on the pressure side and Probe 7 is at the trailing edge. All three probes here lock into the frequency $St \approx 5.2$ in the streamwise velocity component, and Probes 5 and 6 also in the vertical velocity component. They also exhibit a band at $St \approx 10.2$, which is likely a second harmonic of the previous band. Interestingly, the $St \approx 5.2$ peaks are also observed in the endplate vortex. The signatures of Probes 5 and 6 are similar, although they are located on opposite sides of the main element. The suction side and the trailing probe also carry high frequency bands of $14.8$ and $12.4$ respectively. 

Probe 4 is located on the Y250 plane, i.e., 250 mm from the vehicle centreline, at the same streamwise position as Probes 1,2,8 and 9. The Y250 vortex generated in this region is a defining aerodynamic feature of pre-2022 Formula 1 cars \cite{wright2001formula}, and this IFW section was previously studied  \cite{Slaughter2022}. This location shares the common frequency bands with $St \approx 4$ from vortex probes, $St \approx 5.2$ from the surface probes 5, 6 and $St \approx 5.9$ from the endplate.

\section{Scalable dynamic mode decomposition}
\label{sec:dmd}
The results of DMD can be interpreted as structures of a linear tangent approximation to the underlying flow and describe fluid elements that have a dominant dynamic behaviour inside the captured data. The fact that the modes extracted from DMD are part of a dynamic system implies that they are not correlated with a time signal anymore but with a coherent structure and an associated frequency, amplitude and damping ratio. 

The data used for DMD has to be constantly sampled in time ($\Delta t = ct.$) and has to be presented in a snapshot sequence given by the matrix ${\cal{D}}$,
\begin{equation}
    {\cal{D}} = \left[d_1,d_2,d_3,...,d_N\right]
\end{equation}
where $d_i$ stands for the $i$th sample of the flow field. The first step is to assume that there exists a linear mapping $A$ which connects the field $d_i$ to the next field $d_{i+1}$,
\begin{equation}
    d_{i+1} = A d_i\text{,}
\end{equation}
if $A$ is assumed as constant (Koopman assumption), it is possible to  formulate the sequence of flow fields as a Krylov sequence:
\begin{equation}
    \label{eqn:krylov}
    {\cal{D}} = \left[d_1, Ad_1, A^2d_1,...,A^{N-1}d_1\right]
\end{equation}

The goal of DMD is to find the linear operator $A$ and then compute the dynamic characteristics of the dynamical process described by $A$ based on the sequence ${\cal{D}}$. The dynamic information is computed with the eigendecomposition of $A$ and the frequency of the modes is obtained as
\begin{equation}
\label{eqn:freq}
    f_i = \frac{\theta_i}{2\Delta t \pi},
\end{equation}
where $\theta_i$ is the argument of the eigenvalue $\mu_i$. 

The relevance that DMD modes have in the flow is given by their amplitude. Among the different methods presented to rank the DMD modes, the most efficient and the one implemented in pyLOM, is the method introduced by Jovanovic et al. \cite{spdmd}.

DMD is about finding the dynamic response of a linear mapping, $A$, that allows the formulation of a sequence of flow fields as a Krylov sequence (\autoref{eqn:krylov}). One of the most efficient and accurate ways to compute $A$ is presented by Tu et al. \cite{tu_2014} and is known as the exact DMD. In this procedure, the data is first organised in
\begin{equation}
    Y_1 = \left[d_1, ..., d_{N-1}\right]
    \label{eqn:y1}
\end{equation}
\begin{equation}
    Y_2 = \left[d_2, ..., d_{N}\right]
    \label{eqn:y2}
\end{equation}
then the Krylov sequence can be expressed as
\begin{equation}
    Y_2 = A Y_1
\end{equation}
and the Koopman operator is computed as
\begin{equation}
    A = U^TY_2VS^{-1}
\end{equation}
where $U$, $V$ and $S$ are the result of a SVD applied to $Y_1$. As $Y_1$ is distributed across the processors, this step is done using the parallel algorithm presented by Eiximeno et al. \cite{eiximeno2025}. 

The results of the SVD can be truncated to a certain number of modes, $N_r$, to reduce the size of the matrix for the following computations. The truncation residual should be the lowest possible as long as the resulting computations fit in the available memory. 

The computation of the linear mapping $A$ involves the parallel matrix-matrix multiplication $C=U^TY_2$ as the transposed left eigenvectors, $U^T$, and the data array, $Y_2$ are distributed across the processors. In this particular case, the shape of the matrices is $(N_r \times M) \times (M \times N)$. Considering the data distribution in pyLOM, the left array is split column-wise, while the right array is split row-wise. The resulting array of shape $N_r \times N$ can be shared by all processors because $N_r \leq N << M$. 

This hypothesis allows for a simpler approach compared to a standard parallel-parallel matrix product. The operation is simplified to the global sum of the local products between the chunk of matrices stored in each processor. The reduction and sum of the result is done using the \texttt{all\_reduce} function from the MPI library.

Afterwards, the linear mapping $A$ is shared between all the processors and has a size of $N_r \times N_r$. Hence, the rest of the computations can be done in serial by all the processors without concerns for performance or memory usage. The next step is the eigendecomposition of $A$, $A = \mu w$, and then the DMD modes are computed as:
\begin{equation}
    \label{eqn:dmdmodes}
    \varphi = Y_2 V S^{-1}w/\mu
\end{equation}

To compute the amplitudes as in Jovanovic et al. \cite{spdmd}, the reconstruction of the snapshots is taken as
\begin{equation}
    Y_1 = \varphi D_{\alpha}V_{and}
\end{equation}
where $D_{\alpha}$ is a diagonal matrix containing the vector of amplitudes of each mode. $V_{and}$ is the Vandermonde matrix of size $N_r \times N$ and governs the time evolution of the dynamic modes of the first $N_r$ eigenvalues of $A$. Finally, the computation of $D_{\alpha}$ is done by minimising the following Frobenius norm
\begin{equation}
\underset{\alpha}{\operatorname{minimise}}\left\|Y_1- \varphi D_{\alpha} V_{\text {and }}\right\|_{F}^{2}
\end{equation}

In the present work, the DMD computations are done leveraging pyLOM \cite{pyLOM}, a high-performance-computing enabled tool for massively parallel reduced order modelling. pyLOM can handle both classical algorithms such as proper orthogonal decomposition (POD), dynamic mode decomposition (DMD) and spectral proper orthogonal decomposition (SPOD) as well as a set of deep learning architectures for model order reduction and surrogate modelling. The core of all parallel decompositions in pyLOM is a parallel QR factorisation algorithm based on a binary tree reduction algorithm. The scalability and verification of pyLOM on both CPUs and GPUs are reported by \cite{eiximeno2025, Miro2025billionaire}.

\section{Modal analysis}
\label{sec:modal_analysis}
\begin{figure}
    \centering
    \includegraphics[width=0.65\linewidth]{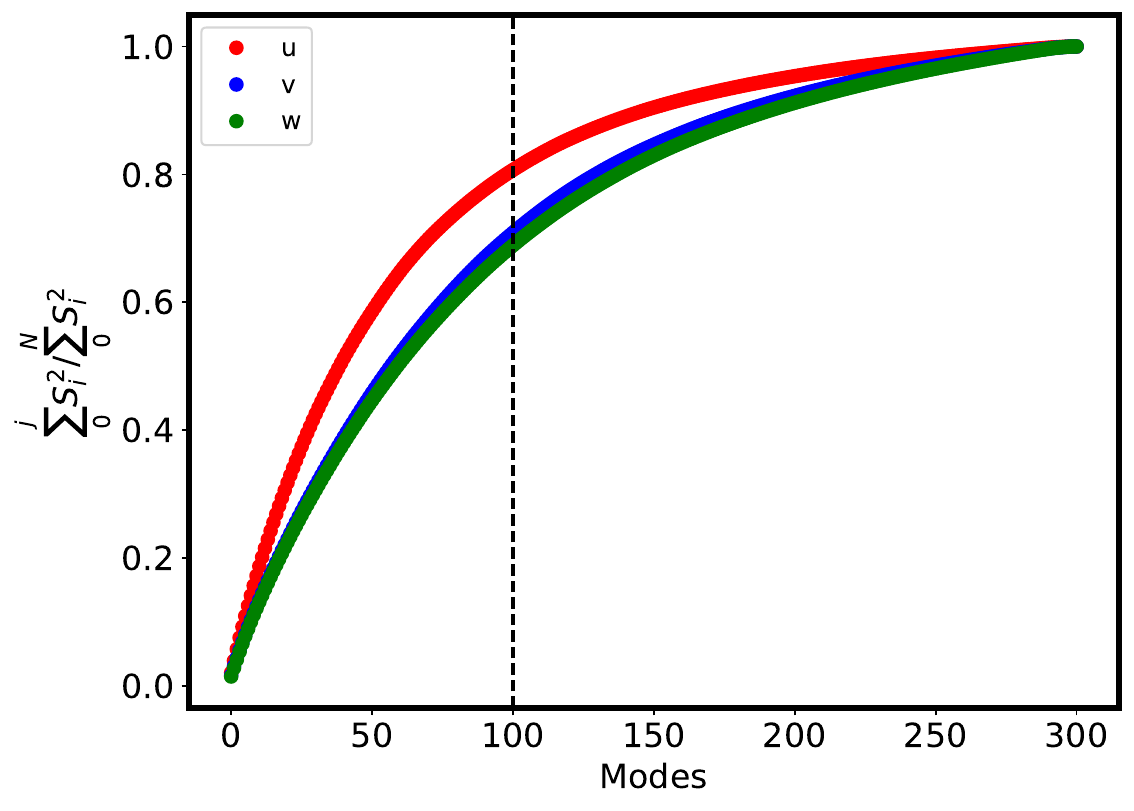}
    \caption{Cumulative energy for each POD mode for the three velocity components}
    \label{fig:energies}
\end{figure}

The temporal mean of each variable is removed from the snapshots before the DMD, so the modes describe the fluctuating fields. The first step in the DMD analysis is computing the SVD of all the snapshots except the last one. Then, the basis can be truncated to reduce the cost of the DMD computation as well as to filter out the smaller and unresolved scales, which correspond to the lower rank SVD modes. There is no clear methodology on how to define the energy threshold for the mode truncation. In this work, 100 SVD modes of each velocity component and of the total pressure coefficient, $C_{p0}$, are selected to fit the DMD model. For the velocity components, these modes comprise between the 65\% and the 80\% of the energy and after this point the singular value spectra, \autoref{fig:energies}, tend to a flatter tail. This truncation also reduces the storage footprint of the dataset: the 300 snapshots, of approximately 20~GB each (about 6~TB in total), are represented by DMD results of approximately 200~GB for the four variables, a reduction by a factor of approximately 30. This compression is not lossless, as the energy beyond the retained SVD modes is discarded.

\begin{figure}
\centering
\subfloat[Total pressure coefficient, $C_{p0}$]{%
\includegraphics[width=0.45\textwidth]{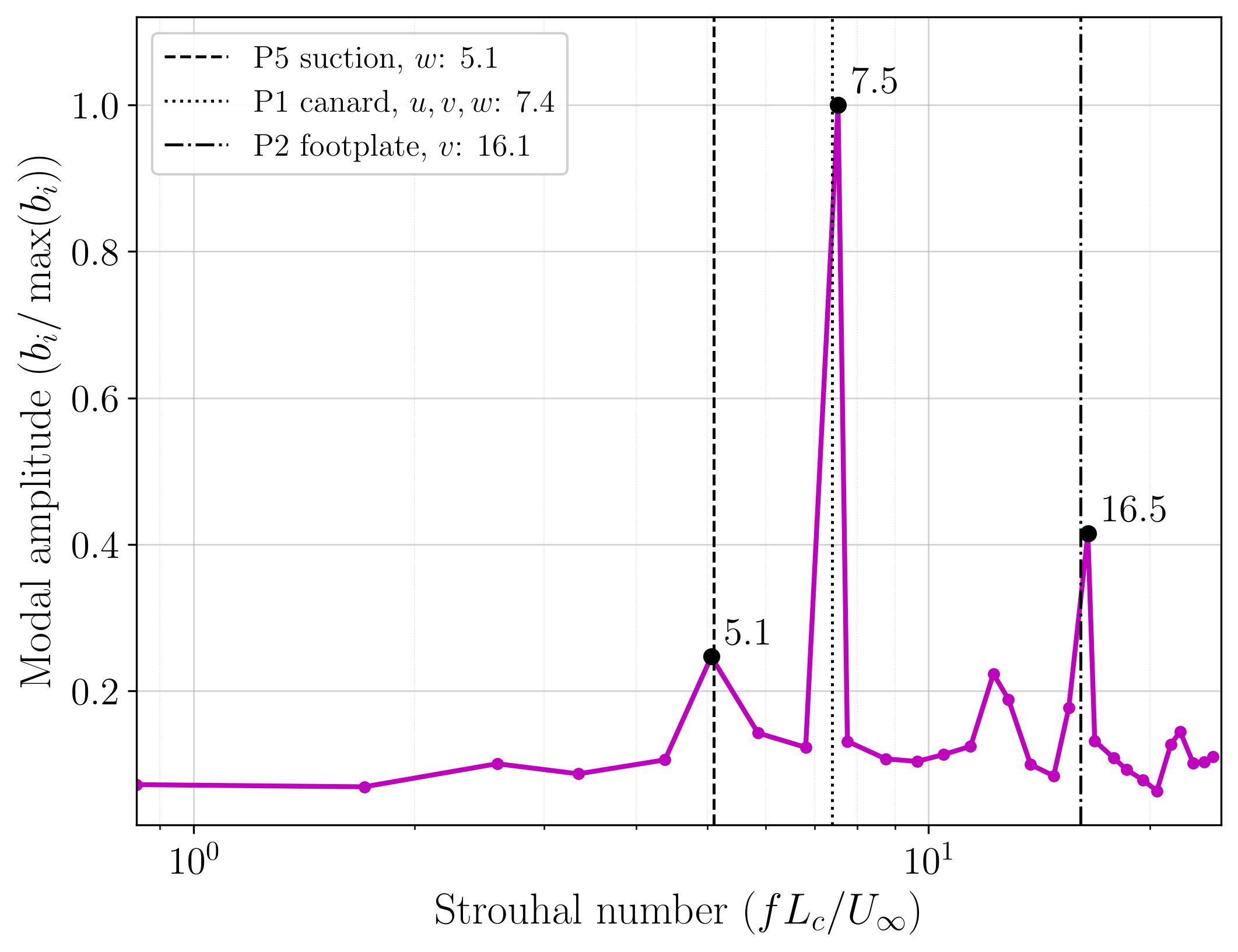}%
\label{fig:spectraCp0}%
}
\subfloat[Streamwise velocity, $u$]{%
\includegraphics[width=0.45\textwidth]{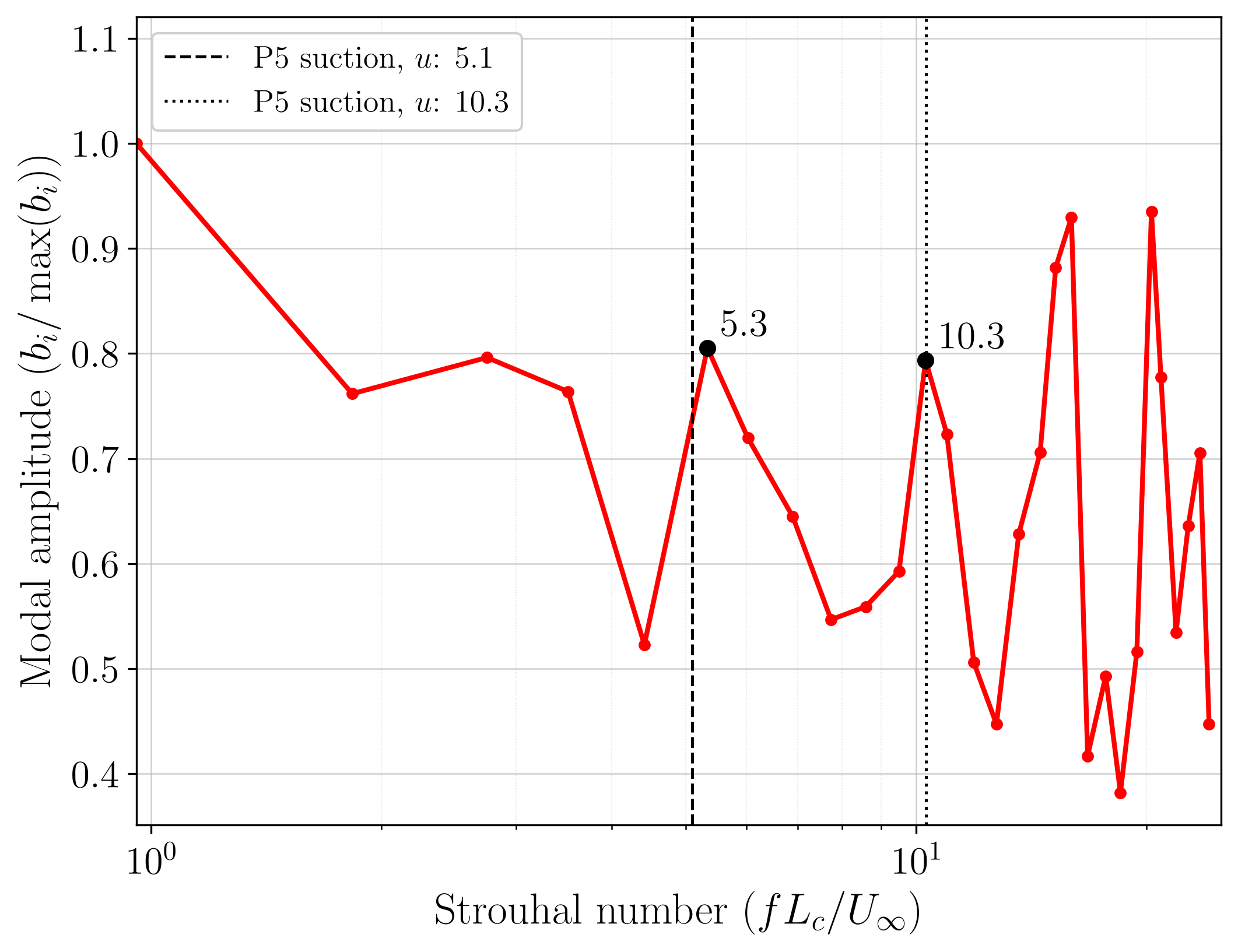}%
\label{fig:spectrau}%
}

\subfloat[Cross-stream velocity, $v$]{%
\includegraphics[width=0.45\textwidth]{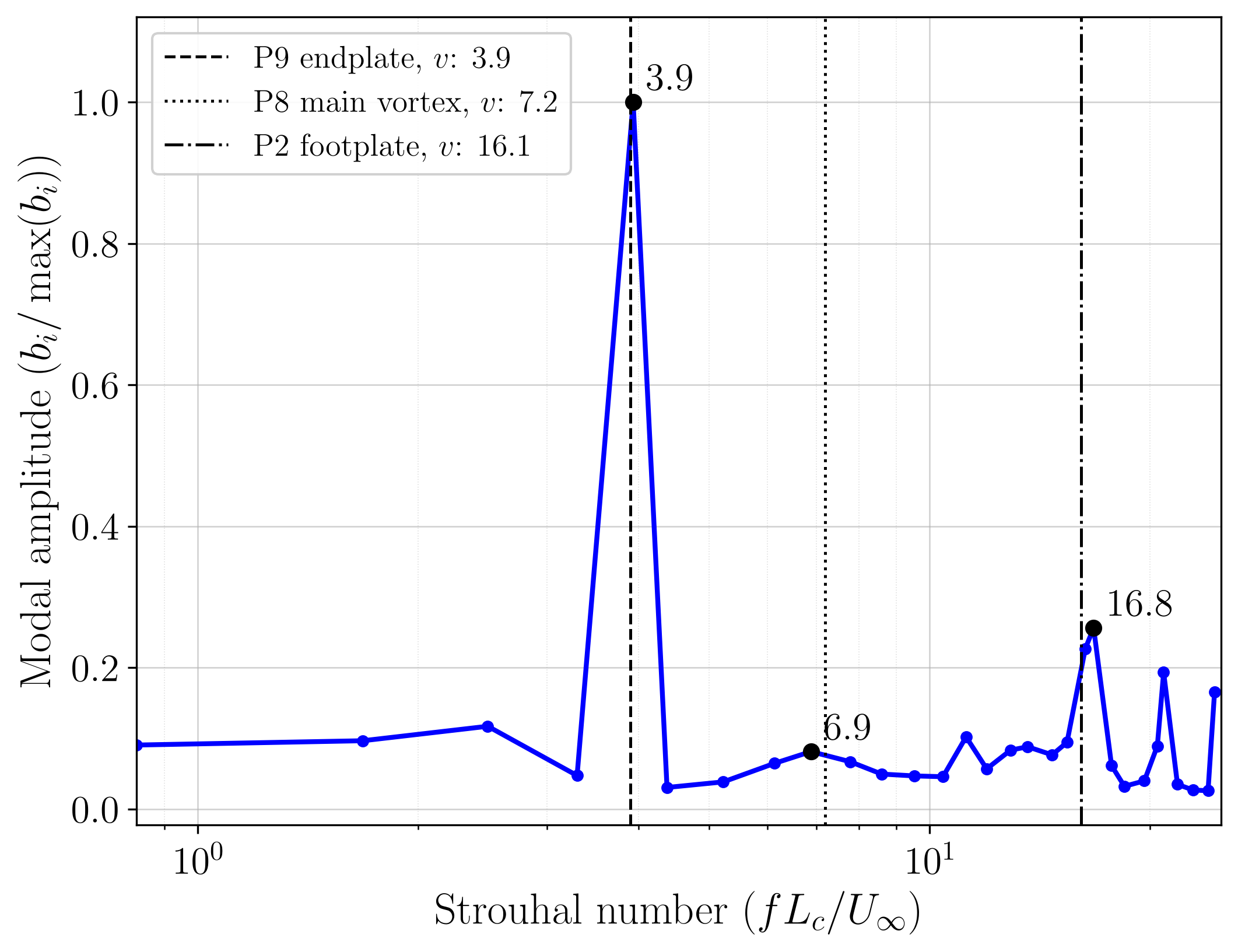}%
\label{fig:spectrav}%
}
\subfloat[Vertical velocity, $w$]{%
\includegraphics[width=0.45\textwidth]{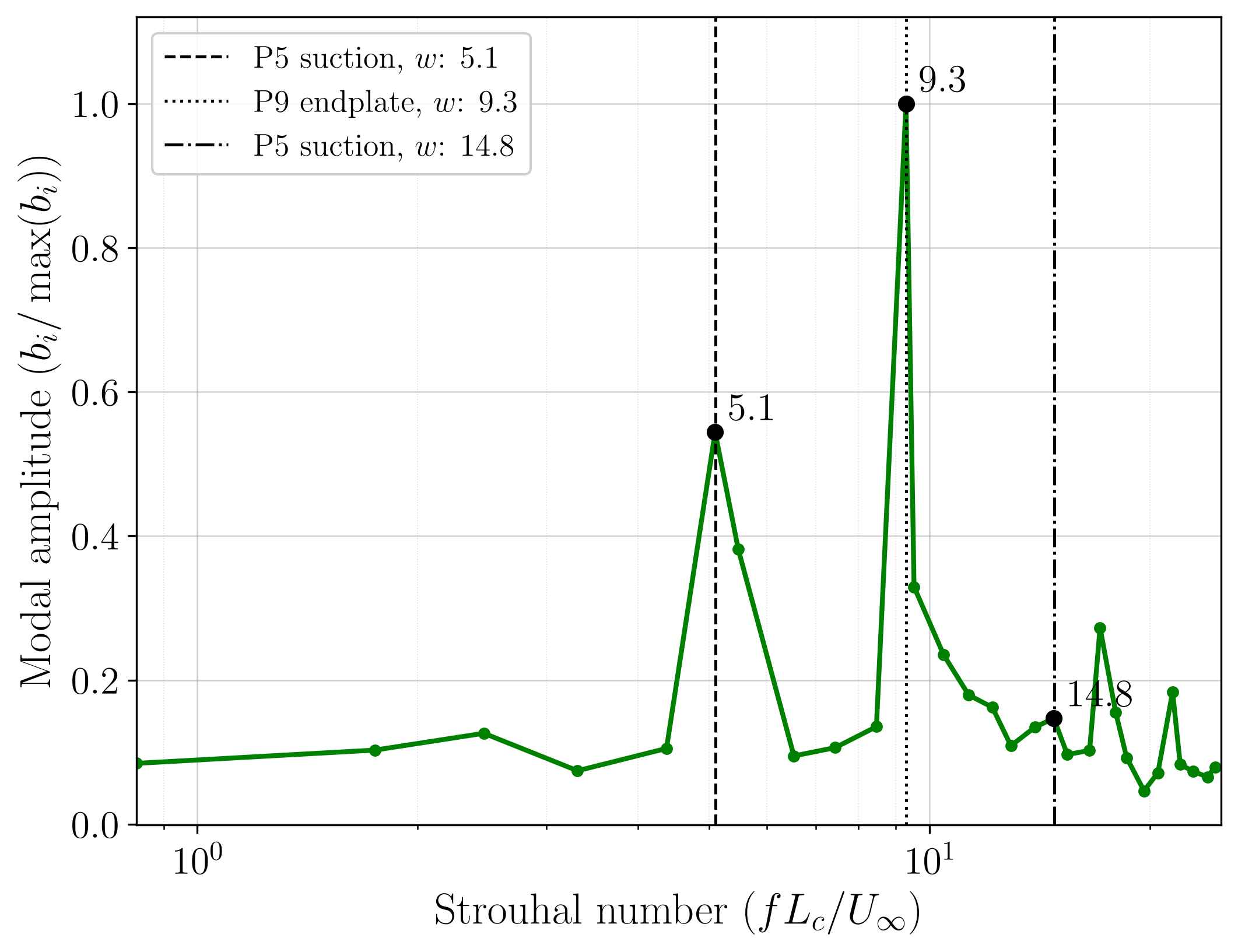}%
\label{fig:spectraw}%
}

\caption{DMD modal-amplitude spectra for $u$, $v$, $w$, and $C_{p0}$. The spectra show the normalised modal amplitude $b_i/\max(b_i)$ as a function of the nondimensional frequency $St = fL_c/U_{\infty}$. Each spectrum uses $N=300$ snapshots sampled at $250$~Hz ($T=1.2$~s), giving $\Delta St \approx 0.21$; reported frequencies are therefore rounded to one decimal.  Peaks are reported for $2 < St \leq 25$, with the lower bound set by the approximately ten oscillation periods and the upper bound set by the $-1$~dB point of the composite decimation filter. (Nyquist limit $St=31.25$). Vertical lines mark selected probe peaks from \autoref{tab:freqs}, with the probe, velocity component and $St$ given in the legend. Black markers denote the largest-amplitude DMD mode within the matching tolerance of each line ($\Delta St \leq 0.5$ for $St < 10$; $\Delta St/St \leq 0.08$ for $St \geq 10$).}
\label{fig:dmdspectra}
\end{figure}

The analysis of the DMD modes begins with the study of the dominant frequencies in the system described by each variable. \autoref{fig:dmdspectra} depicts the amplitude-frequency plot for each velocity component and for $C_{p0}$ before showing the spatial correlation of the most relevant modes. \autoref{fig:dmdspectra} also relates the dominant modes with the energy peaks identified in the probes from \autoref{sec:inst}. The spectrum of the streamwise velocity (\autoref{fig:spectrau}) is broadband, with no single dominant mode, so the discussion is led by the modes of $C_{p0}$, $v$ and $w$.

A DMD mode describes a coherent structure across the flow field at a single oscillation frequency. When a peak in a probe spectrum matches a modal frequency, the mode’s spatial pattern suggests which flow feature may contribute to the local fluctuations. The mode and the probe peak can come from different variables because a single flow feature can modulate several quantities. DMD is applied separately to each variable. Frequency agreement alone leaves the relative phase between variables and the direction and shape of the motion unresolved.

\autoref{tab:dmdmodes} lists the modes discussed in this section and their damping ratios, $\zeta_i=-\sigma_i/\sqrt{\sigma_i^2+\omega_i^2}$. The growth rate is $\sigma_i=\ln|\mu_i|/\Delta t$, and the angular frequency is $\omega_i=\theta_i/\Delta t$, where $\theta_i=\arg(\mu_i)$. All listed modes have negative growth rates. The two modes with the largest initial amplitudes, $C_{p0}$ at $fL_c/U_{\infty}=7.5$ and $v$ at $fL_c/U_{\infty}=3.9$, also have the highest damping ratios. In the DMD reconstruction, $|b_i|$ sets the modal amplitude at the first snapshot, and $\sigma_i$ governs its subsequent decay. Both modal frequencies are close to peaks in the probe spectra computed over the full record (\autoref{tab:freqs}).

\begin{table}[ht]
    \centering
    \small
    \caption{\small DMD modes discussed in this section: variable, nondimensional frequency, normalised amplitude (as in \autoref{fig:dmdspectra}) and damping ratio $\zeta$, together with the matching probe peaks (P1--P9) from \autoref{tab:freqs} and the associated flow feature.}
    \begin{tabular}{ccccp{4.3cm}p{3.8cm}}
        \hline
        \textbf{Variable} & \textbf{$fL_c/U_{\infty}$} & \textbf{$b_i/\max(b_i)$} & \textbf{$\zeta$} & \textbf{Matching probe peaks} & \textbf{Associated flow feature} \\
        \hline
        $C_{p0}$ & 7.5 & 1.00 & 0.28 & P1 $u,v,w$ 7.4; P9 $u,v$ 7.3; P8 $v,w$ 7.2 & Canard, endplate and main vortices \\
        \hline
        $w$ & 5.1 & 0.54 & 0.12 & P5 $u,w$ 5.1; P6 $u,w$ 5.2; P9 $w$ 5.2 & Suction side of the main element \\
        \hline
        $C_{p0}$ & 5.1 & 0.25 & 0.10 & P5 $u,w$ 5.1 & Suction side of the main element \\
        \hline
        $u$ & 5.3 & 0.81 & 0.06 & P5 $u$ 5.1; P6 $u$ 5.2; P7 $u$ 5.4 & Suction side of the main element \\
        \hline
        $v$ & 3.9 & 1.00 & 0.86 & P9 $v$ 3.9, $u$ 4.0, $w$ 3.6; P1, P2 $v$ 4.0 & Endplate vortex; $St \approx 4$ band of the vortex probes \\
        \hline
        $w$ & 9.3 & 1.00 & 0.12 & P9 $w$ 9.3 & Endplate vortex \\
        \hline
        $C_{p0}$ & 16.5 & 0.41 & 0.05 & P2 $v$ 16.1 & Shear layers of the wing elements and footplate \\
        \hline
        $v$ & 16.8 & 0.26 & 0.04 & P2 $v$ 16.1 & Footplate vortex \\
        \hline
    \end{tabular}
    \label{tab:dmdmodes}
\end{table}

The dominant $C_{p0}$ mode occurs at $fL_c/U_{\infty}=7.5$ (\autoref{fig:spectraCp0}), close to the $St=7.4$ peak at the canard vortex probe (Probe 1). This peak is the strongest in $v$ and $w$ and the second strongest in $u$. Peaks in the same band occur at the endplate vortex probe (Probe 9, $St=7.3$ in $u$ and $v$) and at the main vortex probe (Probe 8, $St=7.2$ in $v$ and $w$) (\autoref{tab:freqs}).

\begin{figure}[t]
    \centering
    \includegraphics[width=\textwidth]{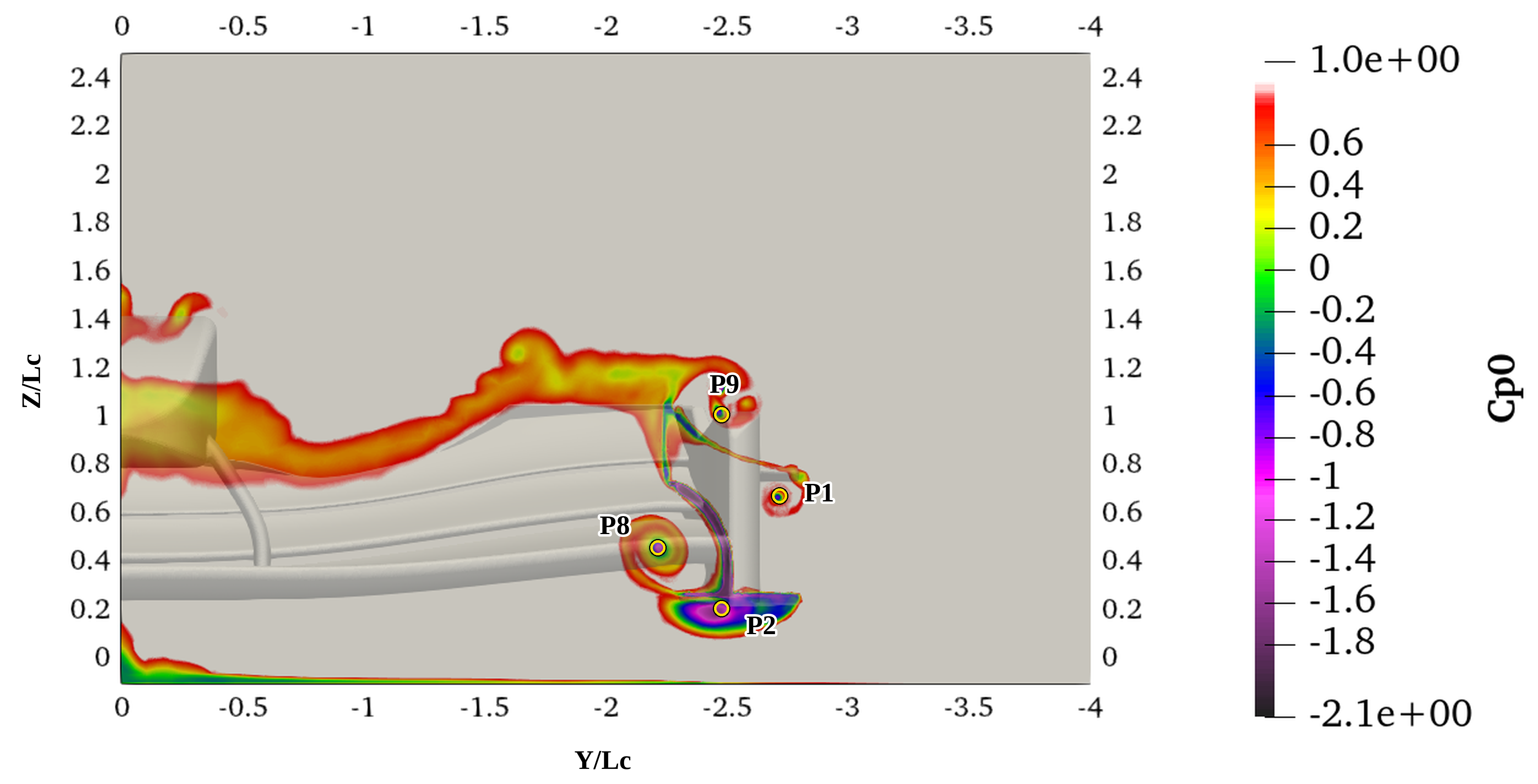}
    \caption{Time-averaged total pressure coefficient, $C_{p0}$, on PIV plane 1 ($x/L_c = -1.176$), looking downstream. The circles mark Probes 1 (canard vortex), 2 (footplate vortex), 8 (main vortex) and 9 (endplate vortex).}
    \label{fig:plane1}
\end{figure}

The probes were placed after the mean flow analysis, but only at approximate locations. On PIV plane 1, Probes 1, 8 and 9 lie approximately $0.5r_c$, $0.15r_c$ and $0.8r_c$ from their respective core centres (\autoref{fig:plane1}). The core radius $r_c$ is defined by the maximum time-averaged azimuthal velocity about the local $C_{p0}$ minimum. It is approximately $0.03L_c$ for the canard vortex and $0.06L_c$ for the main vortex. The latter radius is less sharply defined because the azimuthal velocity remains close to its maximum over a range of radii. For a small translation of an approximately axisymmetric core, the first-order changes in $u$, $C_{p0}$ and swirl speed depend on the radial component of the displacement relative to the probe. A displacement perpendicular to the line joining the core centre and the probe rotates the local swirl without changing its magnitude, $u$ or $C_{p0}$ to first order. At a core-centre extremum, the linear response of $u$ or $C_{p0}$ to translation vanishes. The cross-flow velocity can still respond at first order.

\begin{figure}[t]
    \centering
    \includegraphics[width=\textwidth]{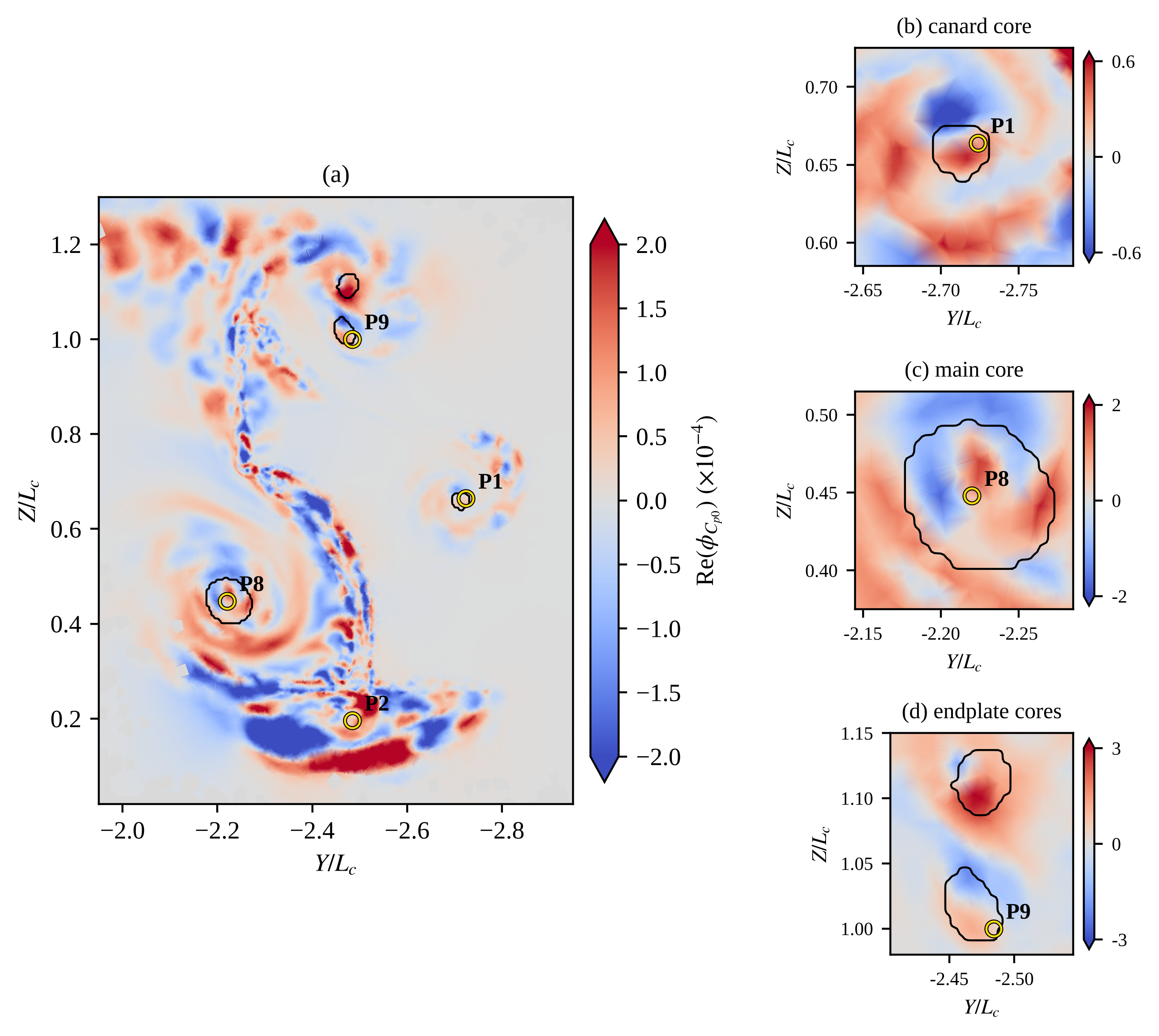}
    \caption{Real part of the DMD mode of the total pressure coefficient, $C_{p0}$, at $St = fL_c/U_{\infty} = 7.5$ on PIV plane 1 ($x/L_c = -1.176$), looking downstream. (a) Vortex region; (b) canard, (c) main and (d) endplate vortex cores. Crosses and dashed circles: centre (minimum of the time-averaged $C_{p0}$) and radius of maximum time-averaged azimuthal velocity of each vortex core. Circles: Probes 1, 2, 8 and 9. Each panel has its own symmetric colour range and is saturated outside it. Only the real part of the mode is available, so a single phase is shown.}
    \label{fig:plane1Cp0}
\end{figure}

A small translation of the total-pressure deficit produces a pair of lobes of opposite sign aligned with the displacement. A pulsation that changes the strength of the deficit can produce a single lobe centred on the core. In the $C_{p0}$ mode slice on PIV plane 1 (\autoref{fig:plane1Cp0}), the lobe pair at the main vortex aligns with $y$, consistent with lateral displacement (\autoref{fig:plane1Cp0}c). The small offset of Probe 8 is consistent with the absence of a detected $u$ peak in this band. The pair at the lower endplate vortex aligns approximately with $z$ (\autoref{fig:plane1Cp0}d). A vertical displacement would change $u$ and the swirl velocity at Probe 9, consistent with the peaks in $u$ and $v$. The upper endplate vortex shows a similar vertical pair with an additional positive lobe at the plotted phase. The real part of the mode is largest in the footplate vortex region.

At the canard vortex, the real part of the mode is below 20\% of its maximum in the plotted region, and the lobe pair is aligned approximately with $z$ (\autoref{fig:plane1Cp0}b). This pattern is consistent with vertical displacement. Probe 1 lies to the side of the core, where such a displacement would have little first-order effect on $u$. A displacement component along the line from the core centre to the probe, or a pulsation of the axial velocity, could account for the $u$ peak at $St=7.4$. The canard association therefore rests mainly on the frequency match at Probe 1, and the core motion remains unresolved.

Only the real part of the mode was stored. The slice thus shows the spatial pattern at one modal phase and leaves the relative phase between cores unresolved. The fitted damping ratio is $\zeta=0.28$ (\autoref{tab:dmdmodes}). It describes the amplitude decay in the DMD reconstruction, while the persistence of the physical motion over the record remains unresolved.

\begin{figure}[t]
    \centering
    \subfloat[]{\includegraphics[width=0.49\textwidth]{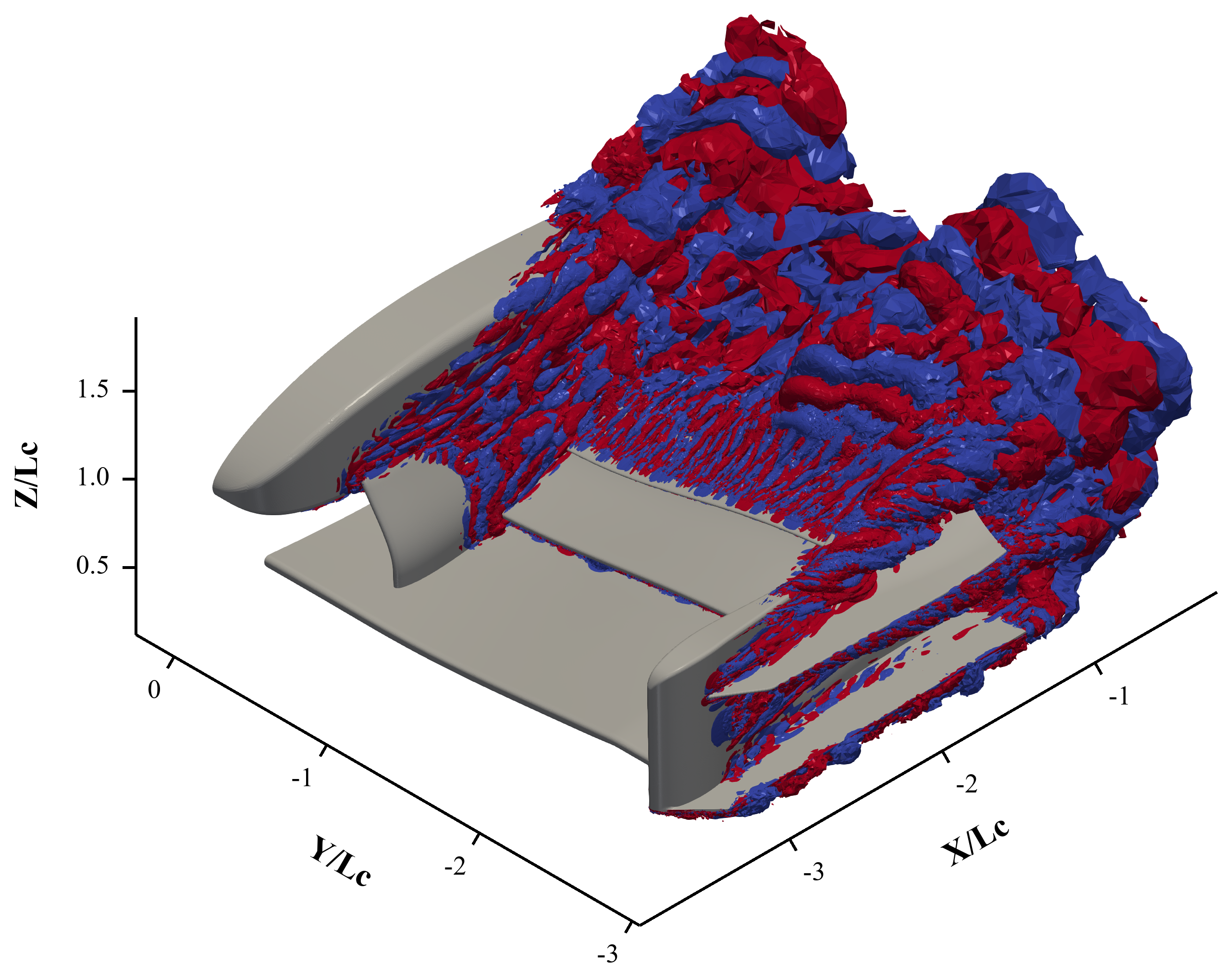}}\hfill
    \subfloat[]{\includegraphics[width=0.49\textwidth]{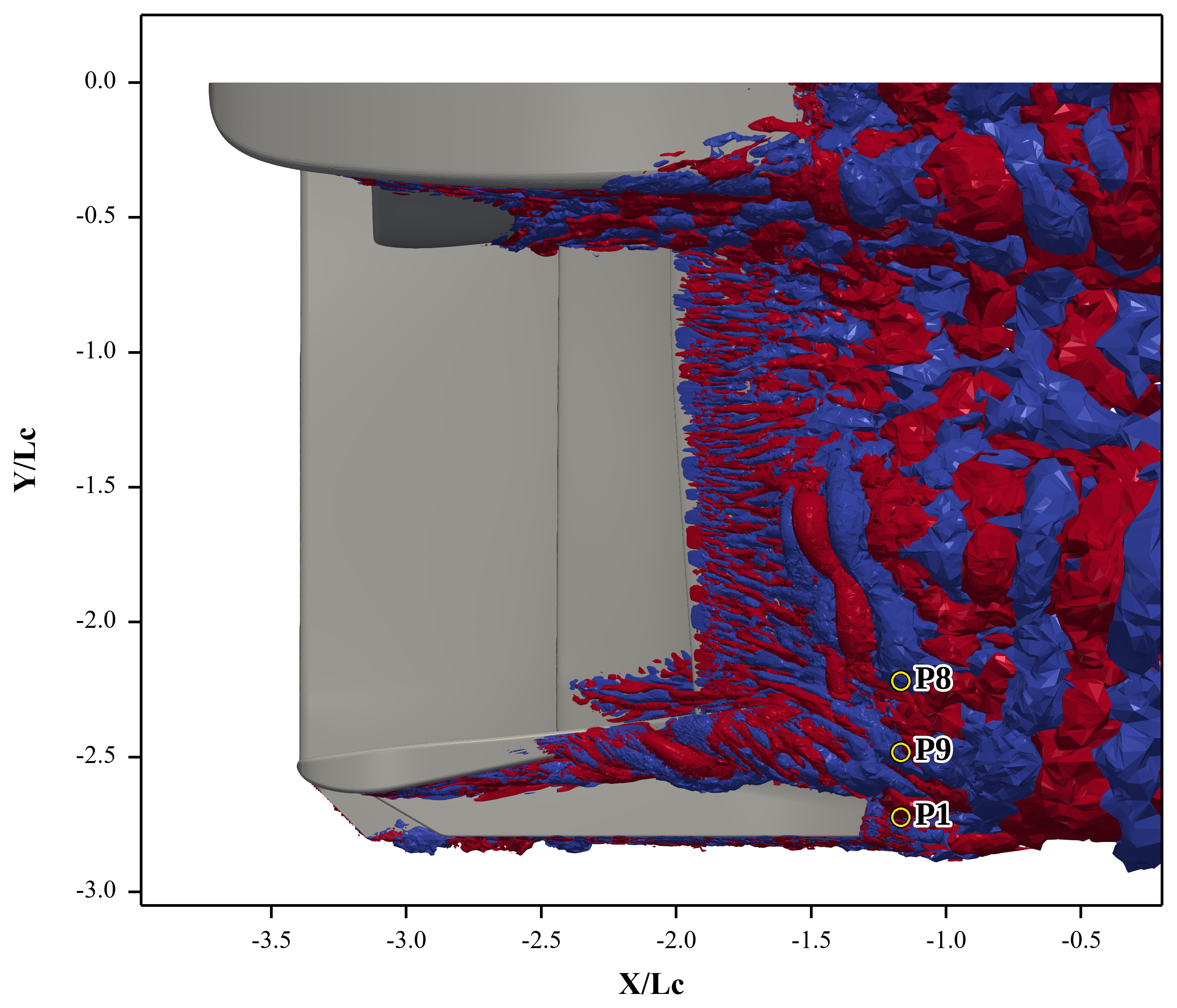}}

    \subfloat[]{\includegraphics[width=0.7\textwidth]{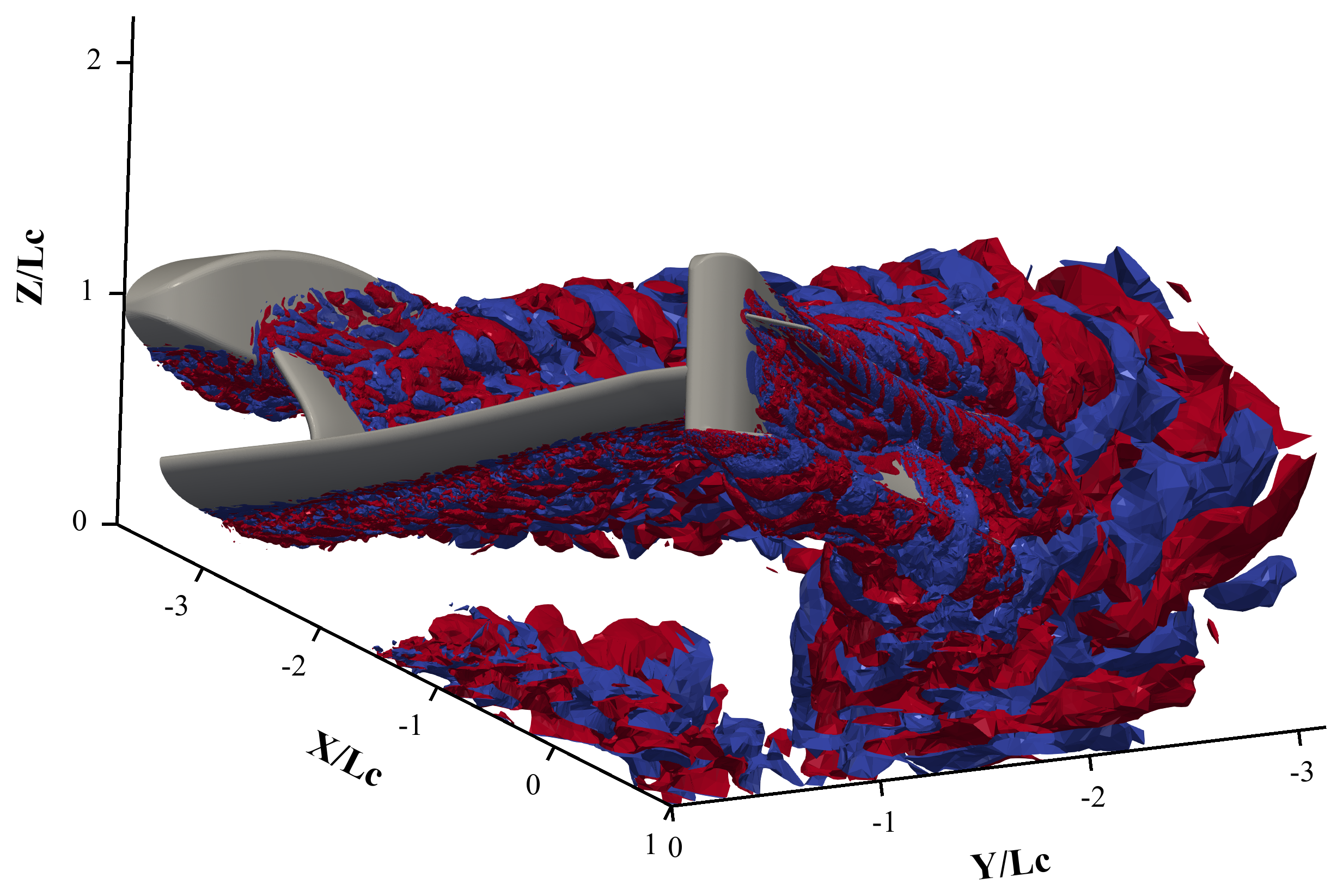}}
    \caption{DMD mode of the total pressure coefficient, $C_{p0}$, at $St = fL_c/U_{\infty} = 7.5$: isosurfaces of the real part of the mode at $\pm 1\times10^{-5}$ (red: positive, blue: negative). (a) Isometric view; (b) top view; (c) rear view from low elevation. The circles in (b) mark Probes 1, 8 and 9 on PIV plane 1 ($x/L_c = -1.176$), projected onto the view.}
    \label{fig:dmdCp0}
\end{figure}

The spatial structure linked to this mode is depicted in \autoref{fig:dmdCp0}. The mode has support across the vortex system downstream of the endplate, where Probes 1, 8 and 9 are located (\autoref{fig:dmdCp0}b), and in the wakes of the wing and the nose, in agreement with the presence of this band at several vortex probes. It also has support near the ground, downstream of the wing and close to the centreline (\autoref{fig:dmdCp0}c).

\begin{figure}[t]
    \centering
    \subfloat[]{\includegraphics[width=0.49\textwidth]{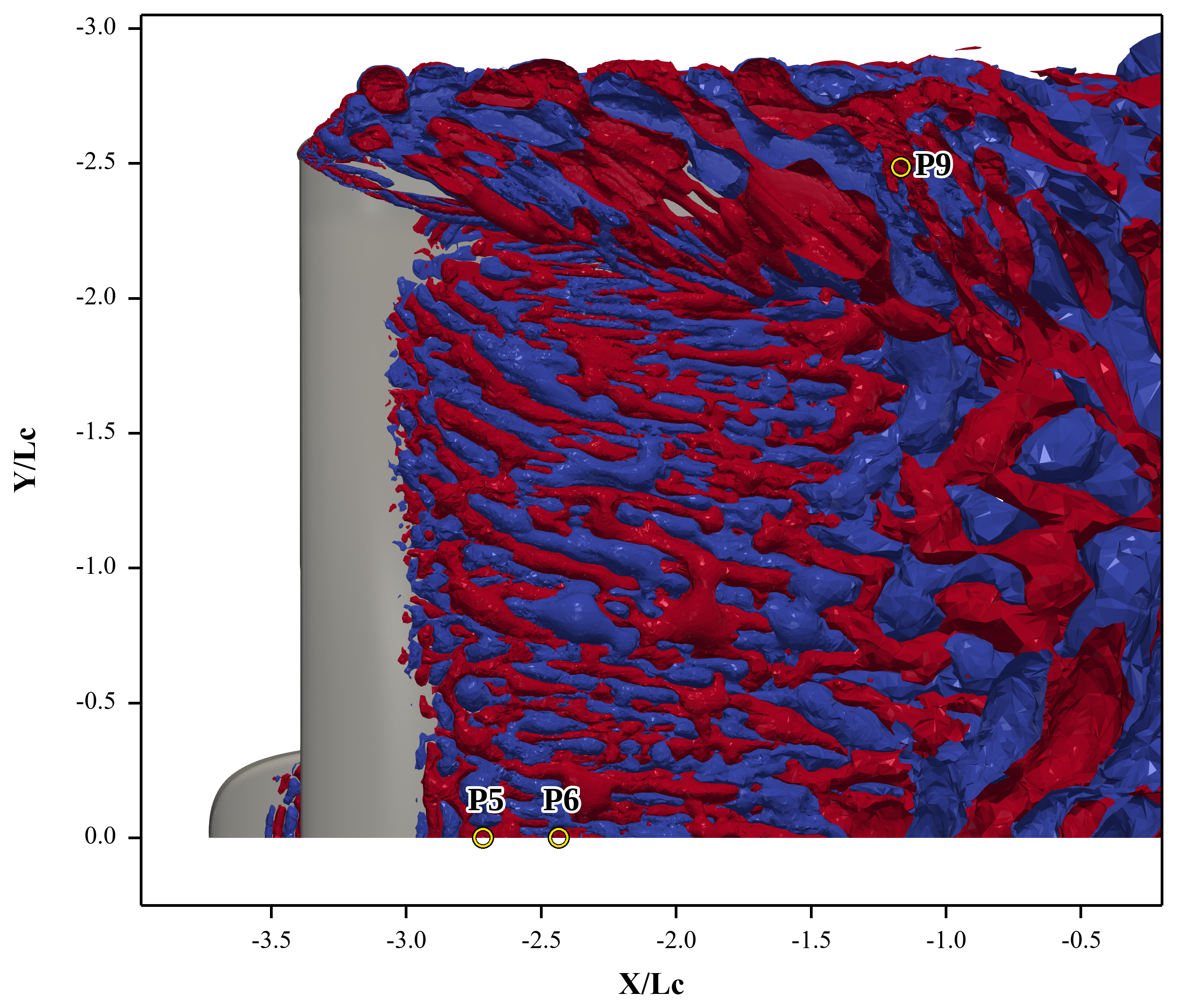}}\hfill
    \subfloat[]{\includegraphics[width=0.49\textwidth]{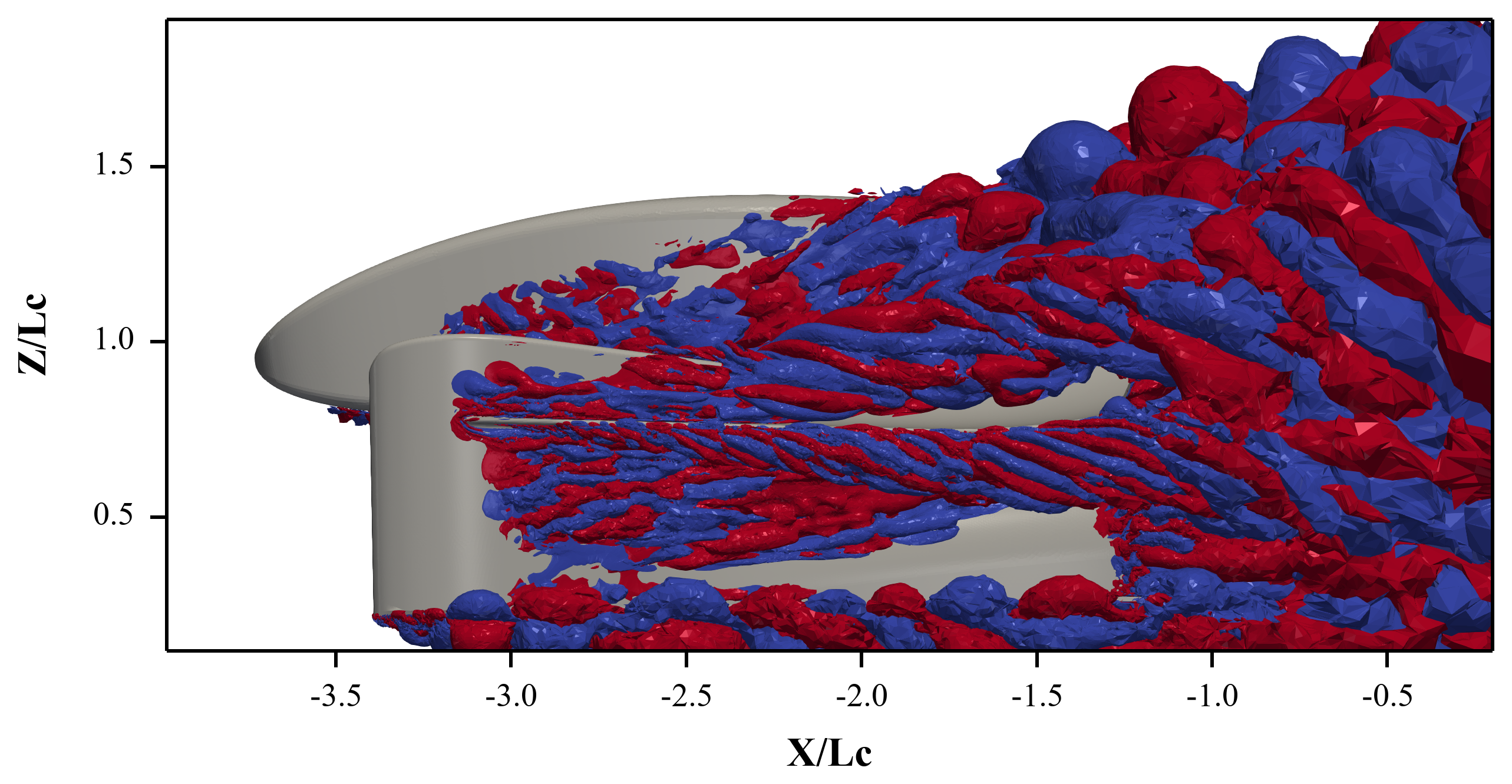}}
    \caption{DMD mode of the vertical velocity, $w$, at $St = fL_c/U_{\infty} = 5.1$: isosurfaces of the real part of the mode at $\pm 1\times10^{-5}$ (red: positive, blue: negative). (a) Bottom view; (b) side view from the outboard side. The circles in (a) mark Probes 5 and 6 on the main element ($y/L_c = 0$) and Probe 9, projected onto the view.}
    \label{fig:dmd2w}
\end{figure}

The third most energetic $C_{p0}$ mode occurs at $fL_c/U_{\infty}=5.1$, and the same frequency is more prominent in the vertical velocity, where it is the second most energetic mode after the $fL_c/U_{\infty}=9.3$ mode discussed below (\autoref{fig:spectraw}). It is also present in the streamwise velocity, whose DMD spectrum has a local maximum at $fL_c/U_{\infty}=5.3$ (\autoref{fig:spectrau}). This band matches the most prominent tone of the probe on the suction side of the main element (Probe 5), located to study the laminar-turbulent transition of the main wing element, which shows $St=5.1$ in both the streamwise and vertical velocity components (\autoref{tab:freqs}). At the Y250 section, the mean separation and reattachment points of the main element are located at $x/L_c=-3.00$ and $x/L_c=-2.56$ (\autoref{App:IFW_mean_surface}), so Probe 5 ($x/L_c=-2.72$) lies within the streamwise extent of the bubble, although these locations vary along the span. The mode is therefore interpreted as an up-and-down oscillation of the flow over the suction side of the wing, consistent with the laminar separation bubble described in \autoref{App:IFW_mean_surface}. As for the vortex cores, a vertical displacement of the separated shear layer moves the strong wall-normal gradient of the mean streamwise velocity across the probe, so the same tone also appears in $u$. The associated spatial structure can be seen in \autoref{fig:dmd2w}. Besides the structures over the suction side of the main element, the mode also has support along the endplate vortex, in agreement with the $St=5.2$ vertical-velocity peak of the endplate vortex probe (Probe 9). The streamwise velocity spectrum also shows a mode at $fL_c/U_{\infty}=10.3$ (\autoref{fig:spectrau}), which matches the $St \approx 10.2$ band of the surface probes and, at twice the frequency of this band, is a candidate second harmonic (\autoref{sec:inst}).

The most energetic mode of the cross-stream velocity dominates the spectrum of \autoref{fig:spectrav} and occurs at $fL_c/U_{\infty}=3.9$. It coincides with the most prominent cross-stream tone of the endplate vortex probe (Probe 9, $St=3.9$), which also shows this band in the streamwise ($St=4.0$) and vertical ($St=3.6$) components (\autoref{tab:freqs}). Probe 9 lies close to the centre of the endplate vortex core, slightly below it (\autoref{fig:plane1}), so the line joining the core centre and the probe is approximately vertical. Following the argument given for the canard vortex, the response of $u$ together with $v$ is consistent with a motion of the endplate vortex core with a vertical (up-and-down) component at this frequency: such a displacement moves the probe across the mean axial-velocity gradient of the core and changes the magnitude of the horizontal swirl velocity measured at the probe. The response of $w$ suggests that this motion also has a lateral component. Cross-stream peaks at $St=4.0$ are also found at the canard and footplate vortex probes (Probes 1 and 2), consistent with the $St \approx 4$ band shared by the vortex-core probes (\autoref{sec:inst}).
The isosurfaces of the $v$ mode at $fL_c/U_{\infty}=3.9$ are omitted because organised structures appear only in the far wake at the level used for the other modes. This mode has the largest fitted damping ratio among those discussed ($\zeta=0.86$, \autoref{tab:dmdmodes}). In the DMD reconstruction, its amplitude falls to $1/e$ of its initial value in about one tenth of an oscillation period. The spectra at Probes 1, 2 and 9 each contain a peak near $St=4$ (\autoref{tab:freqs}). Cross-spectral analysis of the probe signals would be needed to assess coherence and relative phase in this band.

\begin{figure}[t]
    \centering
    \subfloat[]{\includegraphics[width=0.49\textwidth]{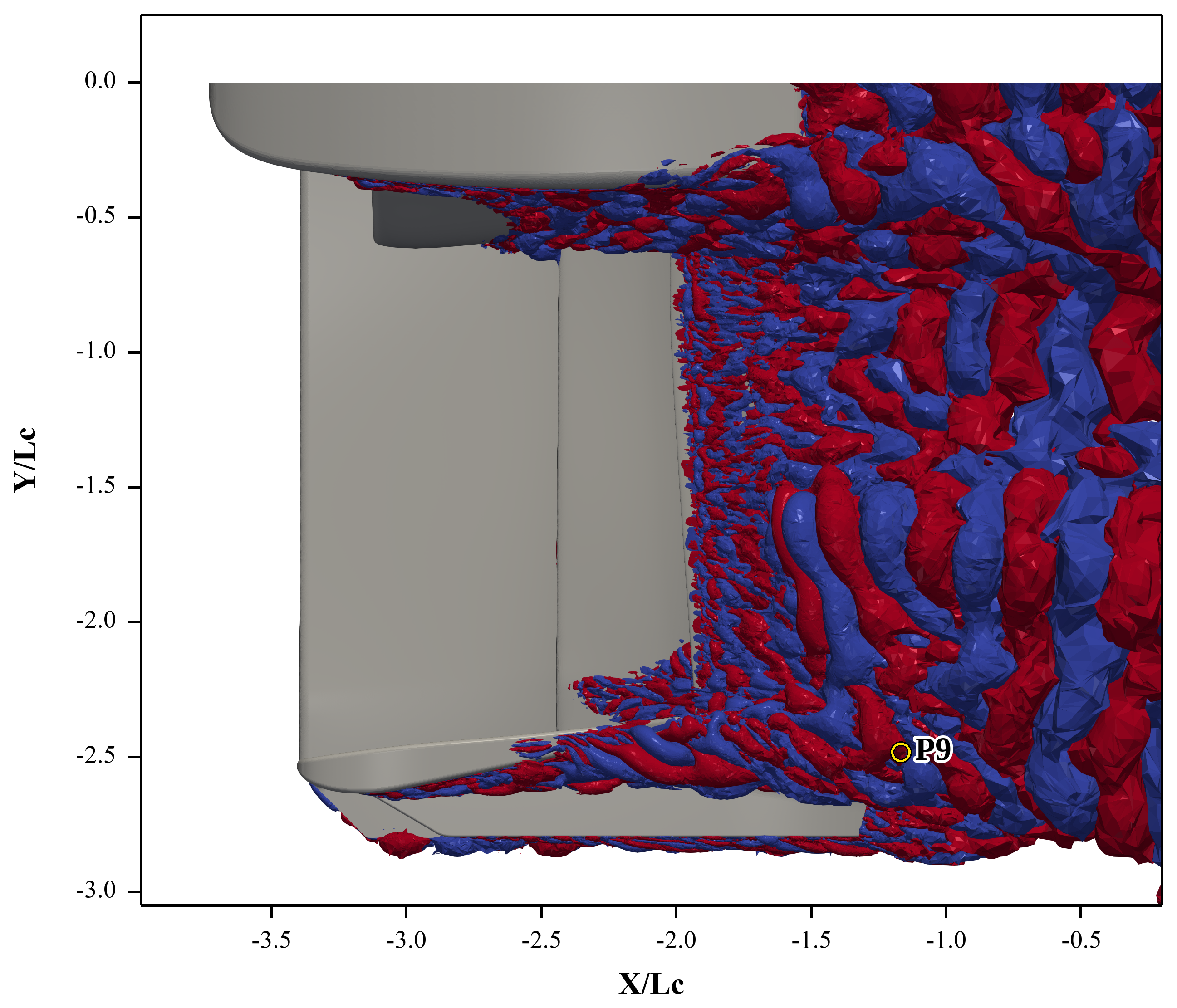}}\hfill
    \subfloat[]{\includegraphics[width=0.49\textwidth]{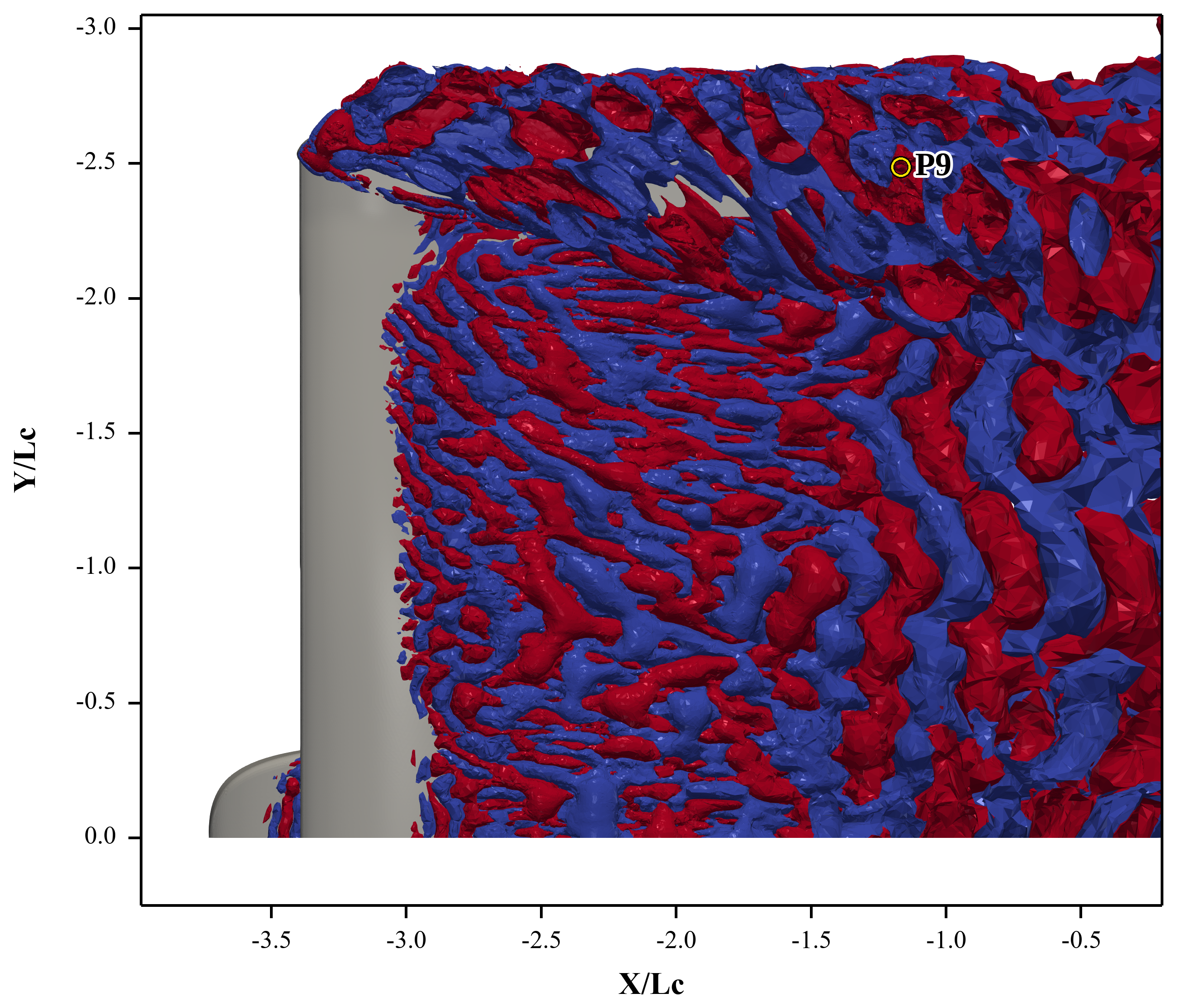}}
    \caption{DMD mode of the vertical velocity, $w$, at $St = fL_c/U_{\infty} = 9.3$: isosurfaces of the real part of the mode at $\pm 1\times10^{-5}$ (red: positive, blue: negative). (a) Top view; (b) bottom view. The circles mark Probe 9, projected onto the view.}
    \label{fig:dmd1w}
\end{figure}
\begin{figure}[t]
    \centering
    \includegraphics[width=\textwidth]{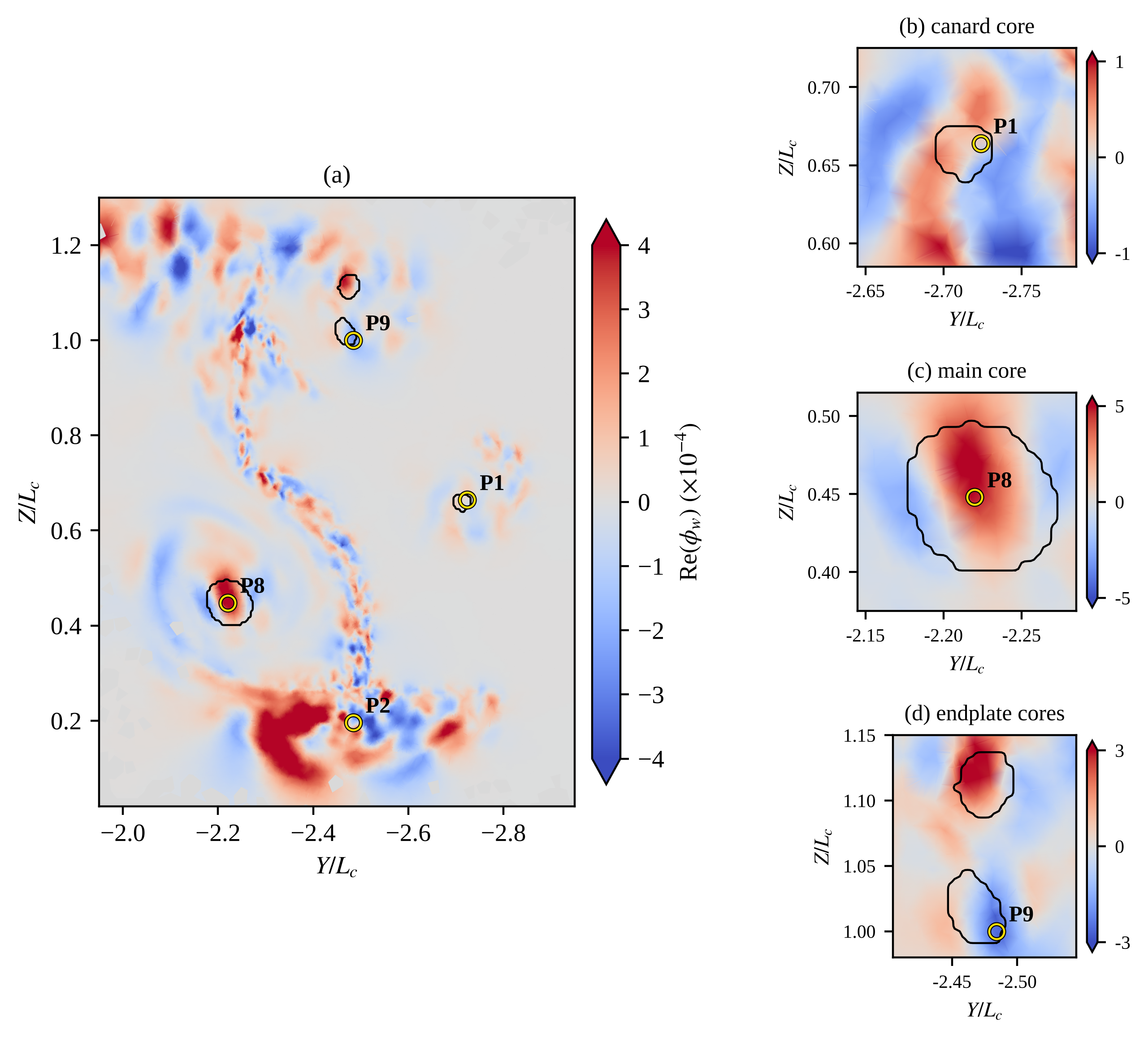}
    \caption{Real part of the DMD mode of the vertical velocity, $w$, at $St = fL_c/U_{\infty} = 9.3$ on PIV plane 1 ($x/L_c = -1.176$), looking downstream. (a) Vortex region; (b) canard, (c) main and (d) endplate vortex cores. Crosses and dashed circles: centre (minimum of the time-averaged $C_{p0}$) and radius of maximum time-averaged azimuthal velocity of each vortex core. Circles: Probes 1, 2, 8 and 9. Each panel has its own symmetric colour range and is saturated outside it. Only the real part of the mode is available, so a single phase is shown.}
    \label{fig:plane1w}
\end{figure}

The dominant $w$ mode occurs at $fL_c/U_{\infty}=9.3$ (\autoref{fig:spectraw}), matching the vertical velocity peak at the endplate vortex probe (Probe 9) to the reported precision. The mode has a fitted damping ratio of $\zeta=0.12$ (\autoref{tab:dmdmodes}). At this frequency, only the $w$ peak meets the prominence criterion at Probe 9 (\autoref{tab:freqs}). The probe lies just below the centre of the lower endplate vortex core (\autoref{fig:plane1}), where a small lateral displacement of an approximately axisymmetric core would rotate the local swirl velocity with little first-order effect on $u$. The isosurfaces show a train of spanwise-coherent structures of alternating sign in the wake of the main element and flap, extending along the endplate to Probe 9 (\autoref{fig:dmd1w}). On PIV plane 1, a single lobe is centred on the upper endplate vortex core, as expected for a small lateral displacement under the solid-body rotation approximation (\autoref{fig:plane1w}d). At the phase shown, the lower endplate lobe is weaker and has the opposite sign. It lies at the lower edge of the core near Probe 9, giving less direct evidence of lateral displacement there. The $St=9.3$ mode is therefore associated with a lateral displacement of the endplate vortex, shown most clearly by the upper endplate core; at Probe 9 the association rests on the frequency match and on the weaker lobe at the edge of the lower core. The strongest centred lobe lies on the main vortex core (\autoref{fig:plane1w}c). Probe 8 lies within this lobe, close to the core centre. Its nearest peak in $w$ is at $St=8.6$, outside the matching tolerance (\autoref{tab:freqs}). A response at $St=9.3$ could fall below the prominence threshold in the broadband spectrum at Probe 8.

\begin{figure}[t]
    \centering
    \subfloat[]{\includegraphics[width=0.49\textwidth]{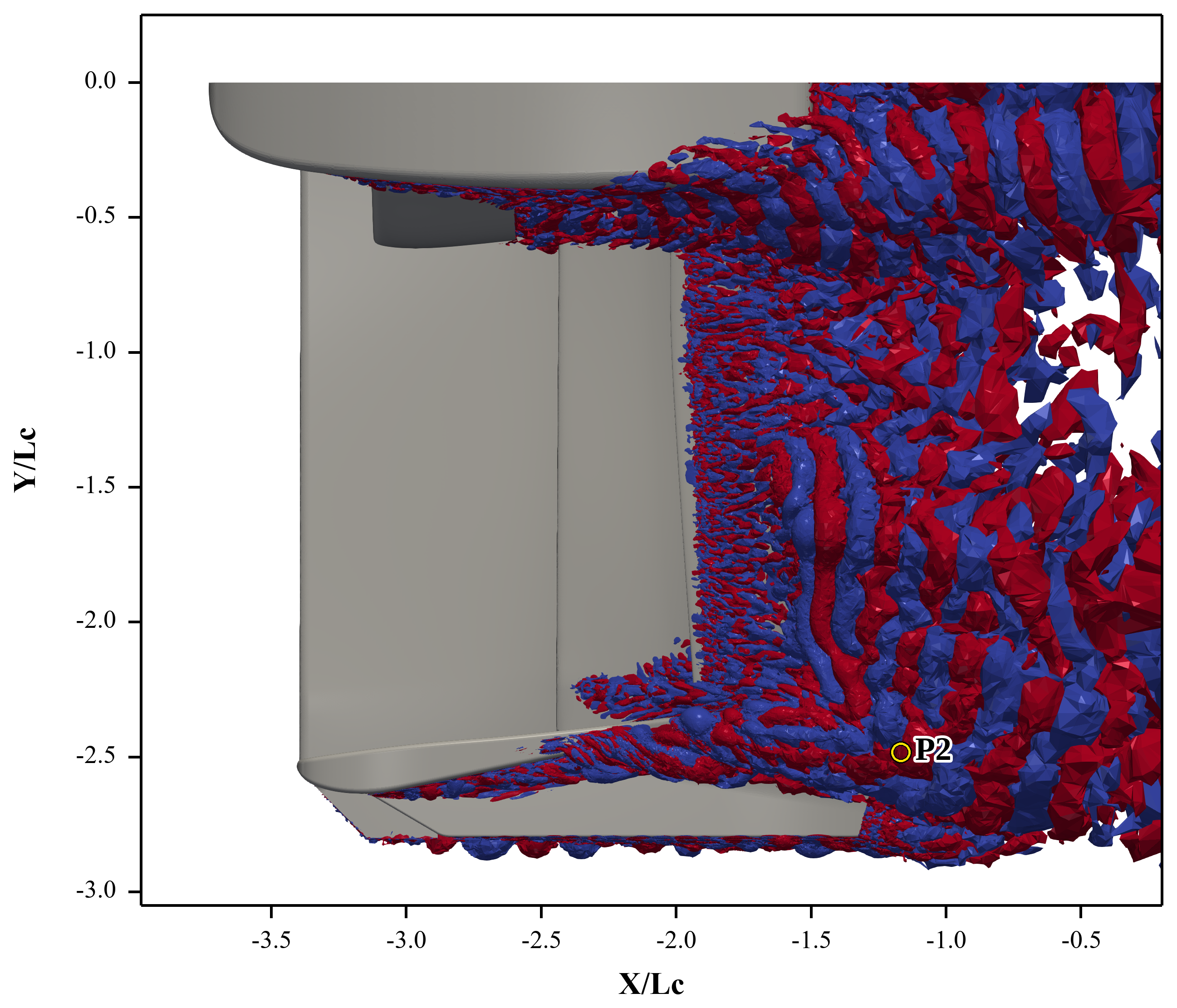}}\hfill
    \subfloat[]{\includegraphics[width=0.49\textwidth]{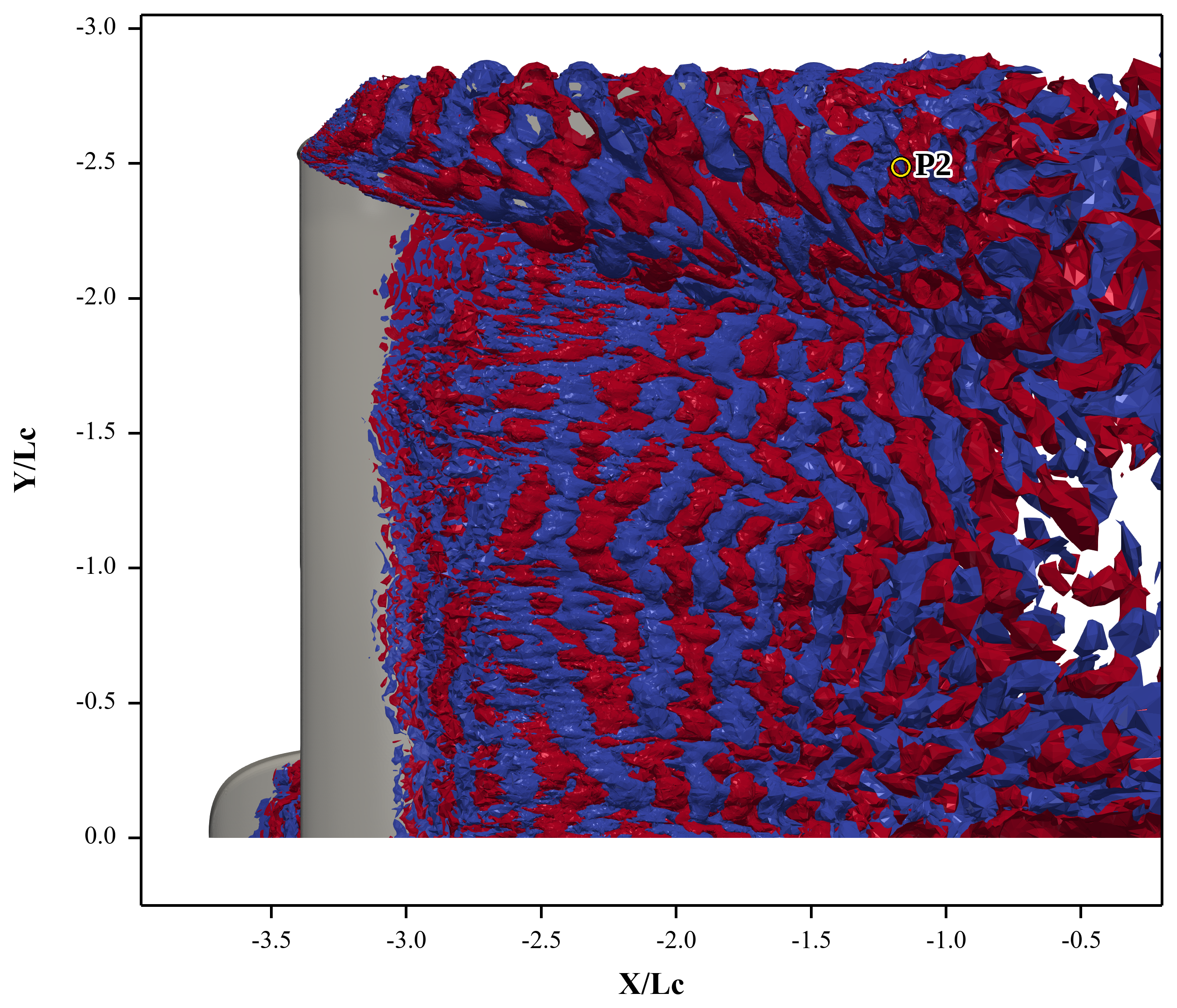}}
    \caption{DMD mode of the total pressure coefficient, $C_{p0}$, at $St = fL_c/U_{\infty} = 16.5$: isosurfaces of the real part of the mode at $\pm 1\times10^{-5}$ (red: positive, blue: negative). (a) Top view; (b) bottom view. The circles mark Probe 2, projected onto the view.}
    \label{fig:dmdu2}
\end{figure}

Finally, the second most energetic $C_{p0}$ mode, at $fL_c/U_{\infty}=16.5$, has weaker probe support. Together with a secondary cross-stream mode at $fL_c/U_{\infty}=16.8$ (\autoref{fig:spectrav}), it lies within the matching tolerance of the $St=16.1$ cross-stream tone of the footplate vortex probe (Probe 2), the only probe peak in this frequency range (\autoref{tab:freqs}). At this frequency the tolerance is approximately $\pm 1.3$, so this correspondence is treated as tentative. Its spatial structure (\autoref{fig:dmdu2}) shows spanwise-coherent bands over the suction side of the wing elements and in their near wake, together with chains of alternating structures along the edges of the endplate and the footplate. This pattern is consistent with the roll-up of the shear layers and with the location of Probe 2 under the footplate. These two modes are also the least damped of those in \autoref{tab:dmdmodes}.

Two weaker correspondences are also marked in \autoref{fig:dmdspectra}: the cross-stream mode at $fL_c/U_{\infty}=6.9$ lies within the matching tolerance of the $St=7.2$ cross-stream peak of the main vortex probe (Probe 8), and the vertical-velocity mode at $fL_c/U_{\infty}=14.8$ matches the $St=14.8$ peak of Probe 5. Both modes have small amplitudes and are not analysed further. The modes at $fL_c/U_{\infty} \approx 20$--$22$, present in all four variables, have no counterpart in the probe spectra (\autoref{tab:freqs}) and are therefore not interpreted.

\section{Conclusions}
\label{sec:conclusions}

This work showcases a comprehensive workflow for the scalable modal analysis of complex motorsport aerodynamics, combining high-fidelity large eddy simulation with dynamic mode decomposition to extract physically meaningful coherent structures from terabyte-scale transient simulation data. The spectral/hp element method implementation in \nekpp successfully resolved the flow around the Imperial Front Wing at Re = 2.2$\times$10$^5$ with 49.72$\times$10$^6$ degrees of freedom, capturing the four dominant vortical structures (endplate, canard, main, and footplate vortices) with time-averaged force coefficients showing excellent agreement with existing literature (2.24\% and 2.87\% differences in $C_L$ and $C_D$, respectively). The DMD analysis, performed using the open-source pyLOM framework, identified coherent flow structures across the complex wake system along with their associated temporal frequencies, amplitudes, and damping ratios, providing systematic insight into the unsteady aerodynamic phenomena that are critical for performance optimisation in motorsport applications.

The modal analysis relates the dominant frequencies of the flow to specific regions of the IFW by comparing the DMD modes of $u$, $v$, $w$ and $C_{p0}$ with the velocity spectra of nine probes. The most dominant $C_{p0}$ mode, at $fL_c/U_{\infty}=7.5$, matches a tone that is present in all three velocity components at the canard vortex probe and is shared with the endplate and main vortex probes; on PIV plane 1, the mode shows a lateral displacement of the main vortex core and a vertical displacement of the endplate vortex core, consistent with the velocity components in which the tone appears at Probes 8 and 9, while at the canard vortex core it shows a weak vertical pattern that does not, on its own, explain the response at the canard vortex probe. The $fL_c/U_{\infty} \approx 5.1$ band, found in $w$, $C_{p0}$ and $u$, is linked to the suction side of the main element and is consistent with the laminar separation bubble. At the endplate vortex probe, the tones matching the most dominant modes of $v$ ($fL_c/U_{\infty}=3.9$) and $w$ ($fL_c/U_{\infty}=9.3$) are consistent with core motions with a vertical component and in the lateral direction, respectively; the $w$ mode itself is dominated by a wavetrain in the wake of the wing that also extends along the endplate, and on PIV plane 1 it shows lateral displacements of the vortex cores, most clearly for the upper endplate core and the main vortex, the latter without a matching peak at the main vortex probe. The $St \approx 4$ band also appears at the canard and footplate vortex probes, although no phase relation between these probes has been established. A further $C_{p0}$ mode at $fL_c/U_{\infty}=16.5$ is tentatively associated with the roll-up of the shear layers of the wing elements and the footplate. These links are associations based on frequency agreement and on the spatial support of the modes. DMD is applied separately to each variable and to a record of 300 snapshots, and the two modes with the largest initial amplitudes also have the highest damping ratios ($\zeta=0.28$ and $\zeta=0.86$), so the mechanism and the phase relation between variables are not established. From a design perspective, these results indicate which component of the wing is associated with a given unsteady frequency in the wake.

The scalable implementation of POD and DMD methods presented here addresses a fundamental challenge in modern computational fluid dynamics: extracting interpretable physical insights from simulations that generate exceptionally high-dimensional state-spaces. While traditional analysis approaches are limited to judiciously selected planes or time-averaged quantities due to computational constraints, the parallel HPC-accelerated framework enables full 3D spatio-temporal analysis of the complete flow field. The DMD  reveals not only the dominant spatial structures but also their temporal evolution characteristics, bridging the gap between the capability to simulate complex unsteady flows and the ability to understand the underlying physics. This integrated workflow offers motorsport engineers a systematic methodology to leverage LES results for design optimisation, moving beyond simple force coefficient comparisons to detailed understanding of vortex dynamics and their contributions to aerodynamic performance.

The industrial relevance of this approach extends beyond the specific case of the Imperial Front Wing. The demonstrated capability to perform modal decomposition on meshes with tens of millions of degrees of freedom opens new possibilities for analysing other complex motorsport geometries where unsteady flow features dominate performance. The DMD additionally serves dual purposes: as both a model reduction technique for understanding dominant flow physics and as a compression scheme for efficient storage of simulation data, here reducing approximately 6~TB of snapshots to approximately 200~GB of modal data for the four variables. Future work will focus on connecting the identified coherent structures and their frequencies to specific design features, enabling targeted geometric modifications for performance enhancement. Furthermore, the integration of this analysis framework into existing industrial CFD pipelines could provide competitive advantages through deeper understanding of transient aerodynamic phenomena that are inaccessible to traditional RANS-based approaches. The open-source nature of both \nekpp and pyLOM facilitates adoption of this methodology across the broader motorsport and automotive industries, democratising access to advanced computational tools that were previously limited to research environments.

The successful validation against experimental PIV data and the clear identification of dominant frequencies associated with different vortical structures demonstrate that this workflow is sufficiently mature for practical engineering applications. The ability to resolve the laminar-to-turbulent transition on wing elements, capture the complex vortex interactions downstream, and systematically decompose these phenomena into interpretable modes represents a significant advancement in motorsport aerodynamics analysis. As computational resources continue to grow and simulation capabilities expand to even higher Reynolds numbers and more complex geometries, scalable modal decomposition methods will become increasingly essential for extracting actionable engineering insights from the wealth of data generated by high-fidelity simulations.

\section*{CRediT authorship contribution statement}
\textbf{Parv Khurana:} Conceptualization, Data curation, Formal analysis, Investigation, Methodology, Software, Validation, Visualization, Writing -- original draft. \textbf{Benet Eiximeno:} Conceptualization, Data curation, Formal analysis, Investigation, Methodology, Software, Validation, Visualization, Writing -- original draft. \textbf{Alexandra Liosi:} Data curation, Investigation, Methodology, Software, Validation, Writing -- review \& editing. \textbf{Spencer Sherwin:} Conceptualization, Funding acquisition, Project administration, Resources, Writing -- review \& editing. \textbf{Athanasios Chatzopoulos:} Funding acquisition, Resources, Supervision, Writing -- review \& editing. \textbf{Julien Hoessler:} Conceptualization, Funding acquisition, Project administration, Resources, Supervision. \textbf{Arnau Miró:} Data curation, Methodology, Software. \textbf{Oriol Lehmkuhl:} Supervision. \textbf{Ivette Rodriguez:} Supervision, Writing -- review \& editing. \textbf{J. Nathan Kutz:} Conceptualization, Funding acquisition, Investigation, Methodology, Project administration, Supervision, Writing -- review \& editing.

\section*{Declaration of competing interest}
The authors declare that they have no known competing financial interests or personal relationships that could have appeared to influence the work reported in this paper.

\section*{Data availability}
The data supporting this study comprise approximately 6~TB of large eddy simulation data (300 snapshots of approximately 20~GB each) and approximately 200~GB of DMD results (four files of approximately 50~GB each, one per variable). The authors are currently working on how to host data of this size. In the meantime, the data are available from the corresponding author upon reasonable request.

\section*{Declaration of generative AI and AI-assisted technologies in the manuscript preparation process}
During the preparation of this work, the authors used Claude (Anthropic) and ChatGPT (OpenAI) to check grammar and references, to write code for the frequency analysis of the probe spectra and DMD modes, and to tidy up figures. After using these tools, the authors reviewed and edited the content as needed and take full responsibility for the content of the publication.

\section*{Acknowledgements}
This work was supported in part by the US National Science Foundation (NSF) AI Institute for Dynamical Systems (dynamicsai.org), grant 2112085. JNK further acknowledges support from the Air Force Office of Scientific Research  (FA9550-24-1-0141). This project received funding from the European Union’s Horizon 2020 research and innovation programme under the Marie Skłodowska-Curie grant agreement No 955923. This work primarily used the ARCHER2 UK National Supercomputing Service (\url{https://www.archer2.ac.uk}) via the UK Turbulence Consortium (Grant no. EP/R029326/1) for the simulations. The authors also acknowledge computational resources and support provided by the Imperial College Research Computing Service (\url{http://doi.org/10.14469/hpc/2232}).
The research leading to this work has been partially funded by project TIFON with reference PLEC2023-010251/ AEI/10.13039/501100011033 and by the Ministerio de Ciencia e Innovación of Spain (PID2023-150408OB-C21/C22). Benet Eiximeno’s work was funded by a contract from the Subprograma de Ayudas Predoctorales given by the Ministerio de Ciencia e Innovación (PRE2021-096927). Oriol Lehmkuhl has been partially supported by a Ramon y Cajal postdoctoral contract (Ref: RYC2018-025949-I). The authors acknowledge the support of Departament de Recerca i Universitats de la Generalitat de Catalunya to the Research Group Large-scale Computational Fluid Dynamics (Code: 2021 SGR 00902) and the Turbulence and Aerodynamics Research Group (Code: 2021 SGR 01051). We also acknowledge the Barcelona Supercomputing Center for awarding us access to the MareNostrum V machine based in Barcelona, Spain.

\bibliographystyle{elsarticle-num}
\bibliography{refs}

\appendix
\numberwithin{figure}{section}
\numberwithin{table}{section}
\numberwithin{equation}{section}
\renewcommand{\theHfigure}{\thesection.\arabic{figure}}
\section{IFW mean surface flow description}
\label{App:IFW_mean_surface}

\begin{figure}[ht]
\minipage{0.32\textwidth}
  \includegraphics[width=\linewidth]{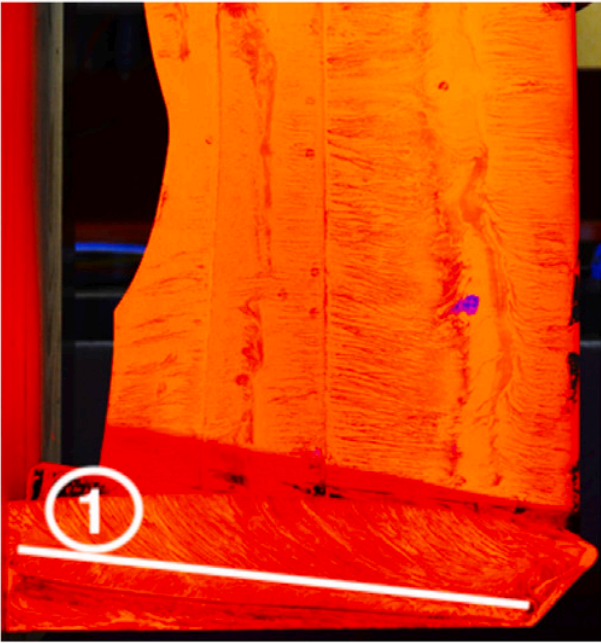}
\endminipage\hfill
\minipage{0.32\textwidth}
  \includegraphics[width=\linewidth]{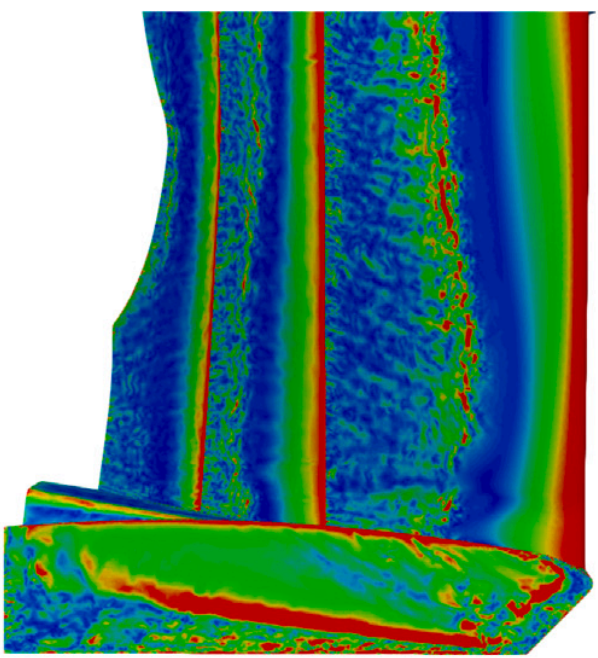}
\endminipage\hfill
\minipage{0.32\textwidth}%
  \includegraphics[width=\linewidth]{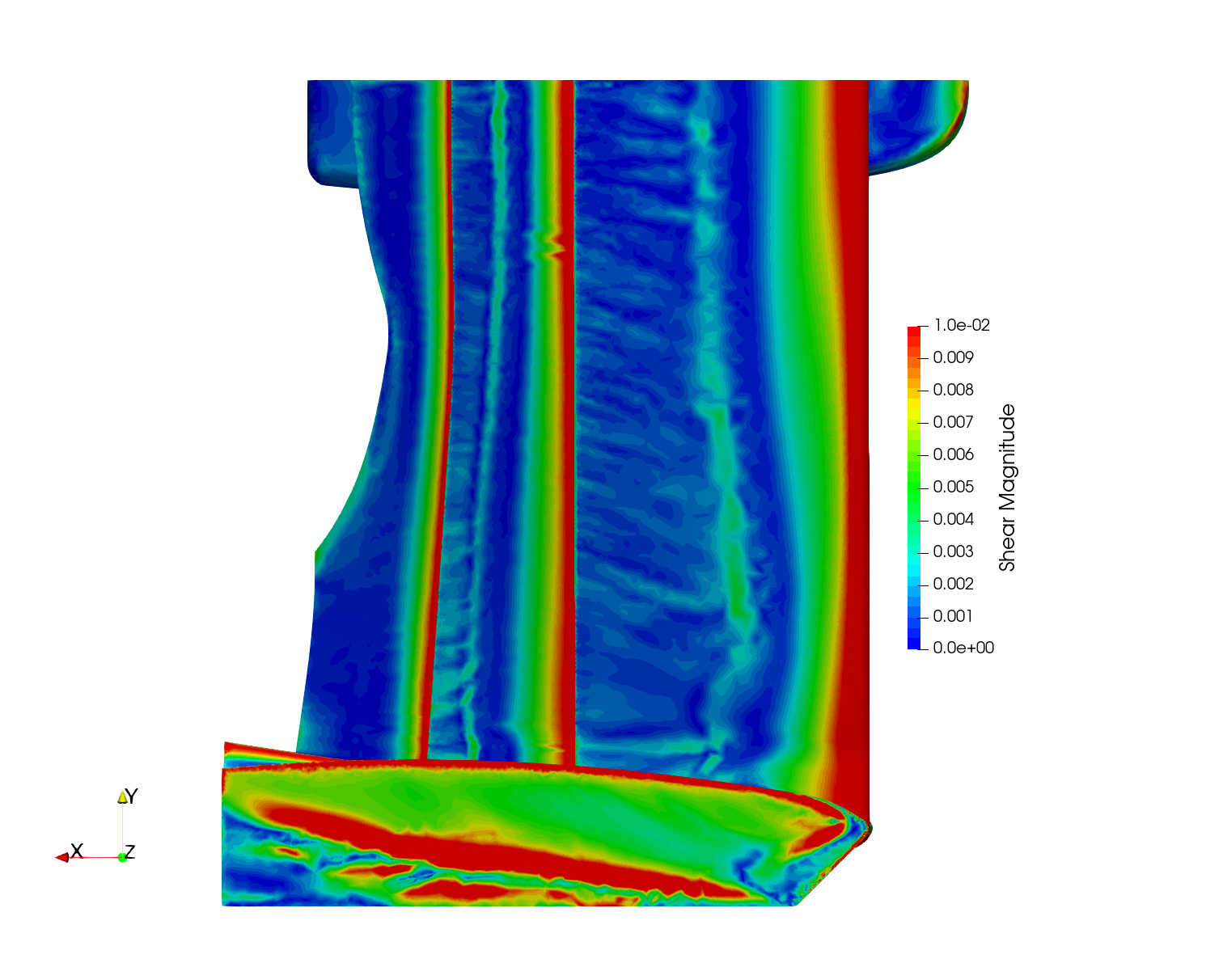}
\endminipage
\caption{Comparison of wall shear-stress magnitude distribution between current 
simulation and \cite{Buscariolo2022} (Bottom View). Left: experimental flow
visualisation; centre: simulation of \cite{Buscariolo2022}; right: current
simulation.}
\label{fig:wss_bottom_view_ifw}
\end{figure}

The surface flow visualisation from \cite{Buscariolo2022} shown in this appendix was obtained at a different Reynolds number from the simulations.
The wall shear-stress magnitude on the lower surface is compared in
\autoref{fig:wss_bottom_view_ifw}. On the mainplane flow visualisation, the flow
appears ordered and streamlined as in laminar flows from the leading edge 
until $25 \% - 30 \%$ of the chord as reported\cite{Buscariolo2022}. Then, 
a separation line is seen, followed by a reattachment line, where the 
streamwise position of the separation and the reattachment points varies 
along the span of the mainplane, highlighting the influence of the varying 
curvature in the airfoil profiles. This area marks the transition of the 
flow from laminar to turbulent and in a time-averaged representation looks 
like a laminar separation bubble (LSB). Then, it follows streamwise streaks 
downstream of the reattachment point. In both numerical models, the laminar 
region of the flow is observed as a smooth and uniform 
region that starts from the stagnation point with large shear-stress 
magnitude until the streamwise location, where the magnitude becomes zero 
and the flow separates. The flow re-attaches further downstream, as indicated
by the increase in the shear-stress magnitude. Both numerical models show
chaotic flow patterns once the flow has transitioned to turbulent, but 
in the present simulation, the flow follows streamwise streaks, as shown
in the experiments. This is not the case for the numerical model in 
\cite{Buscariolo2022} and could be driven by the instantaneous nature
of the shear-stress magnitude presented.

In the flow-visualisation of the first flap, similar patterns are observed
as in the mainplane. Initially the flow appears ordered until the 
presence of the separation and reattachment points, where the flow
transitions from laminar to turbulent. On the areas, where the image is
not well lit, the flow shows streamwise streaks downstream of the 
reattachment point, or signs of separation closer to the endplate 
and the nosecone. The spanwise location of the separated regions aligns
with the hanger and the starting point of the gurney flap on the 
second element. The length of the LSB is smaller when compared to 
the mainplane and its spanwise distribution appears more linear. In
the numerical models, the separation line is straight along the span
of the element and it develops a small curvature closer to the endplate, 
in agreement with the separated region of the flow-visualisation. The
length of the LSB appears similar between them, but longer when compared
to the experiments. This difference could be driven by the different 
Reynolds number between the simulations and the flow-visualisation. 
After the flow has transitioned to turbulent, the numerical model from
\cite{Buscariolo2022} shows chaotic flow patterns of attached flow 
along the span, but the current simulation shows streamwise streaks
and separated regions next to the endplate and in the vicinity of
hanger. The current simulation resembles more the flow-visualisation
experiment and it is driven by the boundary resolution, which is double
when compared to the numerical model in \cite{Buscariolo2022}.

\begin{figure}[ht]
\minipage{0.32\textwidth}
  \includegraphics[width=0.5\linewidth]{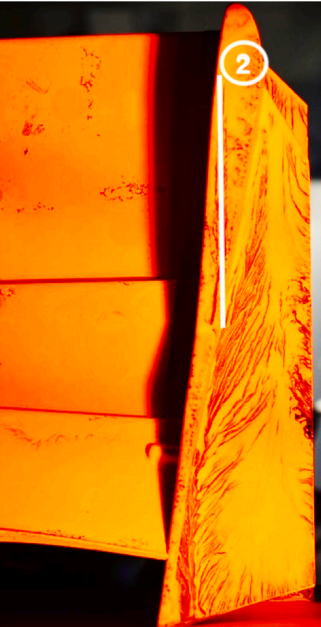}
\endminipage\hfill
\minipage{0.32\textwidth}
  \includegraphics[width=0.6\linewidth]{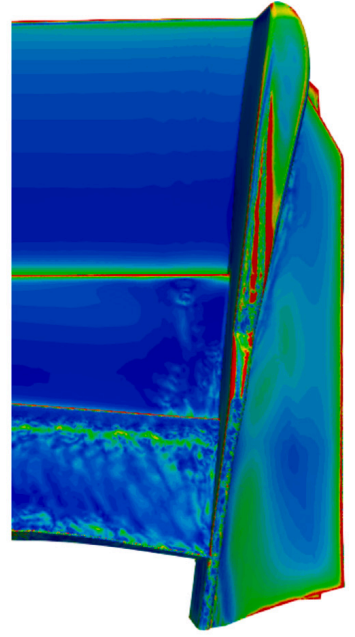}
\endminipage\hfill
\minipage{0.32\textwidth}%
  \includegraphics[width=\linewidth]{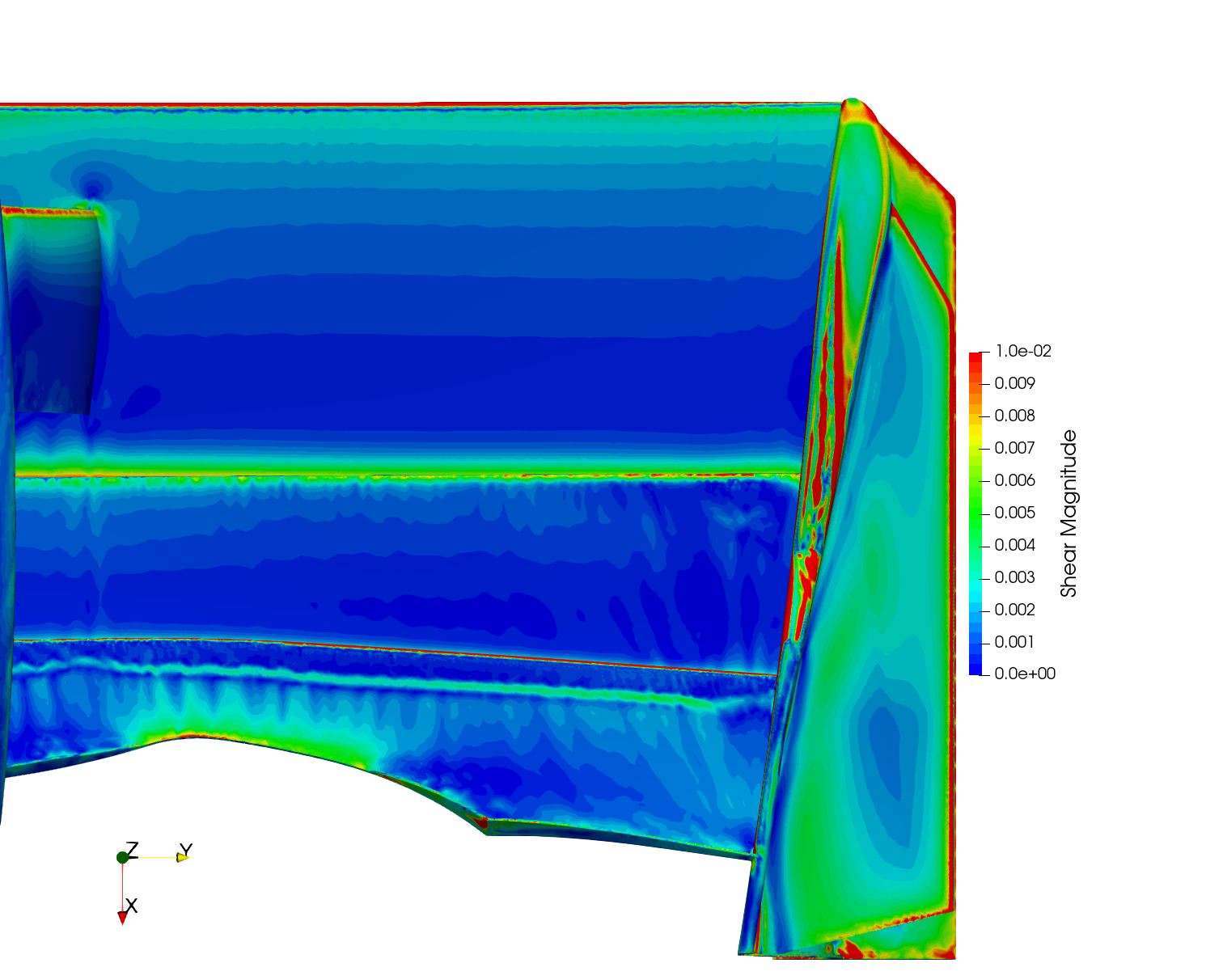}
\endminipage
\caption{Comparison of wall shear-stress magnitude distribution between current 
simulation and \cite{Buscariolo2022} (Top View). Left: experimental flow
visualisation; centre: simulation of \cite{Buscariolo2022}; right: current
simulation.}
\label{fig:wss_top_view_ifw}
\end{figure}

For the second flap, the image resolution is lower in the flow-vis image
when compared to the other elements, but the flow on the suction side
of the airfoil can be decomposed into three regions. From inboard of the
wing, next to the nosecone, there are signs of separated flow that align
with the wake shed from the hanger. Moving further outboard, the paint 
has very high illumination, hence is hard to distinguish any flow features.
Following this region, the flow appears unorganised, where in some parts
it shows streamwise streaks and in others is completely separated. There
isn't evidence of a separation and a re-attachment line, so it is 
expected that the separation point aligns with the stagnation point of 
the airfoil, but the flow does not recover and remains separated. This
is clear in both computational models, as downstream from 20 - 25 \% of
the chord, the flow is fully separated, as indicated by the almost
zero shear-stress magnitude, except for the vicinity around the gurney
flap. The computational model from \cite{Buscariolo2022}
shows more chaotic flow features further outboard of the wing, since it
is an instantaneous field, when compared to the time-averaged surface 
field of the current model. For the outboard region of the second flap,
the computational models agree with the flow-vis for the fully separated
flow, however they predict the separation point downstream of the 
stagnation point, as it was observed from the experiment and could be 
associated with the difference in Reynolds number. 

When looking at the flow-visualisation from the top of the wing in 
\autoref{fig:wss_top_view_ifw}, there are clear flow patterns on the upper 
surface of the endplate and the canard. Unfortunately, the pressure side  
of the wing elements show very uniform dried paint on their surface, which
can be related to using a lot of paint to ensure good coverage in both 
sides of the wing. Nevertheless, on the upper surface of the endplate, 
there are two pathways with large shear magnitude (attachment lines)
and the longest one has been marked by a white line with the number two.
These lines illustrate two thin vortical structures that are shed from the
tip of the endplate, they rotate around each other and form a helical
vortex which advects further downstream. On the canard, the thickness of 
the separation layer that extends from the LE of canard, is clear until 
75 \% of the canard's length. Along this line, the flow is split into
two parts, where the first slips from the edge of the canard, rotates
and forms the canard vortex, while the second flows between the endplate
and the separation line. Downstream from 75 \% of the canard's length, there
is a circular separation region due to strong flow deceleration. In the 
computational models, the streamlines are not visible as in the
experiments. On top of the endplate, the pair of vortical structures
is well-resolved by both models, as indicated by the lines with large
shear magnitude. On the canard, both numerical models show the circular
separation region, but in the model from \cite{Buscariolo2022}, this
region is longer and starts upstream from 75 \% of the canard length. 
In the beginning of this separation region, the flow is connected
with the flow next to the endplate, which is not the case for the 
experiments, nor the present numerical model. Again, both models
show a thick separation line that forms from the LE of the canard 
and a flow component with larger shear magnitude that is shed from the 
tip of the canard to form the canard vortex.

\begin{figure}[ht]
   \centering
      \begin{subfigure}{0.45\textwidth}
      \includegraphics[width=\linewidth]{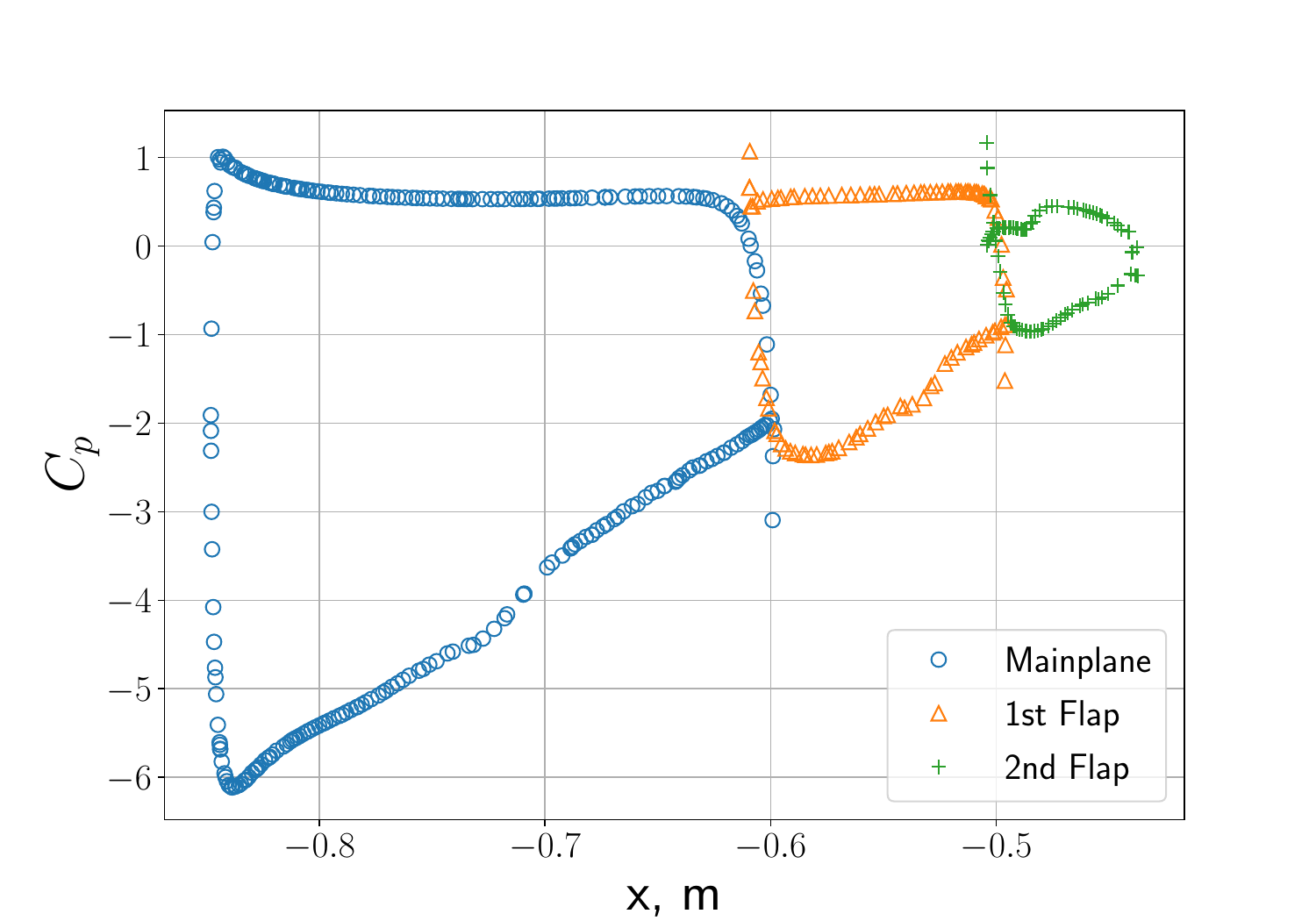}
      \end{subfigure}
      \begin{subfigure}{0.45\textwidth}
      \includegraphics[width=\linewidth]{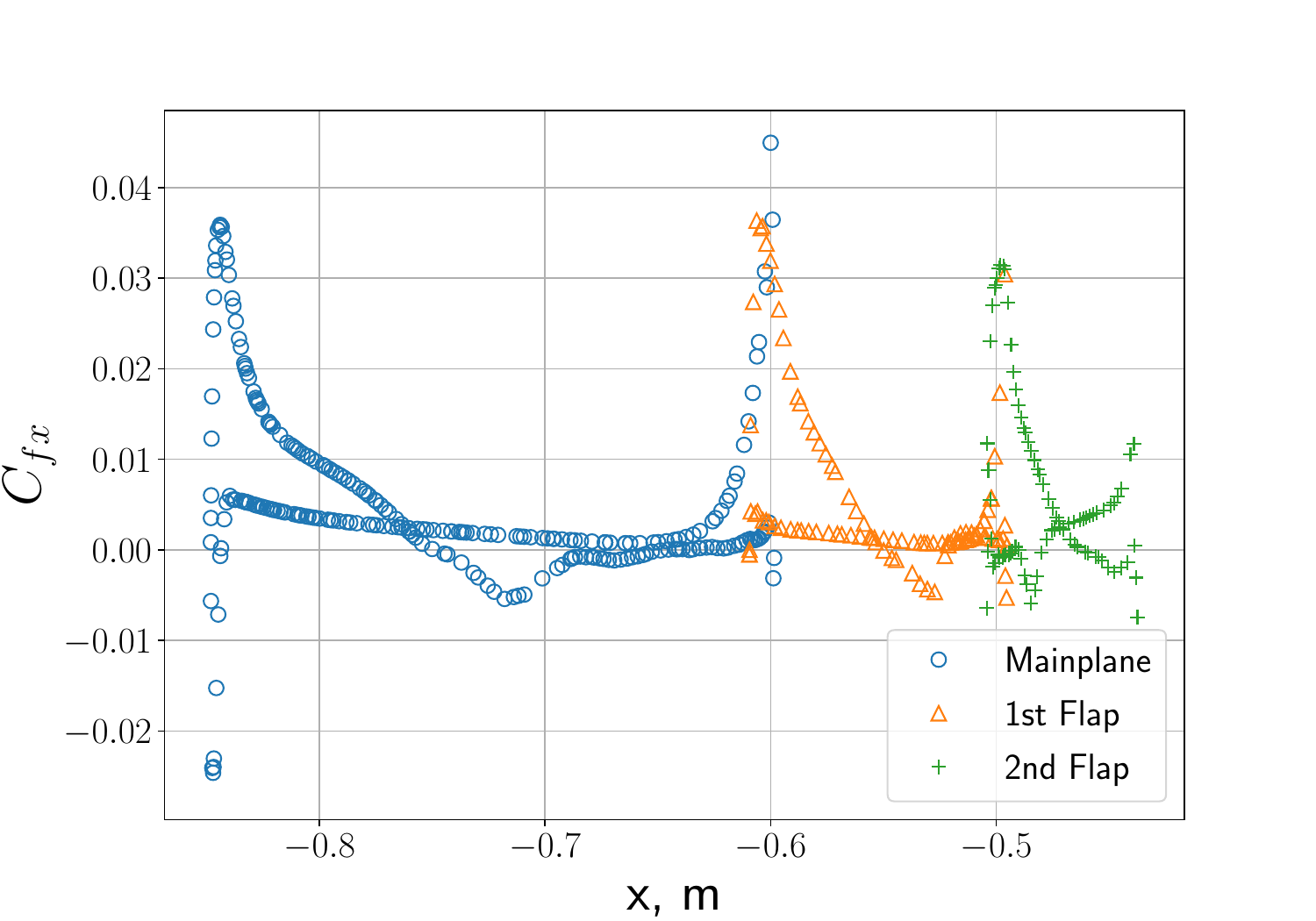}
      \end{subfigure}
      \caption{Streamwise Distribution of the Pressure Coefficient (Left) and the Shear Stress 
      (Right) for the mainplane, first flap and second flap respectively at Y = -0.250 m}
   \label{fig:cp_cfx_ifw_y250}
\end{figure}

To examine the laminar to turbulent transition on the wing elements, the volume mesh 
is sliced at a spanwise location of $Y = -0.25$ m and the streamwise distribution 
for the pressure and skin friction coefficients are shown in 
\autoref{fig:cp_cfx_ifw_y250}. It is highlighted that the height of the gurney flap on
the second flap is zero at this location. The mainplane transitions to turbulence
with a separation point at $X = -0.75$ m and a reattachment point at $X = -0.64$ m. 
This elongated, shallow structure appears as a long laminar separation bubble in the averaged 
shear-stresses fields. Similarly, the first flap transitions to turbulence at 
$X = -0.55$ m, and the reattachment point is found at $X = -0.52$ m. 
The second flap separates from the pressure side at a position $X = -0.50$ m. 
However, these locations change depending on the spanwise location at the wing, 
as each wing element includes a curvature that varies with the spanwise 
position (3D curvature).

\section{IFW instantaneous probe data}
\label{App:IFW_probes}

\begin{figure}
    \centering
    \subfloat[]{\includegraphics[width=0.33\textwidth]{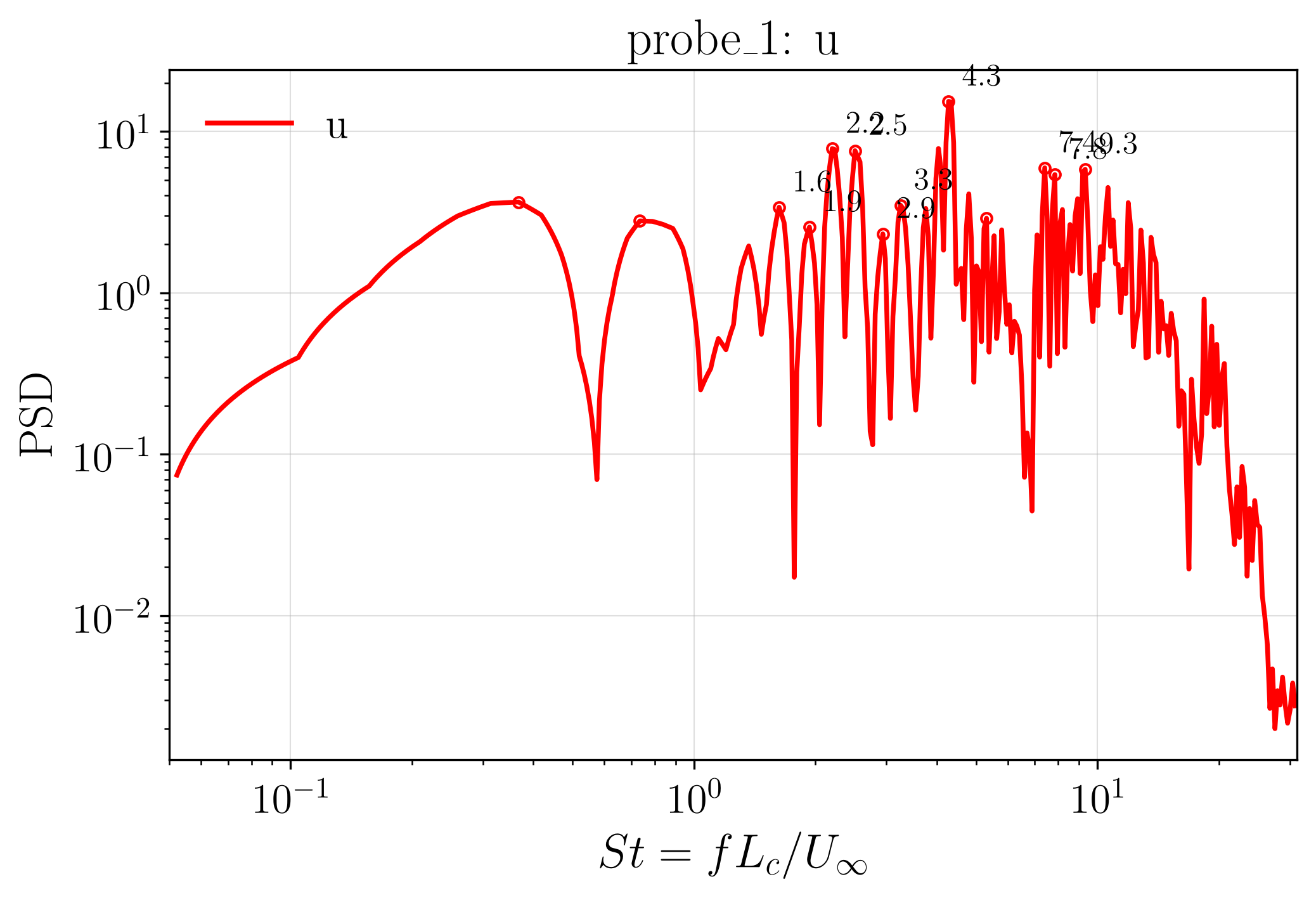}}
    \subfloat[]{\includegraphics[width=0.33\textwidth]{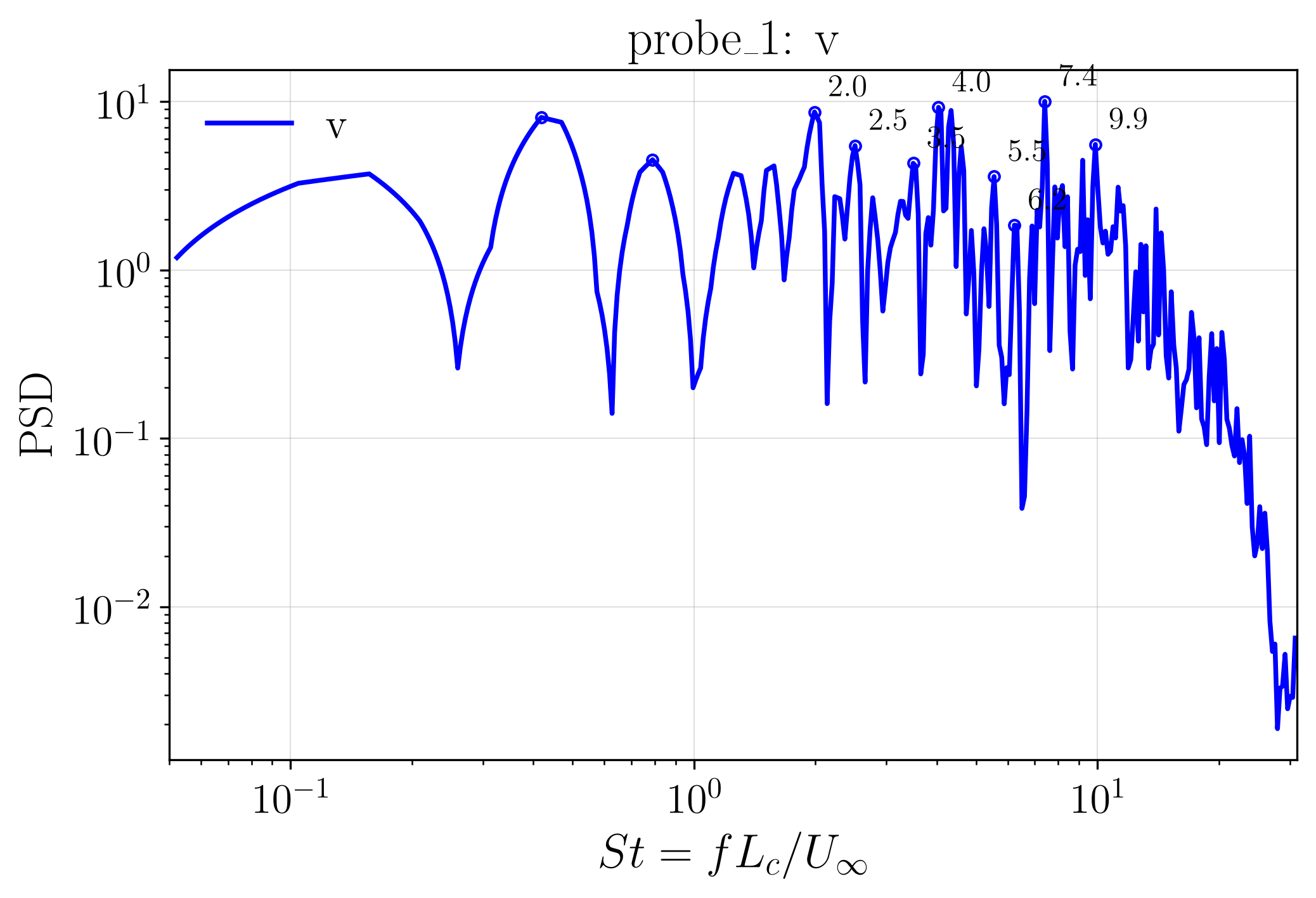}}
    \subfloat[]{\includegraphics[width=0.33\textwidth]{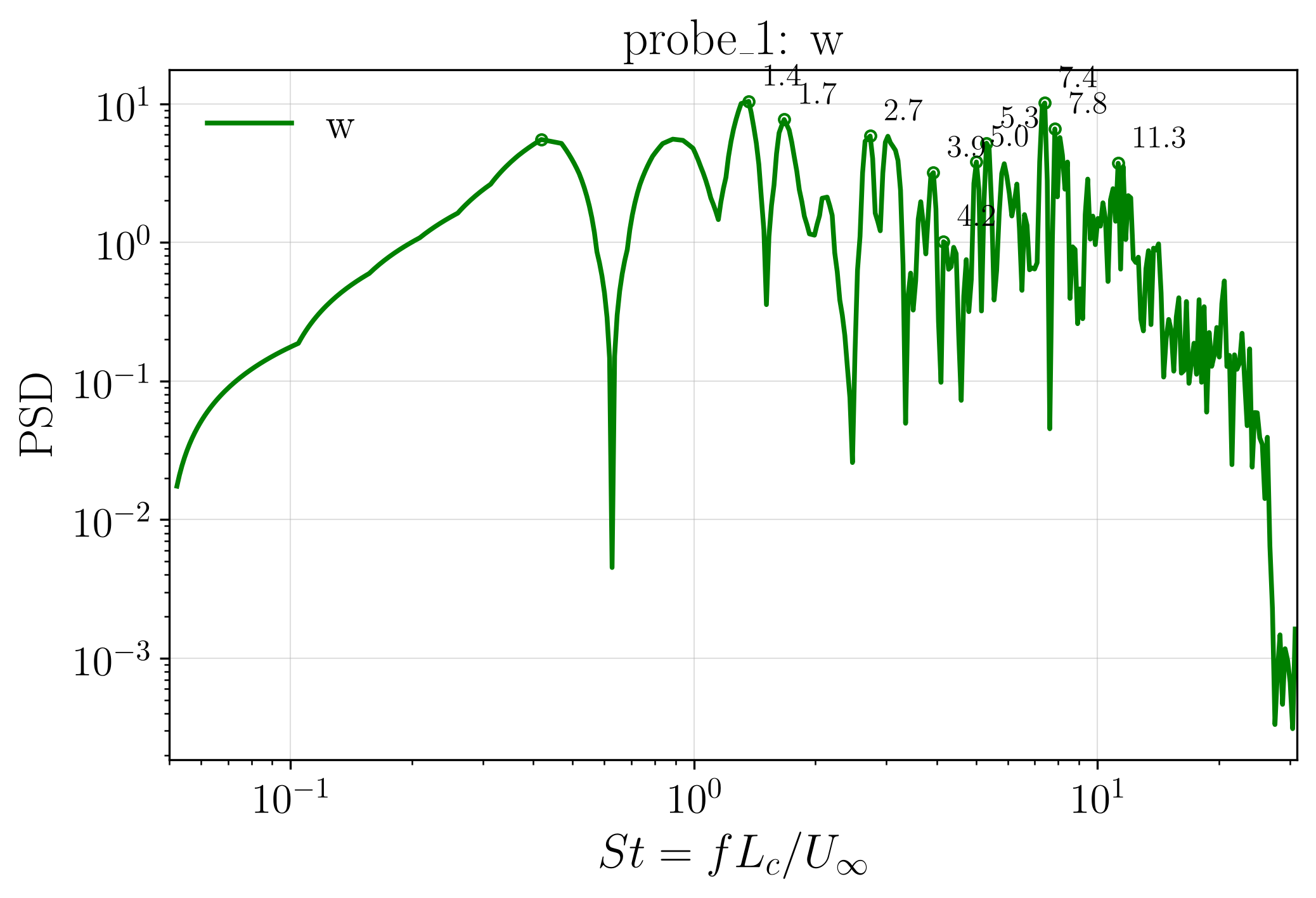}}
    \caption{Spectra of the streamwise (a), cross-stream (b) and vertical (c) velocities at probe 1.}
    \label{fig:spectra_1}
\end{figure}

\begin{figure}
    \centering
    \subfloat[]{\includegraphics[width=0.33\textwidth]{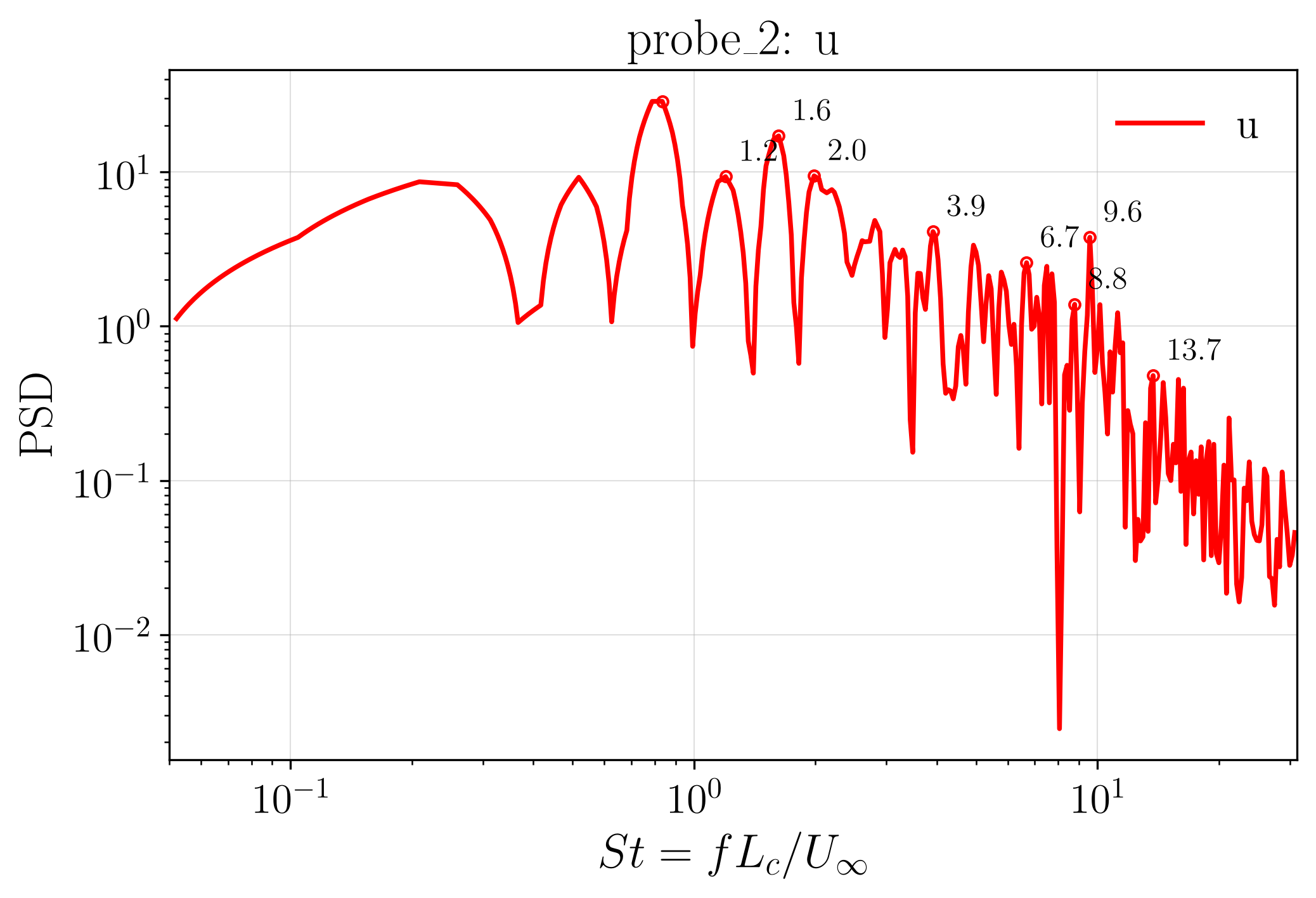}}
    \subfloat[]{\includegraphics[width=0.33\textwidth]{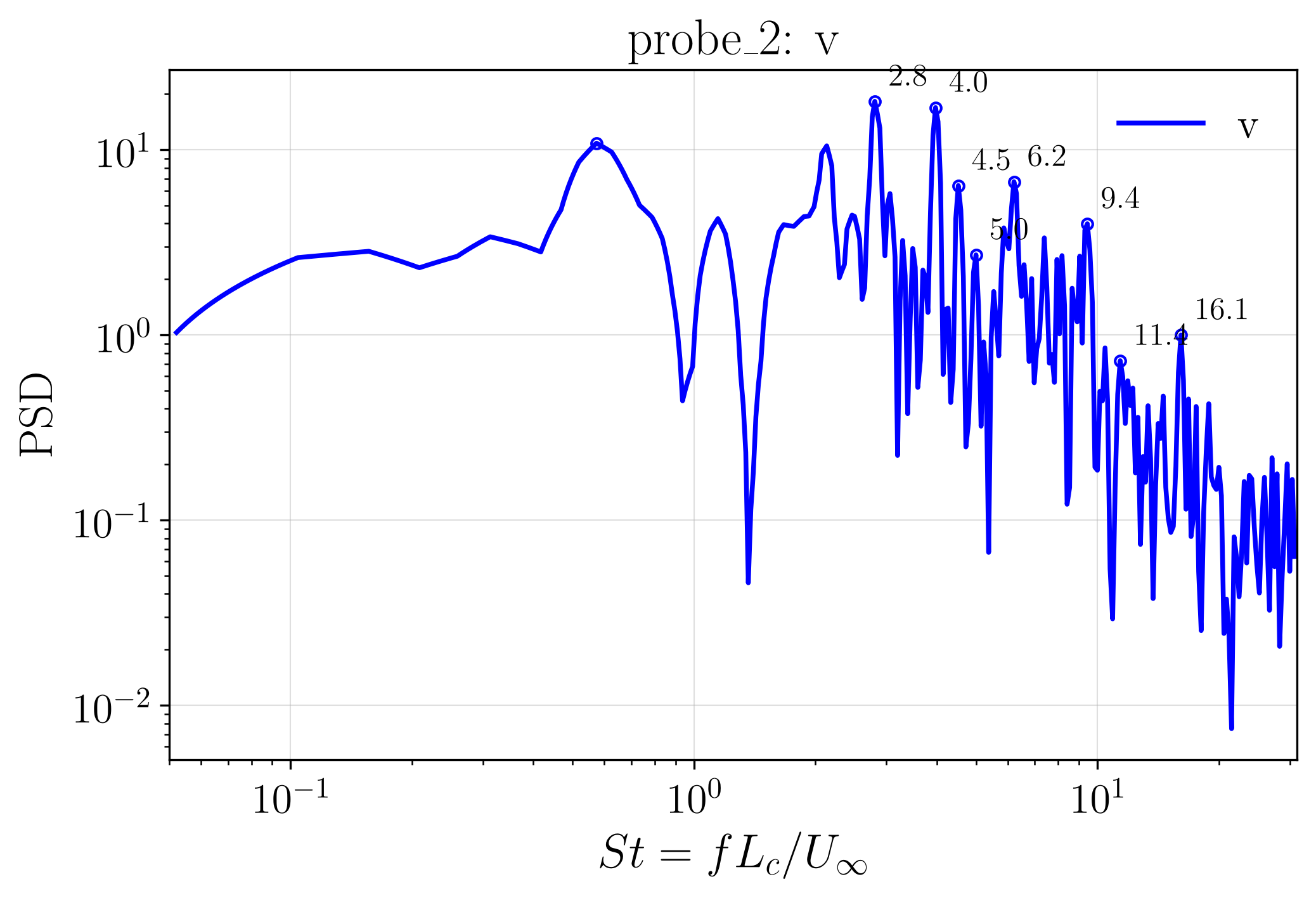}}
    \subfloat[]{\includegraphics[width=0.33\textwidth]{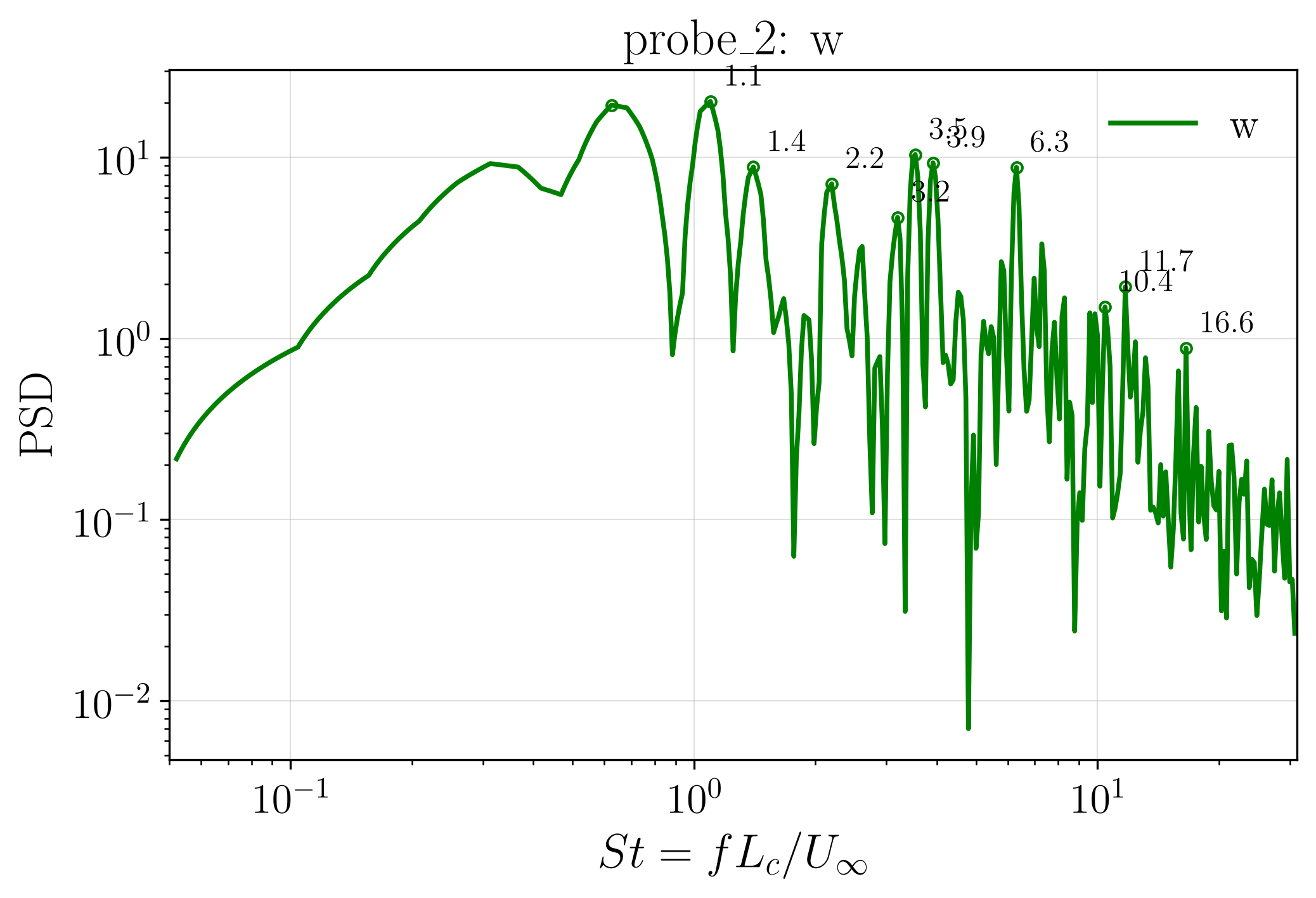}}
    \caption{Spectra of the streamwise (a), cross-stream (b) and vertical (c) velocities at probe 2.}
    \label{fig:spectra_2}
\end{figure}

\begin{figure}
    \centering
    \subfloat[]{\includegraphics[width=0.33\textwidth]{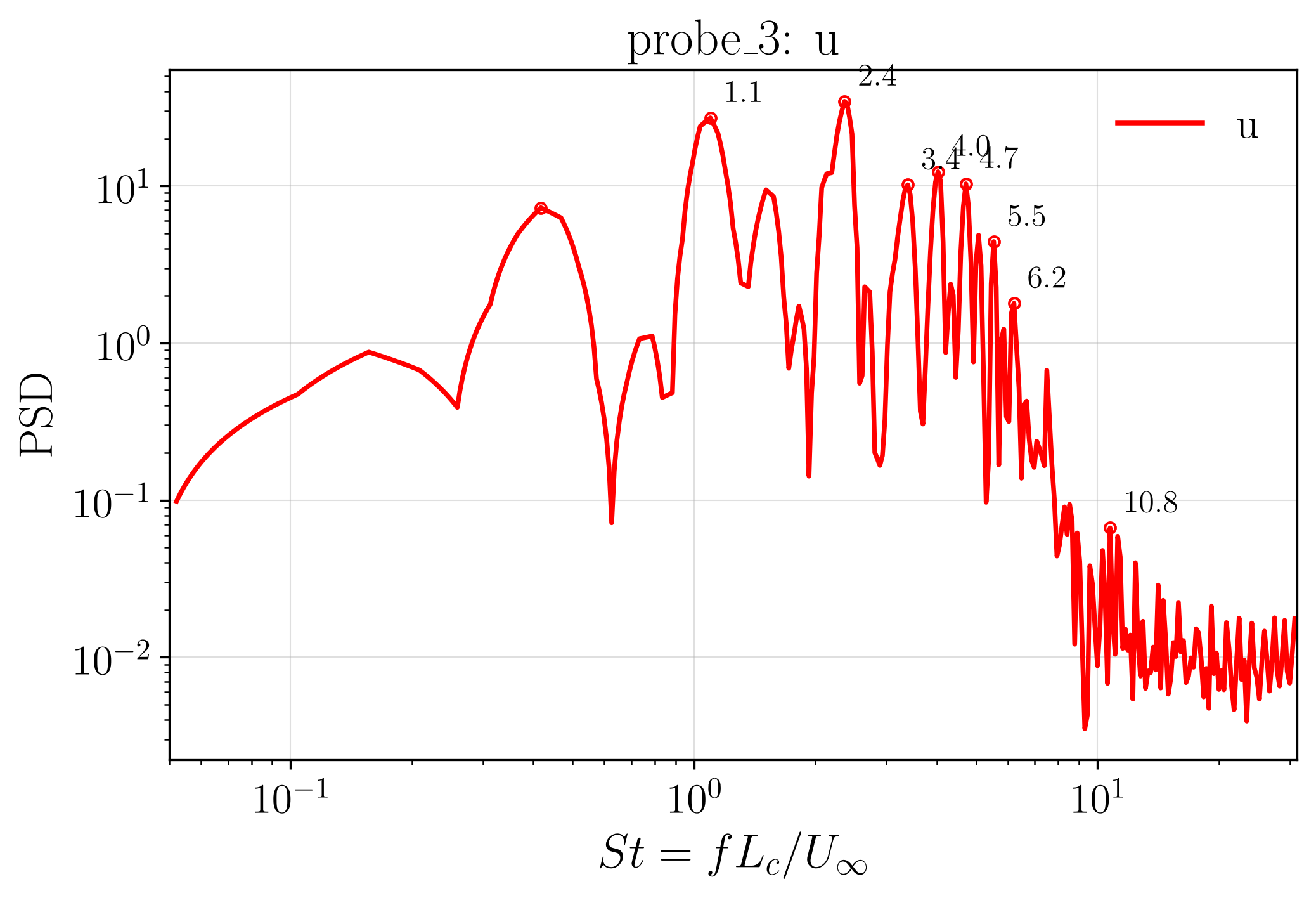}}
    \subfloat[]{\includegraphics[width=0.33\textwidth]{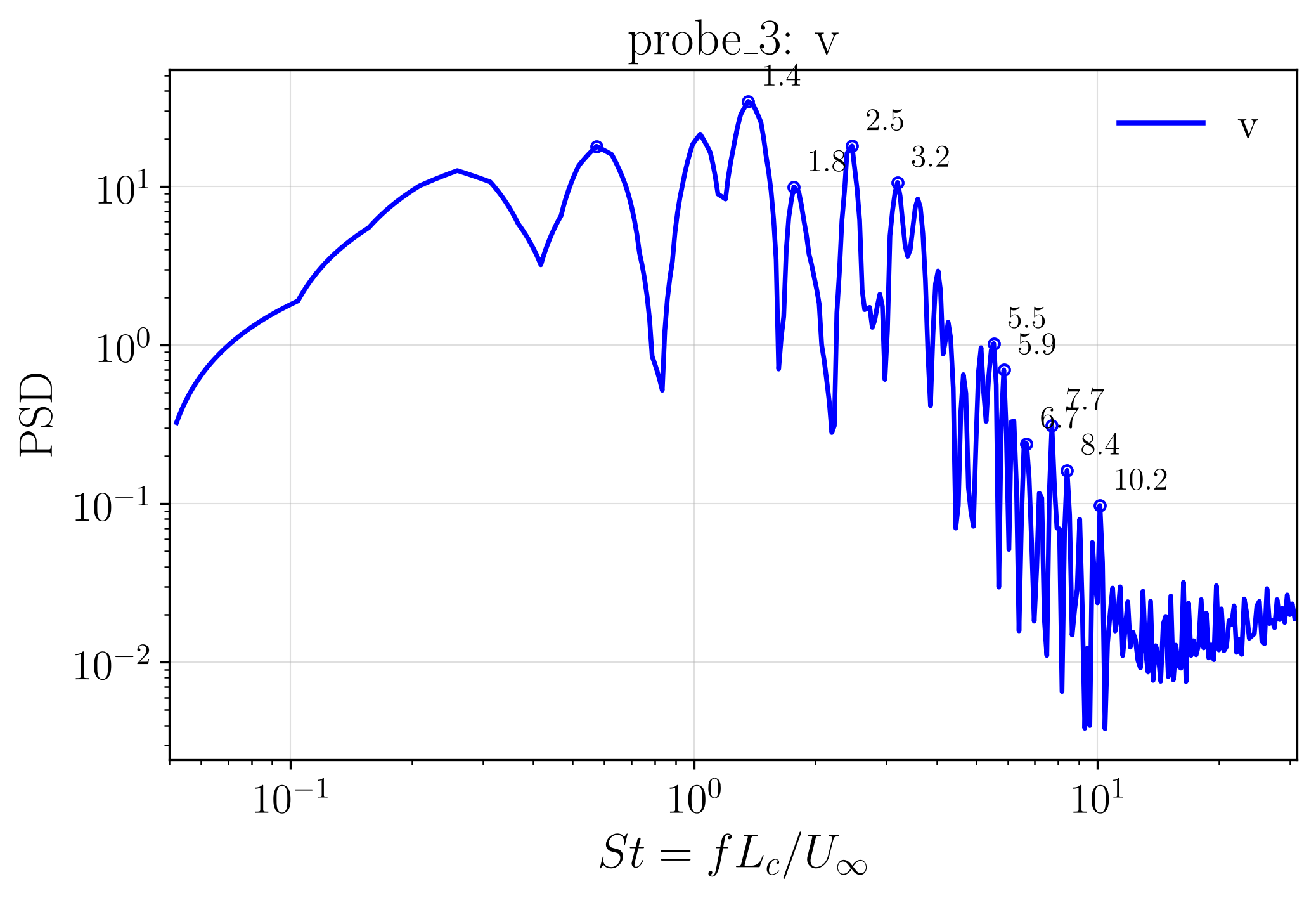}}
    \subfloat[]{\includegraphics[width=0.33\textwidth]{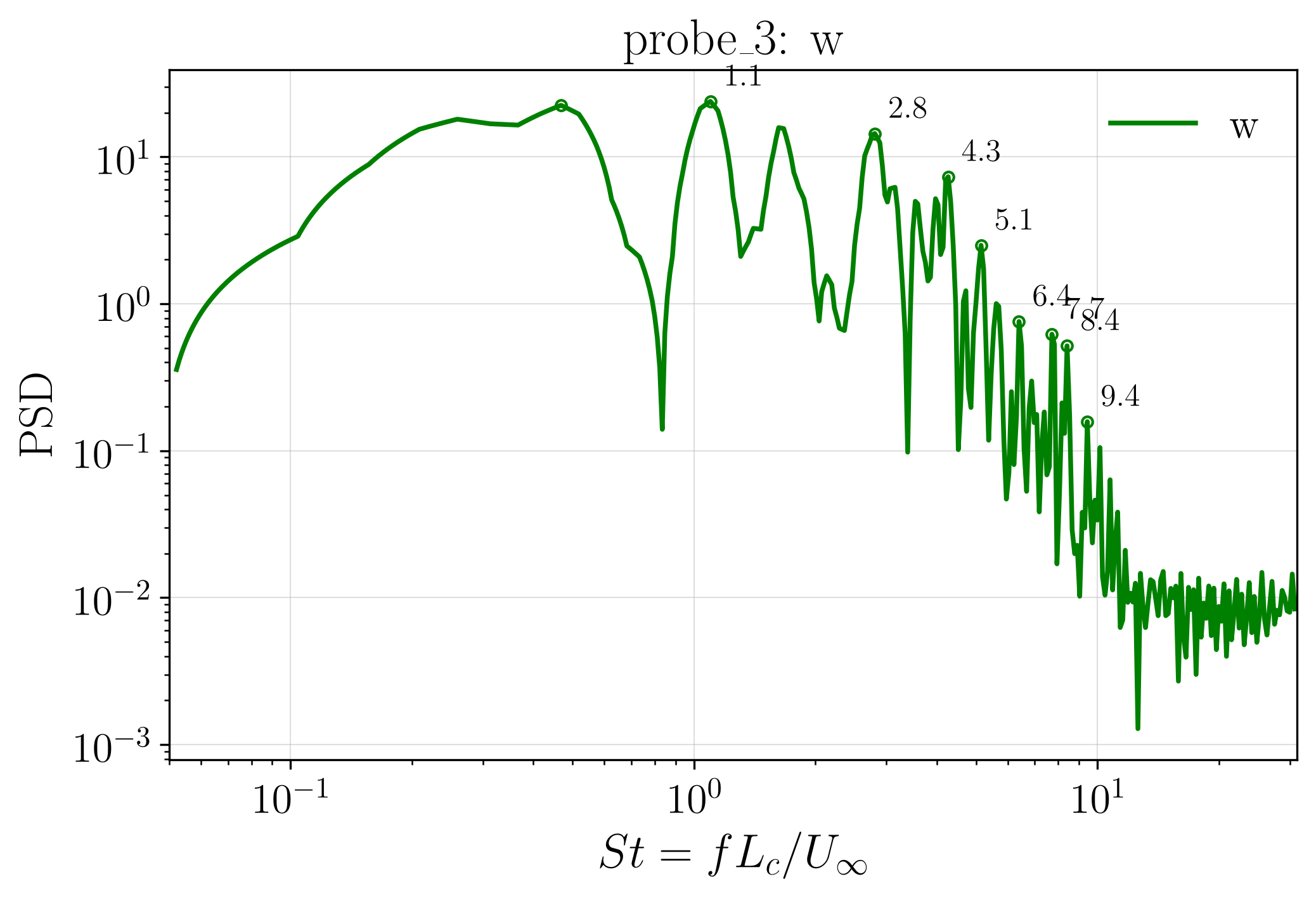}}
    \caption{Spectra of the streamwise (a), cross-stream (b) and vertical (c) velocities at probe 3.}
    \label{fig:spectra_3}
\end{figure}

\begin{figure}
    \centering
    \subfloat[]{\includegraphics[width=0.33\textwidth]{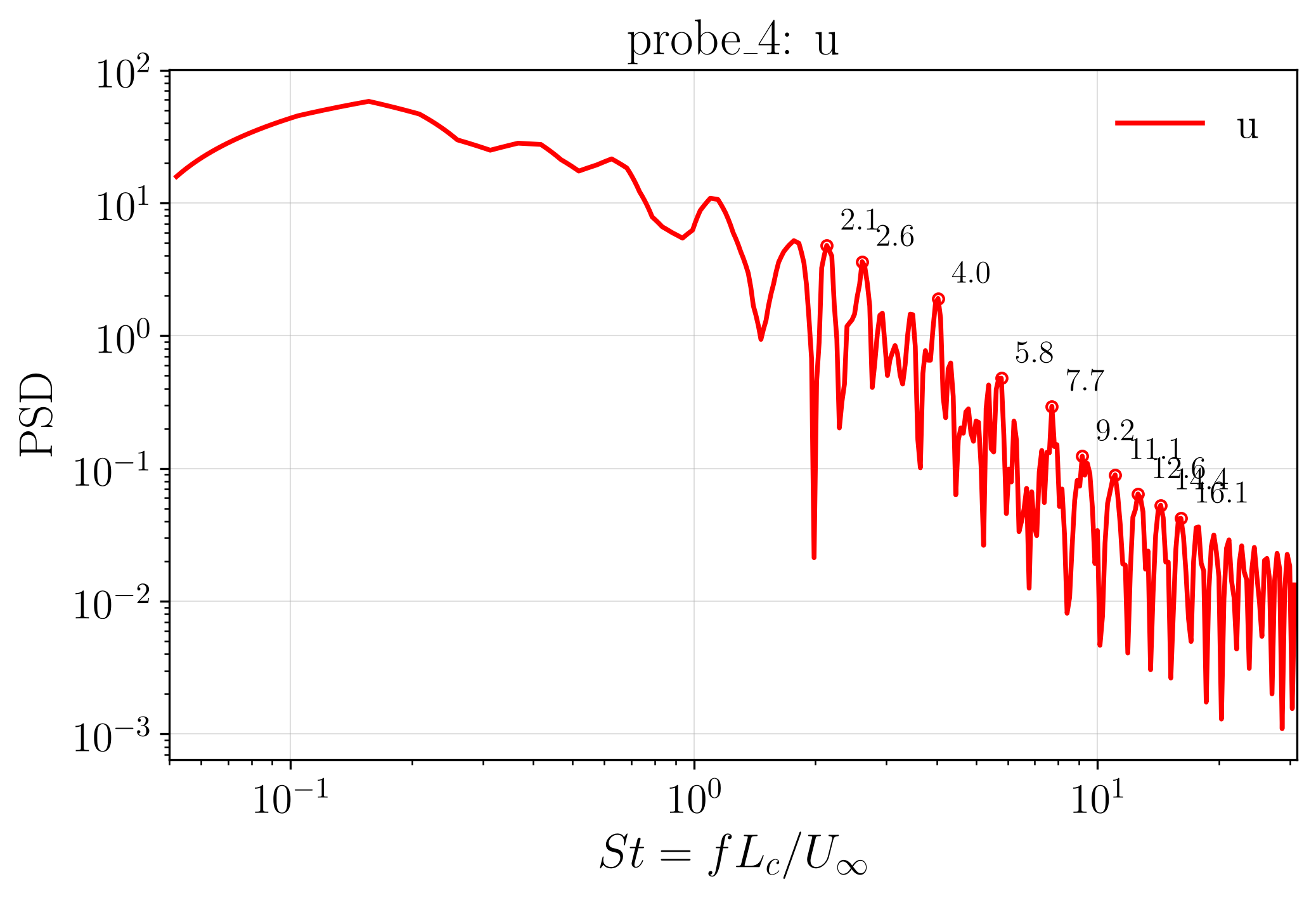}}
    \subfloat[]{\includegraphics[width=0.33\textwidth]{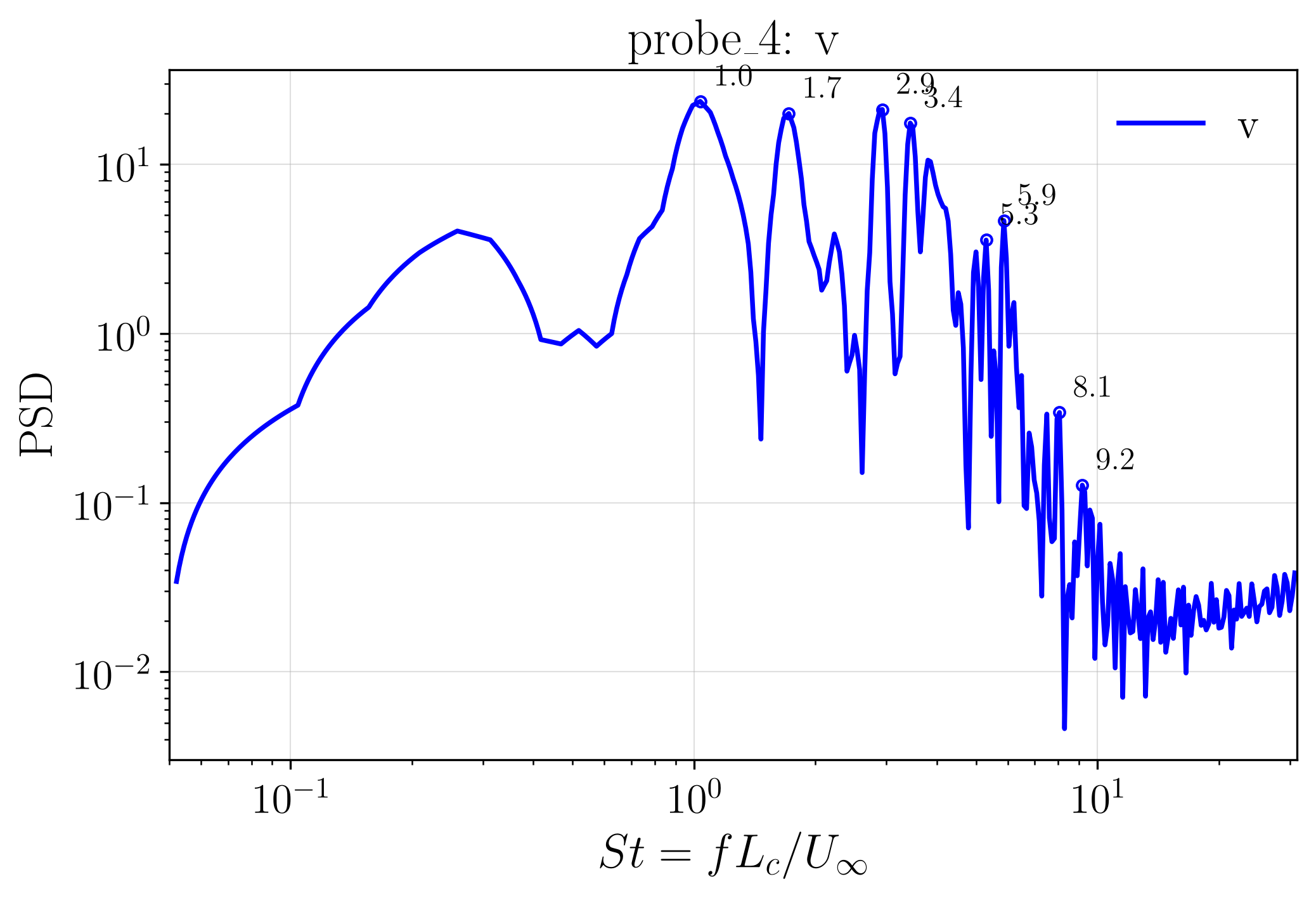}}
    \subfloat[]{\includegraphics[width=0.33\textwidth]{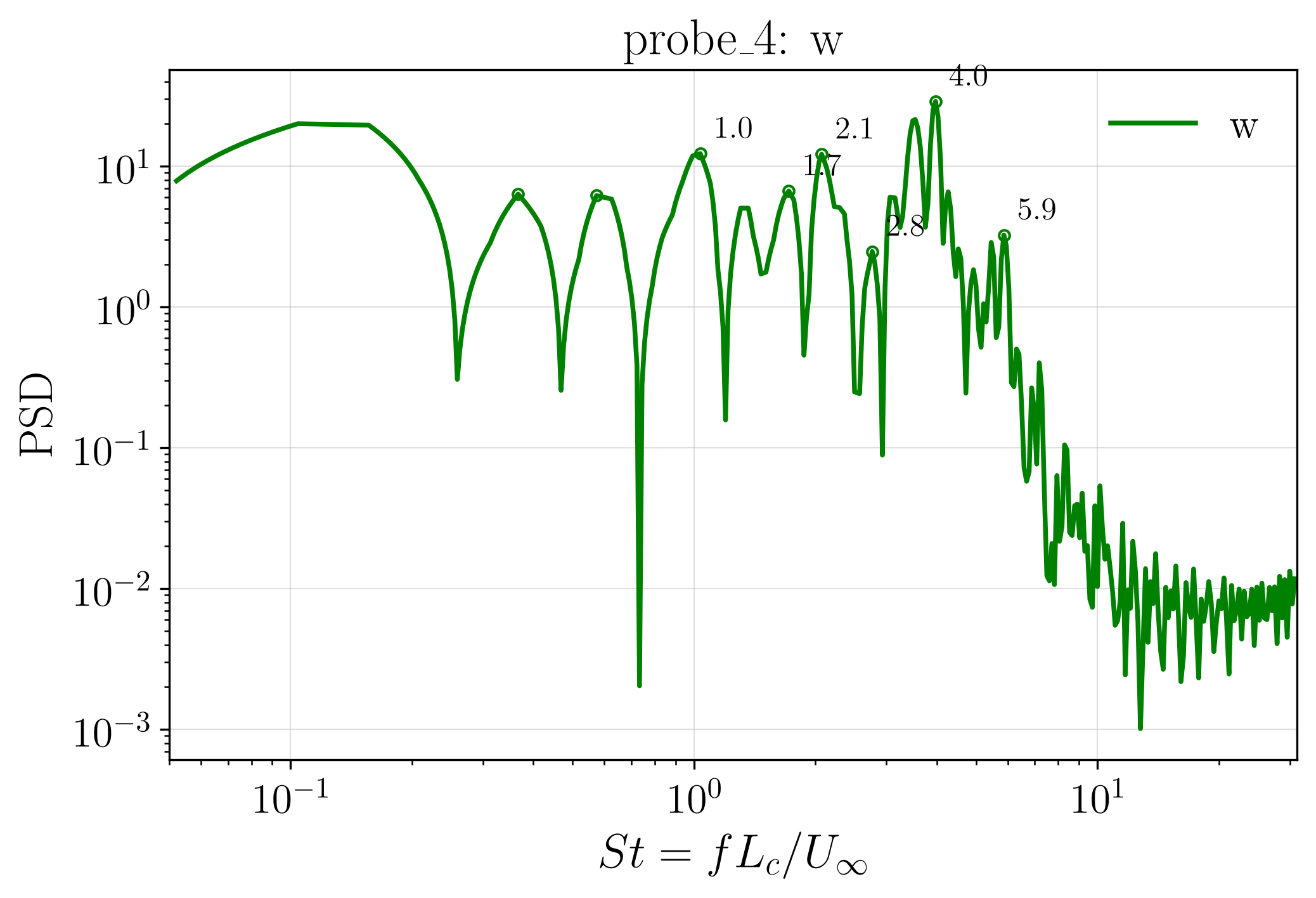}}
    \caption{Spectra of the streamwise (a), cross-stream (b) and vertical (c) velocities at probe 4.}
    \label{fig:spectra_4}
\end{figure}

\begin{figure}
    \centering

    \subfloat[Probe 5, $u$]{\includegraphics[width=0.33\textwidth]{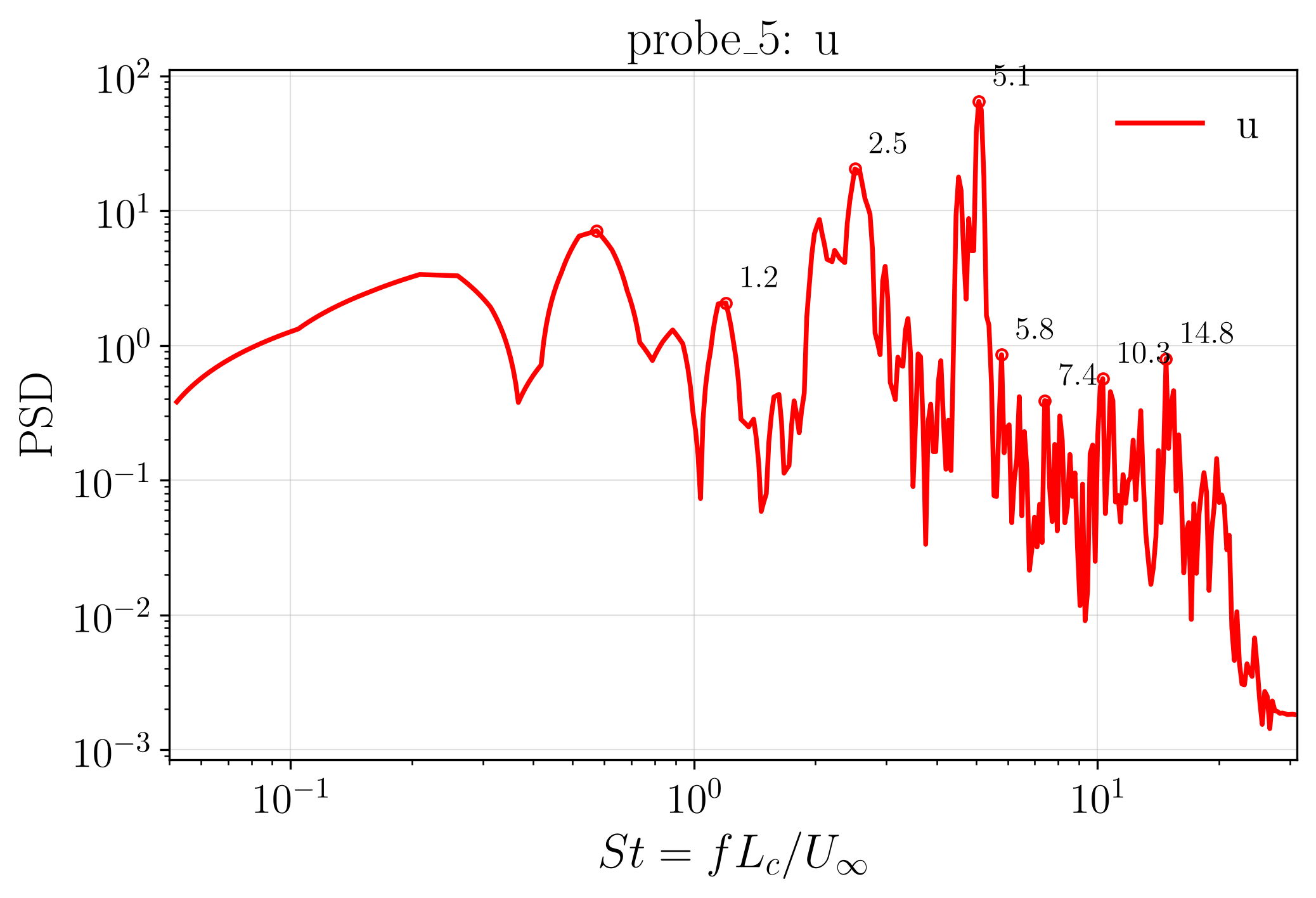}}
    \hspace{0.06\textwidth}
    \subfloat[Probe 5, $w$]{\includegraphics[width=0.33\textwidth]{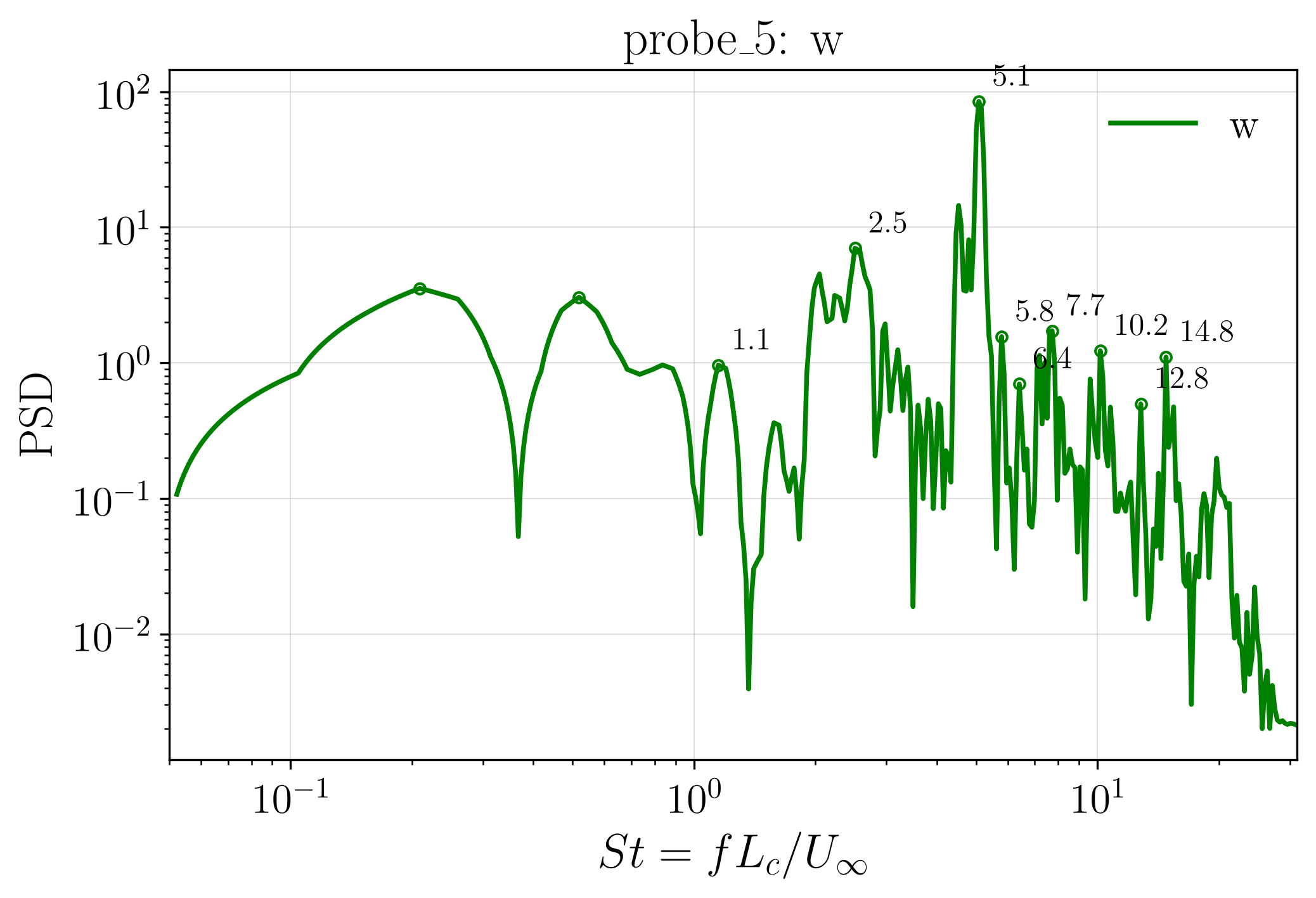}}

    \vspace{0.5em}

    \subfloat[Probe 6, $u$]{\includegraphics[width=0.33\textwidth]{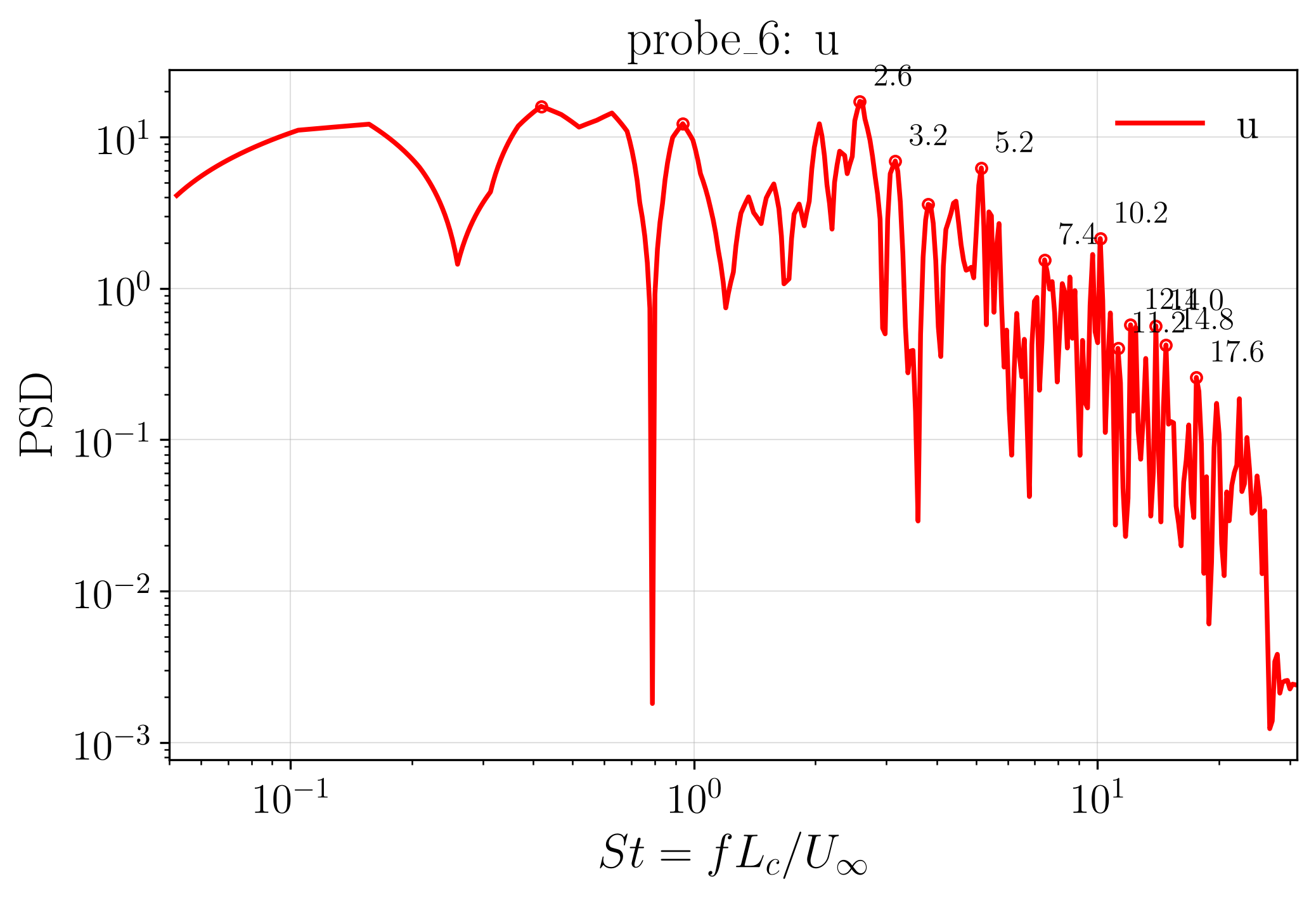}}
    \hspace{0.06\textwidth}
    \subfloat[Probe 6, $w$]{\includegraphics[width=0.33\textwidth]{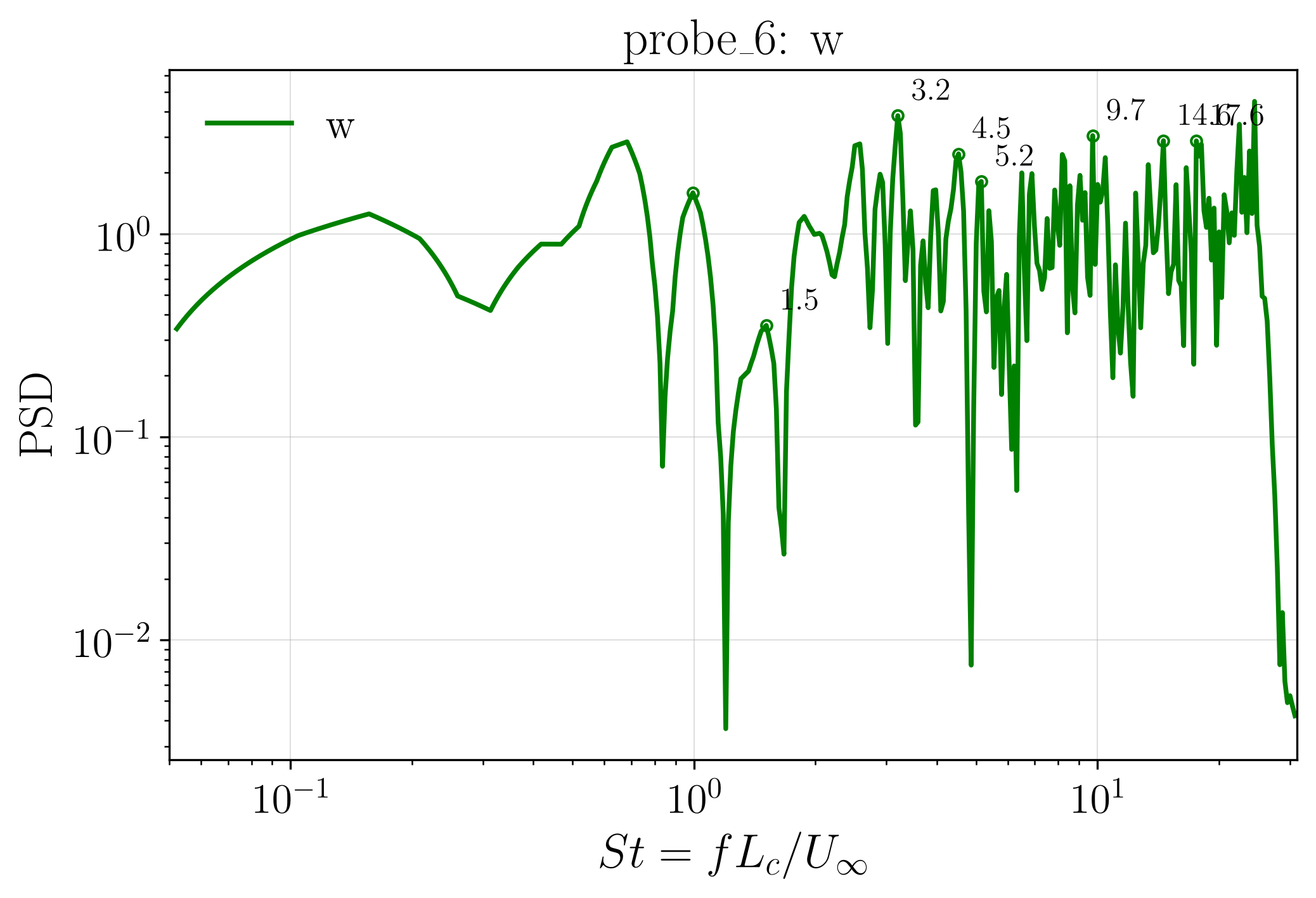}}

    \vspace{0.5em}

    \subfloat[Probe 7, $u$]{\includegraphics[width=0.33\textwidth]{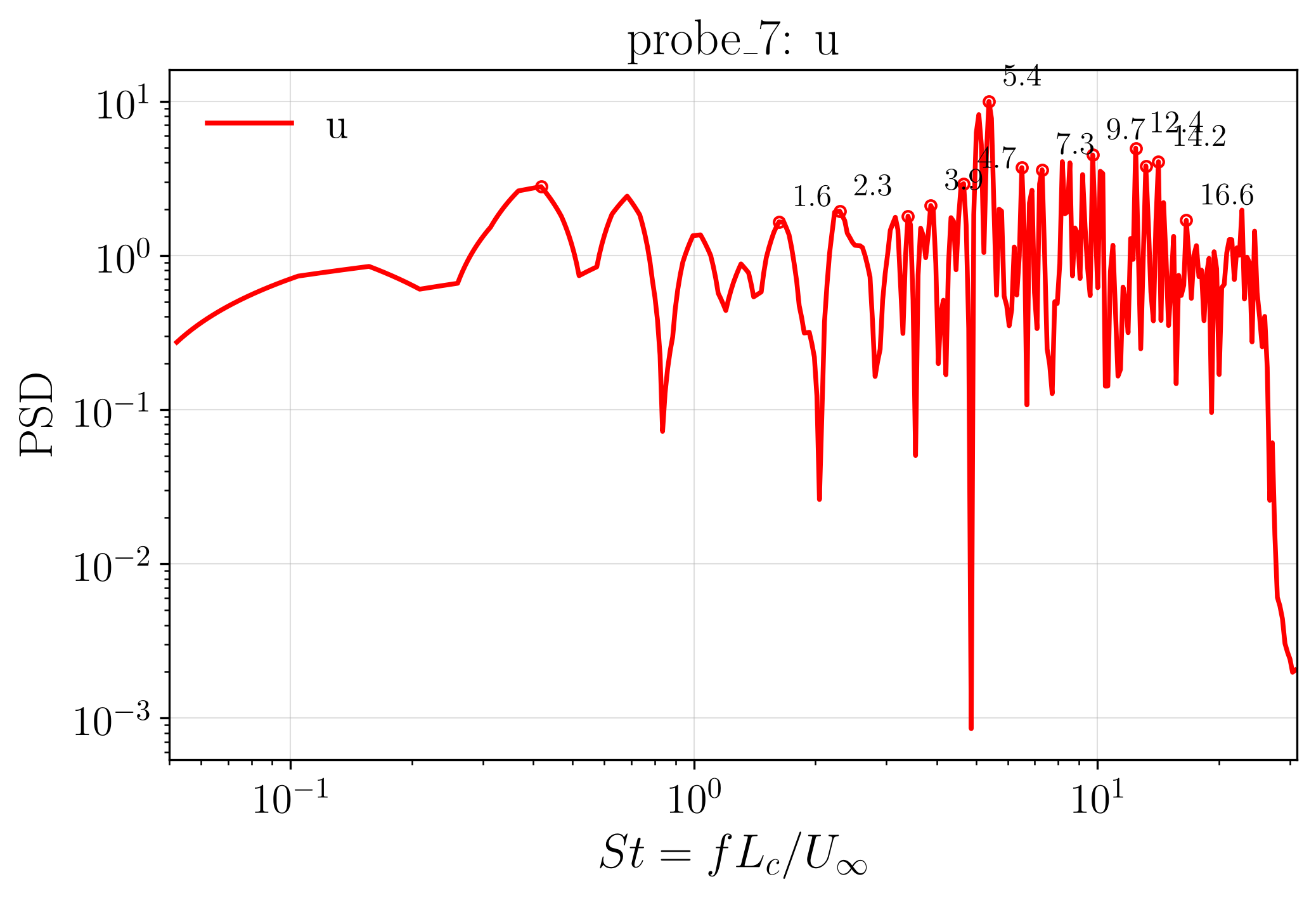}}
    \hspace{0.06\textwidth}
    \subfloat[Probe 7, $w$]{\includegraphics[width=0.33\textwidth]{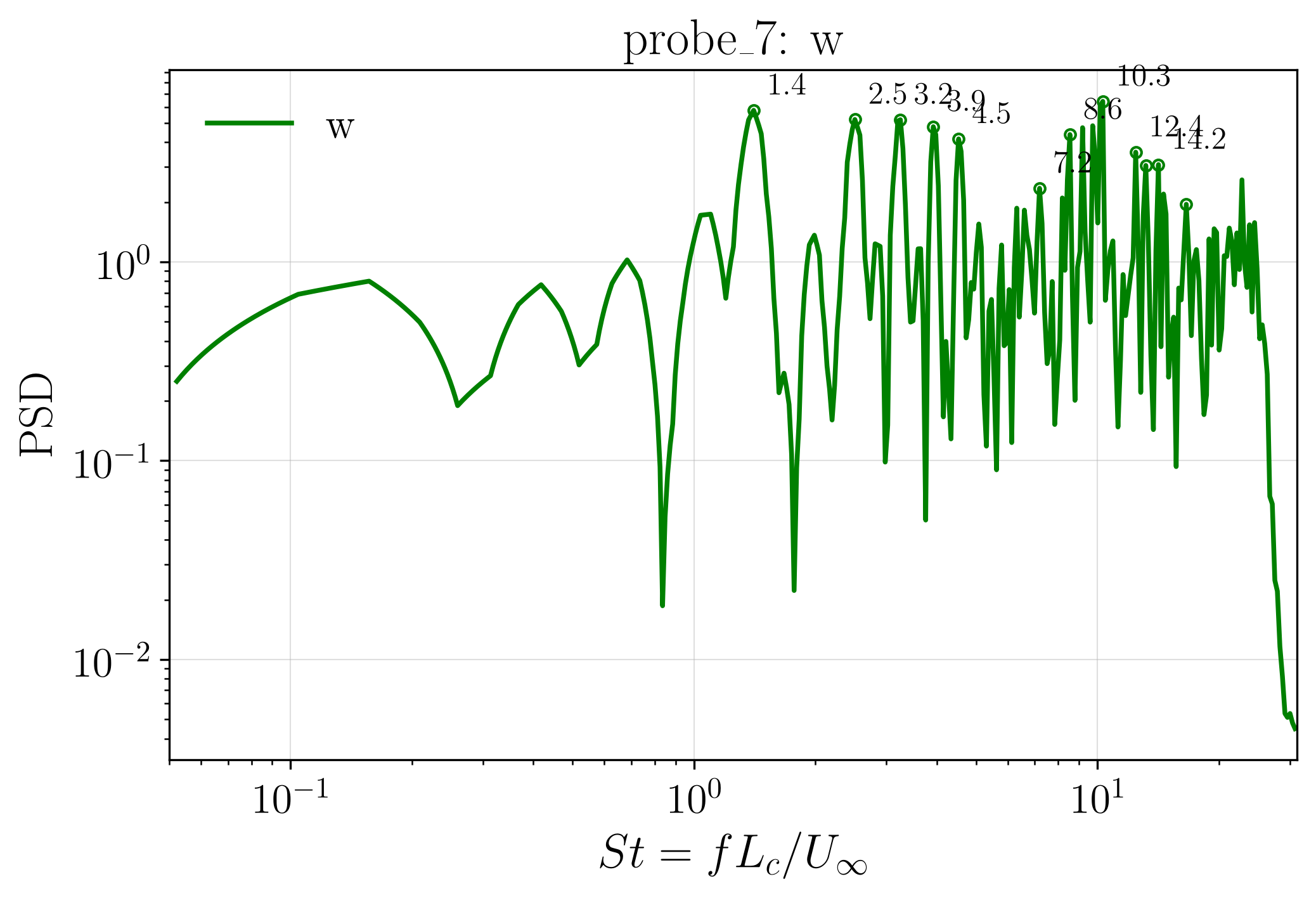}}

    \caption{Spectra of the streamwise and vertical velocities at probes 5, 6 and 7. Rows correspond to probes 5, 6 and 7, respectively; columns correspond to the $u$ and $w$ velocity components.}
    \label{fig:spectra_5_7}
\end{figure}

\begin{figure}
    \centering
    \subfloat[]{\includegraphics[width=0.33\textwidth]{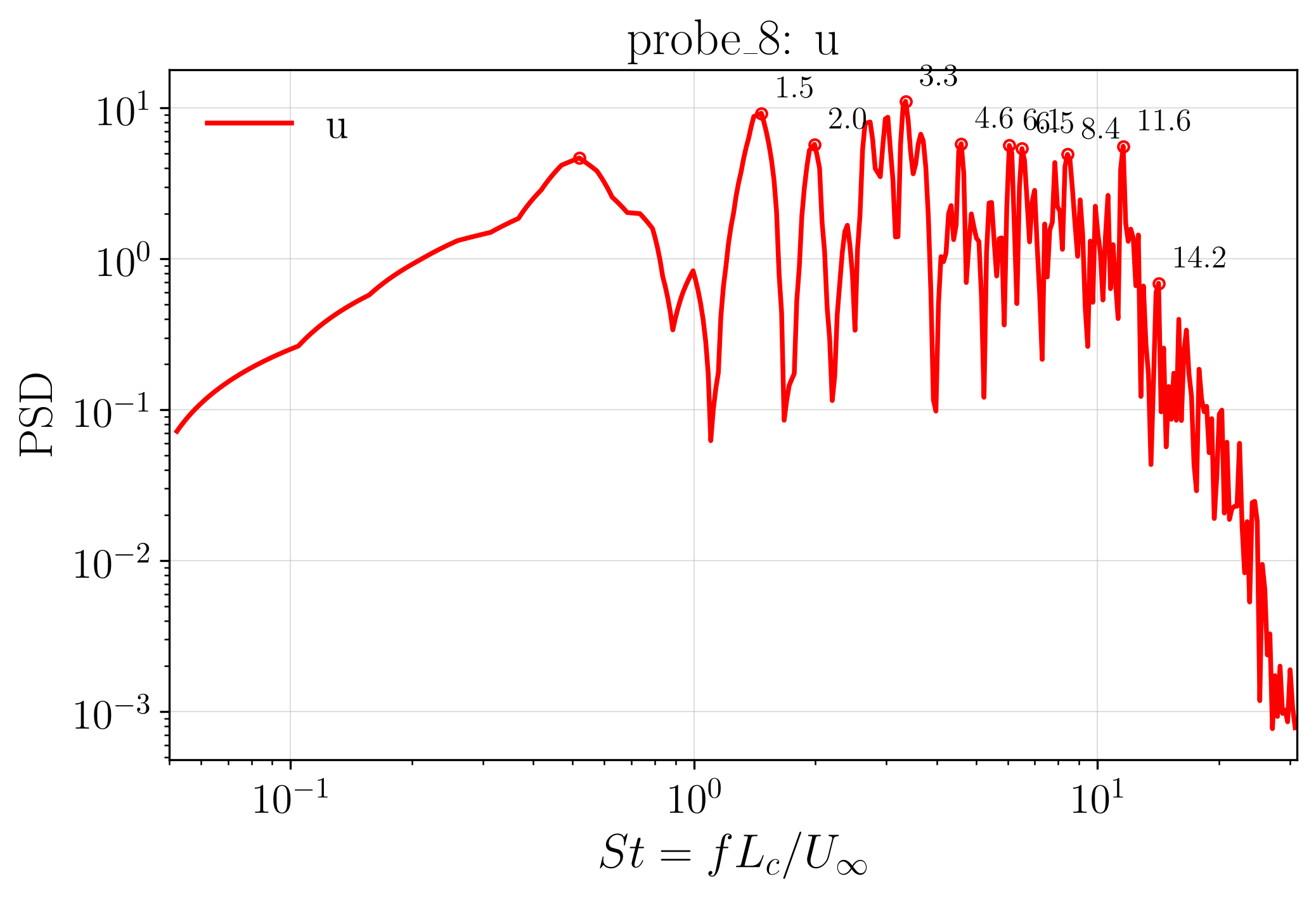}}
    \subfloat[]{\includegraphics[width=0.33\textwidth]{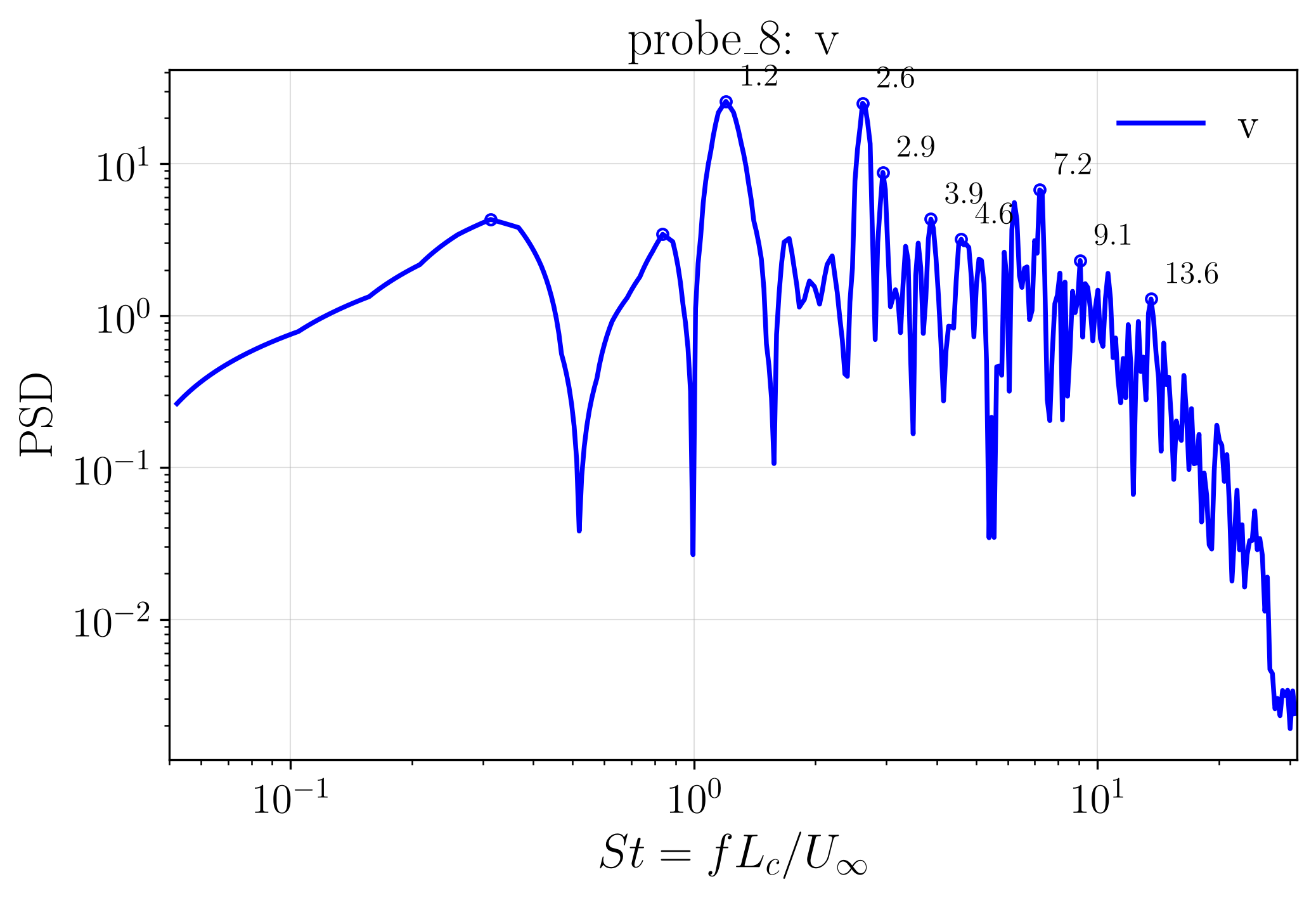}}
    \subfloat[]{\includegraphics[width=0.33\textwidth]{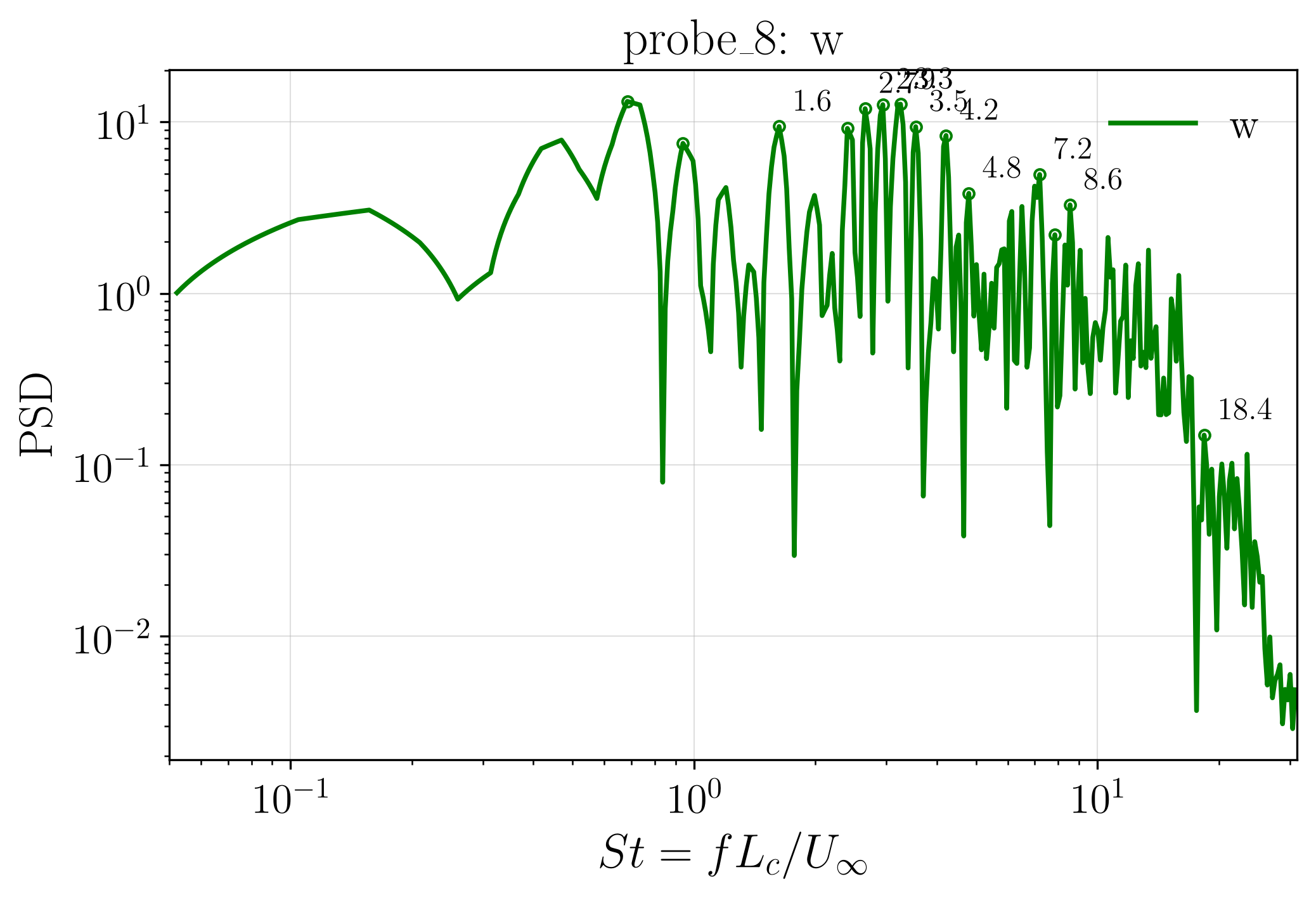}}
    \caption{Spectra of the streamwise (a), cross-stream (b) and vertical (c) velocities at probe 8.}
    \label{fig:spectra_8}
\end{figure}

\begin{figure}
    \centering
    \subfloat[]{\includegraphics[width=0.33\textwidth]{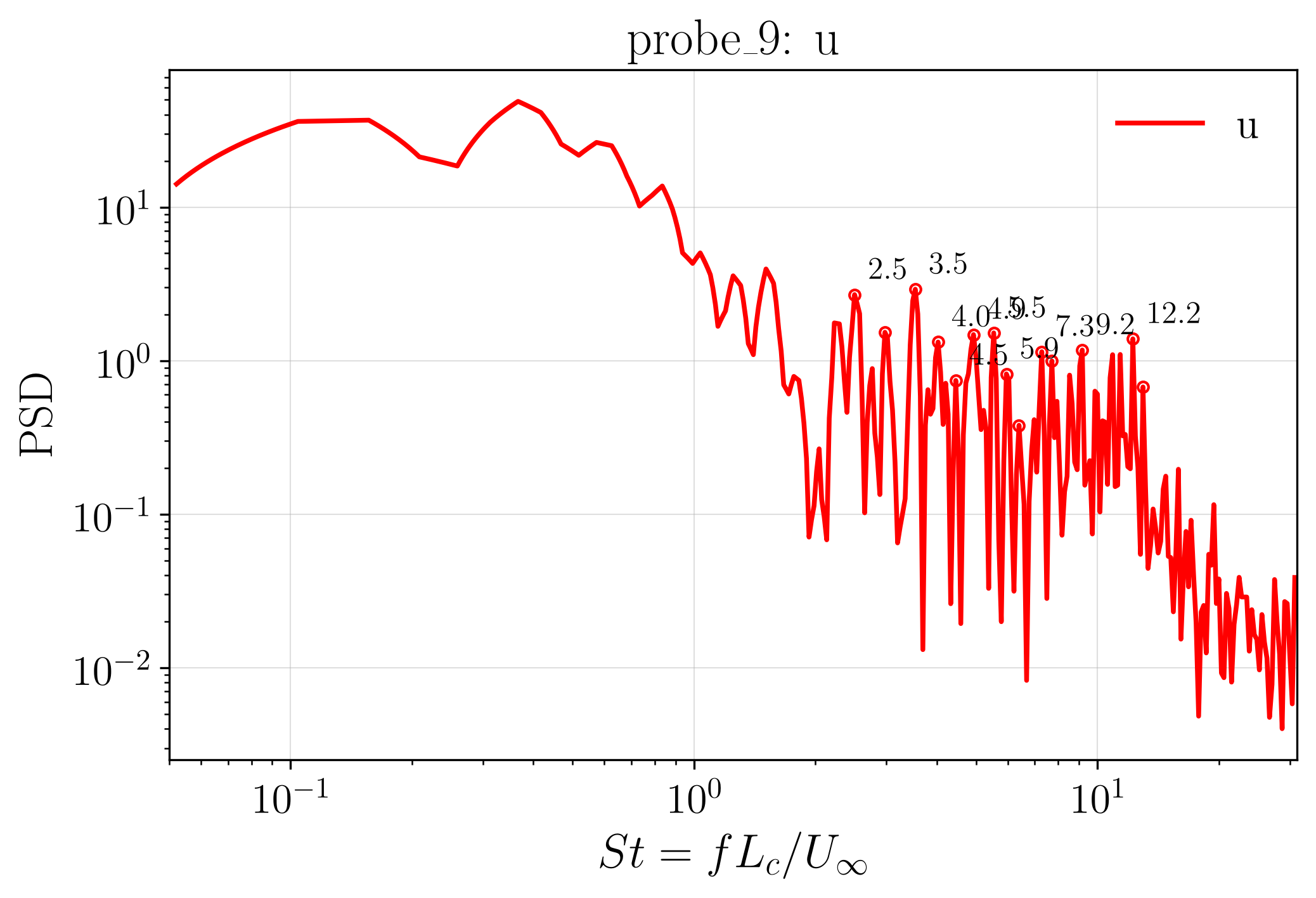}}
    \subfloat[]{\includegraphics[width=0.33\textwidth]{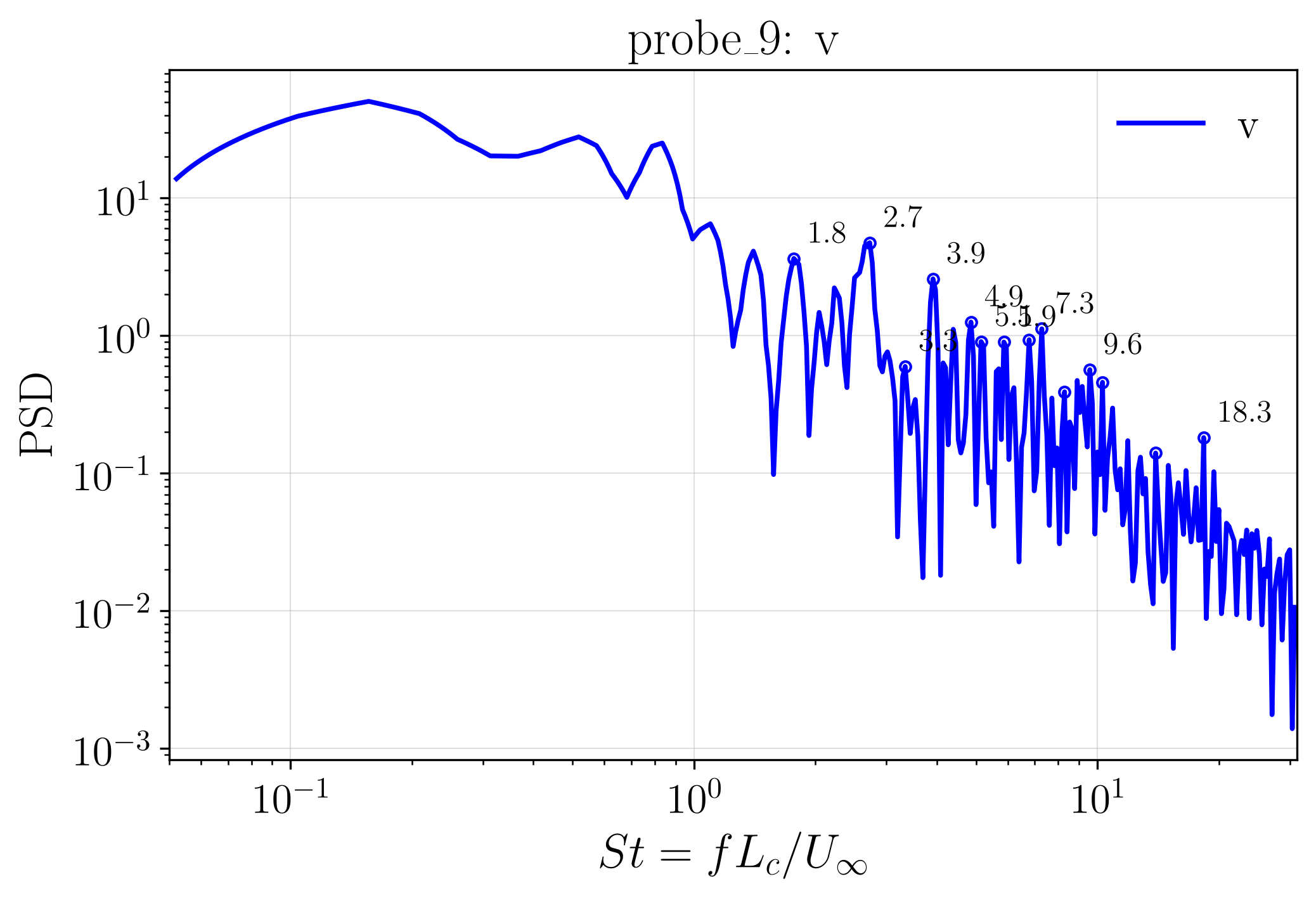}}
    \subfloat[]{\includegraphics[width=0.33\textwidth]{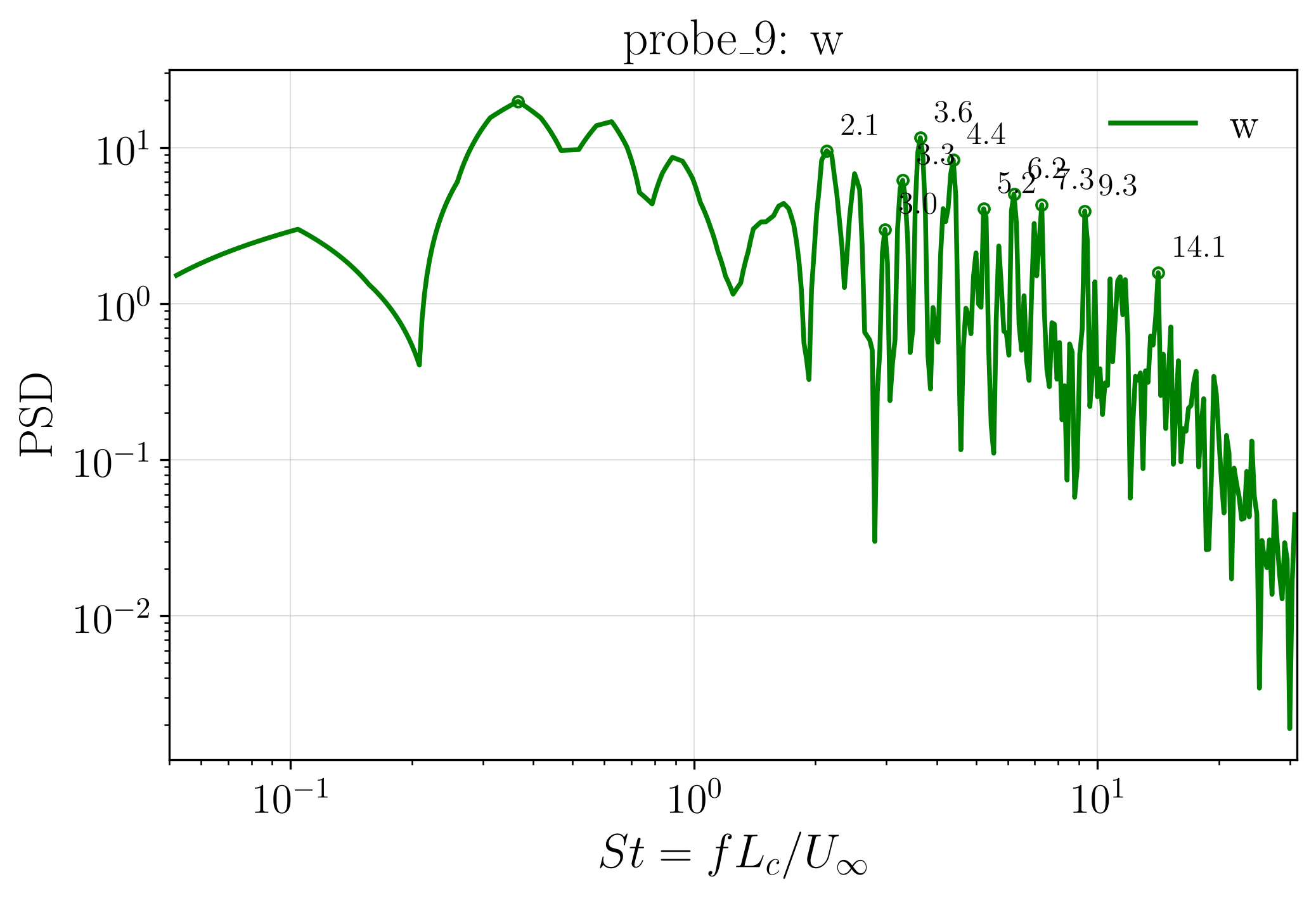}}
    \caption{Spectra of the streamwise (a), cross-stream (b) and vertical (c) velocities at probe 9.}
    \label{fig:spectra_9}
\end{figure}

The velocity spectra at the nine probes are shown in Figures~\ref{fig:spectra_1}--\ref{fig:spectra_9}.

\end{document}